\documentclass[a4paper,12pt]{article}

\usepackage{xcolor}
\definecolor{nosaka}{rgb}{0.0, 0.5, 0.0}
\definecolor{nosaka2}{rgb}{0.6, 0.0, 0.0}
\definecolor{numasawa}{rgb}{0.0, 0.0, 0.6}

\usepackage{listings}
\usepackage{longtable}

\usepackage[utf8]{inputenc}
\usepackage{braket}

\usepackage{url}
\usepackage{amsmath}

\makeatletter
\newcommand{\subscripts}[3]{%
  \@mathmeasure\z@\displaystyle{#2}%
  \global\setbox\@ne\vbox to\ht\z@{}\dp\@ne\dp\z@
  \setbox\tw@\box\@ne
  \@mathmeasure4\displaystyle{\copy\tw@_{#1}}%
  \@mathmeasure6\displaystyle{{#2}_{#3}}%
  \dimen@-\wd6 \advance\dimen@\wd4 \advance\dimen@\wd\z@
  \hbox to\dimen@{}\mathop{\kern-\dimen@\box4\box6}%
}
\makeatother

\usepackage{graphicx}
\usepackage{mathrsfs}

\usepackage{tikz}
\usepackage{bm}
\usepackage{arydshln}
\usepackage{color}
\usepackage{ulem}
\usepackage{amssymb}

\usepackage{framed}
\usepackage{here}
\usepackage{comment}

\newcommand{\be}{\begin{equation}}
\newcommand{\ee}{\end{equation}}
\def\ba#1\ea{\begin{align}#1\end{align}}
\newcommand{\f}{\frac}
\newcommand{\s}{\sqrt}

\DeclareMathOperator{\Tr}{Tr}

\usepackage{geometry}
\numberwithin{equation}{section}
\allowdisplaybreaks

\begin{document}

\newcommand{\hiduke}[1]{\hspace{\fill}{\small [{#1}]}}
\newcommand{\aff}[1]{${}^{#1}$}
\renewcommand{\thefootnote}{\fnsymbol{footnote}}

\begin{titlepage}
\begin{flushright}
% {\footnotesize preprint SISSA 23/2020/FISI,\, MIT-CTP/5242}
\end{flushright}
\begin{center}
{\Large\bf
On SYK traversable wormhole with imperfectly correlated disorders
% Chaos exponents of SYK traversable wormholes
}\\
\bigskip\bigskip
\bigskip\bigskip
{\large Tomoki Nosaka\footnote{\tt nosaka@yukawa.kyoto-u.ac.jp}}\aff{1}
{\large and Tokiro Numasawa\footnote{\tt numasawa@issp.u-tokyo.ac.jp}}\aff{2}\\
\bigskip\bigskip
\aff{1}: {\small
\it
Kavli Institute for Theoretical Sciences, University of Chinese Academy of Sciences,\\
Beijing, China 100190
}\\
% \aff{2}: {\small
% \it RIKEN Interdisciplinary Theoretical and Mathematical Sciences (iTHEMS), Wako, Saitama 351-0198, Japan
% }\\
% \aff{1}: {\small
% \it INFN Sezione di Trieste, Via Valerio 2, 34127 Trieste, Italy
% }\\
% \aff{2}: {\small
% \it International School for Advanced Studies (SISSA), Via Bonomea 265, 34136 Trieste, Italy
% }\\
\bigskip
\aff{2}: {\small
\it Institute for Solid State Physics, University of Tokyo, Kashiwa 277-8581, Japan
}\\
\end{center}
\bigskip
\bigskip
\begin{abstract}
In this paper we study the phase structure of two Sachdev-Ye-Kitaev models ($L$-system and $R$-system) coupled by a simple interaction, with imperfectly correlated disorder.
When the disorder of the two systems are perfectly correlated, $J_{i_1\cdots i_q}^{(L)}=J_{i_1\cdots i_q}^{(R)}$, this model is known to exhibit a phase transition at a finite temperature between the two-black hole phase at high-temperature and the traversable wormhole phase at low temperature.
We find that, as the correlation $\langle J_{i_1\cdots i_q}^{(L)}J_{i_1\cdots i_q}^{(R)}\rangle$ is decreased, the critical temperature becomes lower.
At the same time, the transmission between $L$-system and $R$-system in the low-temperature phase becomes more suppressed, while the chaos exponent of the whole system becomes larger.
Interestingly we also observe that when the correlation is smaller than some $q$-dependent critical value the phase transition completely disappears in the entire parameter space.
At zero temperature, the energy gap becomes larger as we decrease the correlation.
We also use a generalized thermofield double state
% the ``partially entangled thermal state'' (PETS)
as a variational state.
Interestingly, this state coincide with the ground state in the large $q$ limit.

\end{abstract}

\bigskip\bigskip\bigskip

\end{titlepage}

\renewcommand{\thefootnote}{\arabic{footnote}}
\setcounter{footnote}{0}

\tableofcontents

% \newpage

\section{Introduction and Summary}
The Sachdev-Ye-Kitaev (SYK) model \cite{PhysRevLett.70.3339,KitaevTalk} is a useful model to study various aspects of strongly coupled many body systems.
% is an useful model to study various aspects of strongly coupled many body systems.
Moreover, the SYK model is also a toy model of quantum blacks hole  \cite{Maldacena:2016upp}.
% Moreover, the SYK model also plays a toy model   of  quantum blacks hole  \cite{Maldacena:2016upp}. 
%\cite{Hayden:2007cs,Sekino:2008he}
Both theories show the same pattern of the conformal symmetry breaking at low energy and described by the so-called Schwarzian action. 
% Both theories show the same pattern of the conformal symmetry breaking at low energy and described by so called the Schwarzian action. 
This gives a concrete connection between two theories.

Related to black holes, the SYK model also plays an important role to understand wormhole configurations in gravity.
Two kind of wormholes play important roles in the literature.
The first one is the spacial wormhole.
Spacial wormholes are related to entanglement \cite{Israel:1976ur,Maldacena:2001kr,VanRaamsdonk:2010pw}.
% Spacial wormholes are related to entanglement. \cite{Israel:1976ur,Maldacena:2001kr,VanRaamsdonk:2010pw}.
In the context of AdS/CFT correspondence, the area of the wormhole connecting distant regions corresponds to entanglement entropy in CFT \cite{Ryu:2006bv,Hubeny:2007xt,Engelhardt:2014gca}.
Moreover, it is expected that spacial wormholes are dual to entanglement between CFTs \cite{Maldacena:2001kr,VanRaamsdonk:2010pw} and the spacetime is built from entanglement \cite{Maldacena:2013xja,Swingle:2012wq}.
The other kind of wormhole is the spacetime wormholes or Euclidean wormholes.
These are kinds of gravitational instanton and these spacetime wormholes are related to random couplings \cite{Coleman:1988cy,Giddings:1988cx,Giddings:1987cg}.
The SYK model is a model with random couplings and the wormhole configurations associated to a pattern of  random couplings are studied \cite{Saad:2021rcu,Saad:2021uzi}.
These Euclidean wormholes also appears in the context of calculation of (R\'enyi) entanglement entropy.
These are known as replica wormholes \cite{Almheiri:2019qdq,Penington:2019kki} and play important roles in the context of black hole information problems.\footnote{
Recent developments on various aspects of the wormhole geometries are also summarized in a review article \cite{Kundu:2021nwp}.
}

Usually we assume that the random couplings of copies of SYK models have exactly the same couplings, i.e.~in each realization of the random couplings we use the same realization for all of the copies.
% Usually we assume that the random couplings of copies of SYK models have exactly the same couplings i.e. in each realization of the random couplings we use the same realization in each copy.
This is natural setup when we use the replica methods to study the R\'enyi entropy for example.
However, to study entangled states we can also consider the situation where the copies of SYK models have different random couplings.
For example, if we simulate the SYK models on quantum computers it maybe natural to consider different realization because of errors etc.
% For example, if we simulate the SYK models on quantum computers it maybe natural to consider different realization because of errors etc.
Furthermore, we can also consider entangled black holes between different theories. For example, ``Janus black holes'', i.e., two side black holes that are dual to entangled state with different coupling constants, are studied in \cite{Bak:2007jm,Bak:2011ga,Nakaguchi:2014eiu,Goel:2018ubv,Bak:2007qw}.

Motivated by the above questions, we study the  coupled SYK models where the two SYK models have different realization of random couplings.
% \footnote{
% One may also consider different modifications of the random couplings such as imbalanced rescalings $J_{i_1\cdots i_q}^{(R)}=cJ_{i_1\cdots i_q}^{(L)}$ with $c\neq 1$ \cite{Haenel:2021fye} or the sparse couplings \cite{Garcia-Garcia:2020cdo,Xu:2020shn,Caceres:2021nsa} instead of full $J^{(a)}_{i_1\cdots i_q}$.
% Although we do not investigate in this paper, it would be interesting to study the role of the correlation between the random couplings on two sides also in such generalized setups.
% }
% To describe the different random couplings between two sides, we consider the model where the correlation of random couplings are imperfect between two copies of the SYK models.
The two-coupled SYK model was first considered in \cite{Maldacena:2018lmt}, with the two random couplings perfectly correlated, as the holographic dual of the global $\text{AdS}_2$ spacetime (eternal traversable wormhole), which is the static version of the wormhole formation process by the bulk non-local interaction \cite{Gao:2016bin,Maldacena:2016upp}.
In the setup of \cite{Maldacena:2018lmt}, in order the wormhole to become traversable, or in the SYK side the quantum teleportation to be successful, it is crucial for the state of the whole system to be the thermofield double state
% \textcolor{red}{[to be confirmed (find ref.): a noisy state just having a large entanglement does not suffice]}
\cite{Gao:2018yzk}.
It was found that the ground state of the coupled SYK model is close to the thermofield double state \cite{Maldacena:2018lmt,Garcia-Garcia:2019poj,Alet:2020ehp}, which ensures that the low temperature dynamics of the model can be related to the traversable wormhole.
Indeed, the coupled SYK model in the canonical ensemble exhibits a Hawking-Page-like phase transition between the high temperature phase dual to the two-sided $\text{AdS}_2$ black hole and the low temperature phase dual to the global $\text{AdS}_2$.

When the couplings of the two systems are different, it is not clear how to interpret the entanglement structure of the ground state and whether the system is dual to the traversable wormhole at low temperature or not.
It was also found in \cite{Nosaka:2019tcx,Nosaka:2020nuk} that the coupled SYK model does not exhibit a phase transition when the two random couplings are completely independent.
Hence the correlation of the couplings is indeed important for the wormhole formation, and it is a non-trivial question how much correlation would be necessary for the wormhole to be formed.

More concretely, we consider the model where the two random couplings $J_{i_1\cdots i_q}^{(L)}$ and $J_{i_1\cdots i_q}^{(R)}$ obey the same Gaussian distribution while the two realizations are not completely identical, which we quantify by $\langle J_{i_1\cdots i_q}^{(L)}J_{i_1\cdots i_q}^{(R)}\rangle$ normalized by $\langle (J_{i_1\cdots i_q}^{(L)})^2\rangle$ ($=\langle (J_{i_1\cdots i_q}^{(R)})^2\rangle$).
By analyzing this model in the large $N$ limit, we find the following results:
\begin{itemize}
\item [(i)] As the correlation between the two random couplings is decreased, the critical temperature for the Hawking-Page-like phase transition becomes lower.
% This result can also be rephrased that the strength of the LR coupling required for reaching the wormhole phase at fixed temperature becomes higher.
This result can also be rephrased that the strength of the LR coupling required for reaching the wormhole phase at fixed temperature becomes higher, hence both the correlation of random couplings and direct LR coupling make it easy to create a wormhole configuration.
This is also consistent with the fact that the wormhole phase exists even without direct LR coupling if the two random couplings are {\it supercorrelated}, $\langle J_{i_1\cdots i_q}^{(L)}J^{(R)}_{i_1\cdots i_q}\rangle > \langle (J_{i_1\cdots i_q}^{(L)})^2\rangle$ \cite{Garcia-Garcia:2021elz,Garcia-Garcia:2022xsh,Garcia-Garcia:2022zmo,Sorokhaibam:2020ilg,Cai:2022wmd}.
\item [(ii)] We also observe that the phase transition completely disappears when the correlation between $J_{i_1\cdots i_q}^{(L)}$ and $J_{i_1\cdots i_q}^{(R)}$ is smaller than some non-zero finite value.
Technically this occurs in the following way.
Already in the original setup where the two random couplings are identical, there are no phase transition when the LR coupling is larger than some critical value: when the LR coupling is too large, even at a high temperature the dynamics is approximately same as that for the model without SYK interaction which does not exhibit phase transition.
We find that this critical value of the LR coupling becomes smaller as the correlation between $J_{i_1\cdots i_q}^{(L)}$ and $J_{i_1\cdots i_q}^{(R)}$ is decreased, and reaches zero before the two random coupling become completely independent.
At large $q$ limit, we also estimate when the phase transition disappears as we decrease the correlation of random couplings between two sides.
\item [(iii)] We also evaluate the transmission amplitude $T_{LR}$ between the $L$-site and the $R$-site in the low temperature wormhole phase, and found that for the same temperature and the strength of the LR coupling, $T_{LR}$ becomes smaller as the correlation between $J_{i_1\cdots i_q}^{(L)}$ and $J_{i_1\cdots i_q}^{(R)}$ is decreased.
On the other hand, the chaos exponent $\lambda_L$, which is non-zero even in the wormhole phase, becomes larger as the correlation of the random couplings is decreased.
% From these results it would be reasonable to interpret $\lambda_L$ as a measure of the speed that a simple initial excitation spread within single site, which would be suppressed if the excitation leaks to the other site.
These two results are reasonable if $\lambda_L$ of this model measures the speed that a simple initial excitation spread within single site, which would be suppressed if the excitation leaks to the other site.
\end{itemize}
We observe the results (i-iii) numerically for $q=4$, and also confirm the results (i,ii) analytically in the large $q$ limit.

The organization of this paper is as follows.
% In section \ref{sec_JLneqJRmodel}, we clarify the model we study in this paper.
In section \ref{sec_JLneqJRmodel}, we clarify the model we study in this paper, and write the partition function in the large $N$ limit with the bilocal field formalism.
By using the bilocal field formalism, we analyze how the large $N$ phase structure and various properties of each phases are modified by the imperfect correlation of disorders for finite $q$ in section \ref{sec_finiteqlargeN} and in the large $q$ limit in section \ref{sec_largeq}.
In section \ref{sec_structureofgs} we study the structure of the ground state of the coupled system for $\langle J^{(L)}_{i_1\cdots i_q}J^{(R)}_{i_1\cdots i_q}\rangle < \langle (J_{i_1\cdots i_q}^{(L)})^2\rangle$ which generalizes the structure of the thermofield double state for $J^{(L)}_{i_1\cdots i_q}=J^{(R)}_{i_1\cdots i_q}$.
In section \ref{sec_discuss} we summarize the our results and list possible future directions of research.
Some technical details of the calculation in the large $q$ limit are collected in appendix \ref{app_largeqderivation}.

\section{$J_{i_1\cdots i_q}^{(L)} \neq J_{i_1\cdots i_q}^{(R)}$ model}
\label{sec_JLneqJRmodel}
In this paper we consider a one-dimensional quantum mechanics with the following disordered Hamiltonian:\footnote{
Although we do not investigate in this paper, it would be also interesting to study how the correlation between the disorders affects in the two-coupled SYK model with Dirac fermions (so called complex SYK model) \cite{Sachdev:2015efa}
% 1506.05111,
% cond-mat/9212030
whose phase structure was analyzed in \cite{Garcia-Garcia:2020vyr,Rathi:2021mla} when the random couplings on two sites are set to be identical.
}
\begin{align}
H&=H_{SYK}^{(L)}+H_{SYK}^{(R)}+\mu H_{int},
% H&=H_{SYK}^{(L)}+H_{SYK}^{(R)}+i\mu\sum_{i=1}^N\psi_i^L\psi_i^R,
\label{220221_HMQwithJLJRindepmodel}
\end{align}
where
\begin{align}
&H_{SYK}^{(L)}=i^{\frac{q}{2}}\sum_{i_1<i_2<\cdots<i_q}^N
J_{i_1i_2\cdots i_q}^{(L)}
\psi_{i_1}^{L}
\psi_{i_2}^{L}
\cdots \psi_{i_q}^{L},\quad
H_{SYK}^{(R)}=i^{\frac{q}{2}}(-1)^{\frac{q}{2}}\sum_{i_1<i_2<\cdots<i_q}^N
J_{i_1i_2\cdots i_q}^{(R)}
\psi_{i_1}^{R}
\psi_{i_2}^{R}
\cdots \psi_{i_q}^{R},\nonumber \\
&H_{int}=i\sum_{i=1}^N\psi_i^L\psi_i^R,\label{Hint}
\end{align}
$\{\psi_i^a,\psi_j^b\}=\delta_{ab}\delta_{ij}$ ($a=L,R$), and $J_{i_1i_2\cdots i_q}^{(a)}$ are random couplings drawn from the Gaussian distribution with the following mean and variance:
\begin{align}
\langle J_{i_1i_2\cdots i_q}^{(a)}\rangle=0,\quad
\langle J_{i_1i_2\cdots i_q}^{(a)}J_{j_1j_2\cdots j_q}^{(a)}\rangle=\frac{{\cal J}^2\cdot 2^{q-1}(q-1)!}{q\cdot N^{q-1}}
\delta_{i_1j_1}
\delta_{i_2j_2}
\cdots
\delta_{i_qj_q}.
\label{randomcoupling}
\end{align}
Here $J_{i_1i_2\cdots i_q}^{(a)}$ are drawn independently for different set of subscripts $i_1i_2\cdots i_q$.
On the other hand, with respect to $a=L,R$, we consider the case where $J_{i_1i_2\cdots i_q}^{(L)}$ and $J_{i_1i_2\cdots i_q}^{(R)}$ are imperfectly correlated with each other:
\begin{align}
\langle J_{i_1i_2\cdots i_q}^{(L)}J_{j_1j_2\cdots j_q}^{(R)}\rangle=\frac{{\tilde {\cal J}}^2\cdot 2^{q-1}(q-1)!}{q\cdot N^{q-1}}
\delta_{i_1j_1}
\delta_{i_2j_2}
\cdots
\delta_{i_qj_q},
\end{align}
with $0\le {\tilde {\cal J}}\le {\cal J}$.\footnote{
One may also consider the case ${\tilde {\cal J}}>{\cal J}$, where the Hamiltonian is non-Hermitian \cite{Garcia-Garcia:2021elz,Garcia-Garcia:2022xsh,Garcia-Garcia:2022zmo,Sorokhaibam:2020ilg,Cai:2022wmd}.
}
This partial correlation can be realized by drawing two independent random variables $J_{i_1i_2\cdots i_q}^{(\alpha)}$ ($\alpha=1,2$) from the same distribution as $J_{i_1i_2\cdots i_q}^{(L)}$ \eqref{randomcoupling} and writing $J_{i_1i_2\cdots i_q}^{(a)}$ as
\begin{align}
J_{i_1i_2\cdots i_q}^{(L)}&=
J_{i_1i_2\cdots i_q}^{(1)}\sqrt{\frac{1+{\tilde {\cal J}}^2/{\cal J}^2}{2}}
+J_{i_1i_2\cdots i_q}^{(2)}\sqrt{\frac{1-{\tilde {\cal J}}^2/{\cal J}^2}{2}},\nonumber \\
J_{i_1i_2\cdots i_q}^{(R)}&=
J_{i_1i_2\cdots i_q}^{(1)}\sqrt{\frac{1+{\tilde {\cal J}}^2/{\cal J}^2}{2}}
-J_{i_1i_2\cdots i_q}^{(2)}\sqrt{\frac{1-{\tilde {\cal J}}^2/{\cal J}^2}{2}}.
\end{align}
Consider the Euclidean partition function (annealed average) of this theory at finite temperature $\beta^{-1}$:
\begin{align}
Z(\beta)&=\left\langle \int {\cal D}\psi_i^{(a)}(\tau) \text{exp}\biggl[-\int d\tau (\sum_{a=L,R}\sum_{i=1}^N\frac{1}{2}\psi_i^a\partial_\tau\psi_i^a+H)\biggr]\right\rangle_{J_{i_1i_2\cdots i_a}^{(\alpha)}}\nonumber \\
&={\cal N}^{-1}\int \left(\prod_{\alpha=1,2}\prod_{i_1<i_2<\cdots<i_q}^NdJ_{i_1i_2\cdots i_q}^{(\alpha)}e^{-\frac{1}{2}\frac{q\cdot N^{q-1}}{{\cal J}^2\cdot 2^{q-1}(q-1)!}(J_{i_1i_2\cdots i_q}^{(\alpha)})^2}\right)\nonumber \\
&\quad \int {\cal D}\psi_i^{(a)}(\tau) \text{exp}\biggl[-\int d\tau (\sum_{a=L,R}\sum_{i=1}^N\frac{1}{2}\psi_i^a\partial_\tau\psi_i^a+H)\biggr].
\end{align}
After the same manipulation as \cite{Nosaka:2020nuk} we can rewrite the partition function in terms of the bilocal fields $G_{ab}(\tau,\tau')$ and $\Sigma_{ab}(\tau,\tau')$ as
\begin{align}
Z(\beta)=\int {\cal D}G_{ab}(\tau,\tau'){\cal D}\Sigma_{ab}(\tau,\tau')e^{-S_E}.
\label{220502ZbetainGSigma}
\end{align}
% In this paper we consider a one-dimensional quantum mechanics with the following disordered Hamiltonian:
% \begin{align}
% H&=i^{\frac{q}{2}}\sum_{i_1<i_2<\cdots<i_q}^N\left[
% J_{i_1i_2\cdots i_q}^{(1)}(
% \psi_{i_1}^{L}
% \psi_{i_2}^{L}
% \cdots \psi_{i_q}^{L}
% +(-1)^{\frac{q}{2}}
% \psi_{i_1}^{R}
% \psi_{i_2}^{R}
% \cdots \psi_{i_q}^{R}
% )\sqrt{\frac{1+{\widetilde {\cal J}}^2/{\cal J}^2}{2}}\right.\nonumber \\
% &\quad \left.+
% J_{i_1i_2\cdots i_q}^{(2)}(
% \psi_{i_1}^{L}
% \psi_{i_2}^{L}
% \cdots \psi_{i_q}^{L}
% -(-1)^{\frac{q}{2}}
% \psi_{i_1}^{R}
% \psi_{i_2}^{R}
% \cdots \psi_{i_q}^{R}
% )\sqrt{\frac{1-{\widetilde {\cal J}}^2/{\cal J}^2}{2}}
% \right]
% +i\mu\sum_{i=1}^N\psi_i^L\psi_i^R,
% \end{align}
% where
% \begin{align}
% \{\psi_i^a,\psi_j^b\}=2\delta_{ab}\delta_{ij}
% \langle J_{i_1i_2\cdots i_q}^{(\alpha)}\rangle=0
% \langle J_{i_1i_2\cdots i_q}^{(1)} J_{i_1i_2\cdots i_q}^{(2)} \rangle=0
% \langle (J_{i_1i_2\cdots i_q}^{(\alpha)})^2\rangle=\frac{{\cal J}^2\cdot 2^{q-1}(q-1)!}{q\cdot N^{q-1}}
% \end{align}
%%%%%%%%%%%%%%%%%%%%%%%%%%%%%%%%%%%%%%%%%%%%%%%%%%%%%%%%%%%%
\begin{figure}
\begin{center}
\includegraphics[width=12cm]{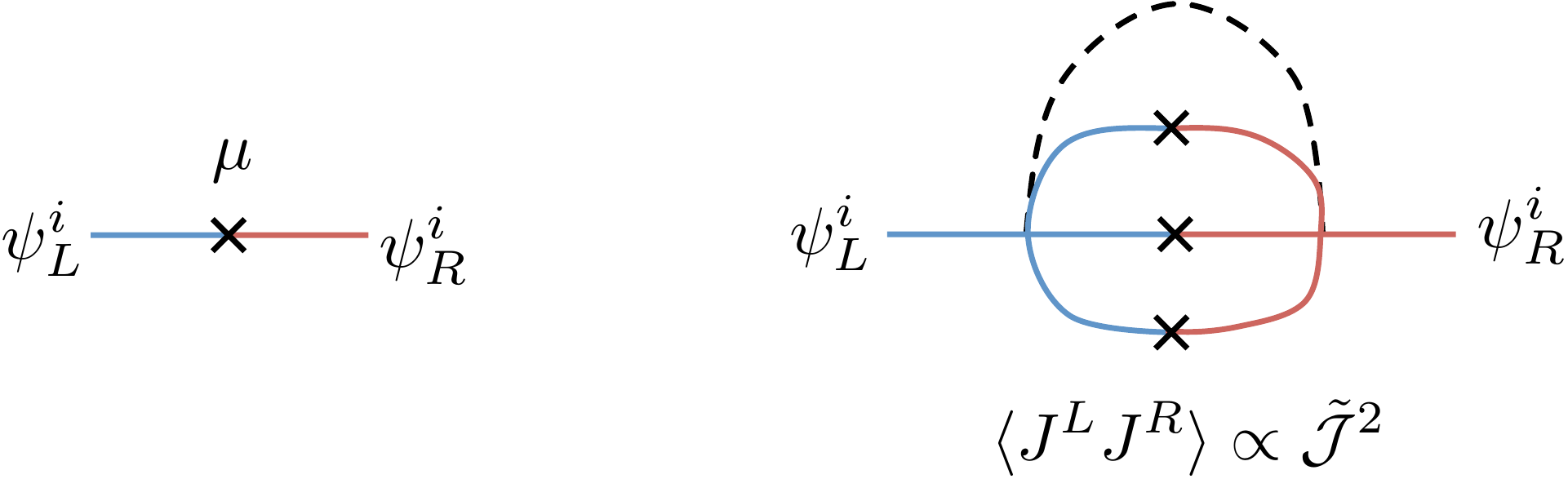}
\caption{
{\bf Left:} A diagram that is not affected by the random couplings.
{\bf Right:} A typical diagram that is reduced when we decrease the correlation of left and right random couplings.
}
\label{fig:DiagramLR}
\end{center}
\end{figure}

The effective action is 
\ba
-S_E/N &= \log \text{Pf}\Bigl(-\delta(\tau-\tau')\partial_{\tau'} \delta_{ab} +\frac{\Sigma_{ab}(\tau,\tau')-\Sigma_{ba}(\tau',\tau)}{2}\Bigr)\nonumber \\
&- \f{1}{2} \int d\tau d\tau' \sum_{a,b} \Big[ \Sigma_{ab}(\tau,\tau')G_{ab}(\tau,\tau')  - s_{ab}\f{\mathcal{J}_{ab}^2}{2q^2} [2G_{ab}(\tau,\tau')]^q\Big] \notag \\
&+ \f{i\mu}{2} \int d\tau [-G_{LR}(\tau,\tau) + G_{RL}(\tau,\tau)].
\label{220201_ZinGSigma}
\ea
% \ba
% -S_E/N &= \log \text{Pf}(-\partial_\tau \delta_{ab} - \Sigma_{ab}) - \f{1}{2} \int d\tau_1 d\tau_2 \sum_{a,b} \Big[ \Sigma_{ab}(\tau_1,\tau_2)G_{ab}(\tau_1,\tau_2)  - s_{ab}\f{\mathcal{J}_{ab}^2}{2q^2} [2G_{ab}(\tau_1,\tau_2)]^q\Big] \notag \\
% &+ \f{i\mu}{2} \int d\tau_1 [-G_{LR}(\tau_1,\tau_1) + G_{RL}(\tau_1,\tau_1)].
% \label{220201_ZinGSigma}
% \ea
Here $s_{LL}=s_{RR}=1$, $s_{LR}=s_{RL}=(-1)^{\frac{q}{2}}$ and $\mathcal{J}_{LL} = \mathcal{J}_{RR} = \mathcal{J}$, $\mathcal{J}_{LR} = \mathcal{J}_{RL} = \tilde{\mathcal{J}}$.
The Schwinger-Dyson equations $\frac{\delta S_E}{\delta G_{ab}(\tau,\tau')}=\frac{\delta S_E}{\delta \Sigma_{ab}(\tau,\tau')}=0$ are
\ba
&\partial_{\tau}G_{ab}(\tau,\tau') - \sum_c\int d\tau''\frac{\Sigma_{ac}(\tau,\tau'')-\Sigma_{ca}(\tau'',\tau)}{2}G_{cb}(\tau'',\tau') = \delta_{ab}\delta(\tau-\tau'), \notag \\
&\Sigma_{ab}(\tau,\tau')=\frac{s_{ab}{\cal J}_{ab}^2}{q}(2G_{ab}(\tau,\tau'))^{q-1}+i\mu(-\delta_{aL}\delta_{bR}+\delta_{aR}\delta_{bL})\delta(\tau-\tau').
% &\partial_{\tau_1}G_{LL} - \Sigma_{LL} * G_{LL} - \Sigma_{LR} *G_{RL} = \delta, \notag \\
% & \partial _{\tau_1}G_{LR} - \Sigma_{LL}* G_{LR}- \Sigma_{LR} *G_{RR} = 0, \notag \\
% & \Sigma_{LL} = \f{\mathcal{J}^2}{q} (2G_{LL})^{q-1}, \qquad  \Sigma_{LR} = (-1)^{\f{q}{2}} \f{\tilde{\mathcal{J}}^2}{q}(2G_{LR})^{q-1}- i\mu \delta(\tau_{12}),
\label{220201_EuclideanSDeq}
\ea
From the two equations it follows that
\begin{align}
G_{ab}(\tau,\tau')=-G_{ba}(\tau',\tau).
% \Sigma_{ab}(\tau,\tau')=-\Sigma_{ba}(\tau',\tau).
\label{antisym}
\end{align}
% with which we can write the first equations in \eqref{220201_EuclideanSDeq} simply as
% \begin{align}
% &\partial_{\tau}G_{ab}(\tau,\tau') - \sum_c\int d\tau'' \Sigma_{ac}(\tau,\tau'')G_{cb}(\tau'',\tau') = \delta_{ab}\delta(\tau-\tau').
% \label{220201_EuclideanSDeqsimplified}
% \end{align}
By identifying $G_{ab}(\tau,\tau')$ with ($\alpha=1,2$)
\begin{align}
G_{ab}(\tau,\tau')&=\frac{1}{N}\sum_{i=1}^N\langle {\cal T}\psi_i^a(\tau)\psi_i^b(\tau')\rangle_\beta\\
&=
\begin{cases}
\frac{1}{N\langle \text{tr}e^{-\beta H}\rangle_{J_{i_1\cdots i_q}^{(\alpha)}}}\sum_{i=1}^N\langle \text{tr}e^{\tau H}\psi_i^a e^{-H(\tau-\tau')}\psi_i^be^{-(\tau'+\beta) H}\rangle_{J_{i_1\cdots i_q}^\alpha},\quad (\tau>\tau')\\
-\frac{1}{N\langle \text{tr}e^{-\beta H}\rangle_{J_{i_1\cdots i_q}^{(\alpha)}}}\sum_{i=1}^N\langle \text{tr}e^{\tau' H}\psi_i^b e^{-H(\tau'-\tau)}\psi_i^be^{-(\tau+\beta) H}\rangle_{J_{i_1\cdots i_q}^\alpha},\quad (\tau<\tau')
\end{cases},
\label{Gabvsoperator}
\end{align}
we also find that $G_{ab}(\tau,\tau')$ obeys the following conditions
\footnote{
These argument can be generalized to complex $\tau,\tau'$ and we obtain $G_{ab}(u_1,u_2)^*=-G_{ab}(-u_1^*,-u_2^*)$, which we will use later in section \ref{realtime}.
}
\begin{align}
G_{ab}(\tau,\tau')^*=-G_{ab}(-\tau,-\tau'),\quad
G_{ab}(\tau+\beta,\tau')=-G_{ab}(\tau,\tau').\label{ccandKMS}
\end{align}
From \eqref{Gabvsoperator} and \eqref{220201_EuclideanSDeq} it also follows that $G_{ab}(\tau,\tau'),\Sigma_{ab}(\tau,\tau')$ depends on $\tau,\tau'$ only through $\tau-\tau'$, hence we may denote $G_{ab}(\tau,\tau')$ and $\Sigma_{ab}(\tau,\tau')$ respectively as $G_{ab}(\tau-\tau'),\Sigma_{ab}(\tau-\tau')$.
Taking these into account, the Schwinger-Dyson equations \eqref{220201_EuclideanSDeq} and the symmetry properties \eqref{antisym},\eqref{ccandKMS} are written in a simpler way as
\begin{align}
&\partial_\tau G_{ab}(\tau)-\sum_c\int d\tau'\Sigma_{ac}(\tau-\tau')G_{cb}(\tau')=\delta_{ab}\delta(\tau),\nonumber \\
&\Sigma_{ab}(\tau)=\frac{s_{ab}{\cal J}_{ab}^2}{q}(2G_{ab}(\tau))^{q-1}+i\mu(-\delta_{aL}\delta_{bR}+\delta_{aR}\delta_{bL})\delta(\tau),\label{simplifiedSD} \\
&G_{ab}(\tau)=-G_{ba}(-\tau),\quad
G_{ab}(\tau)^*=-G_{ab}(-\tau),\quad
G_{ab}(\tau+\beta)=-G_{ab}(\tau).
\label{simplifiedsymproperty}
\end{align}

Note that when we set ${\tilde {\cal J}}=\mathcal{J}$ the effective action and the Schwinger-Dyson equation coincide with those in the Maldacena-Qi model \cite{Maldacena:2018lmt} since the Hamiltonian reduces to that model.
Also note that when we set ${\tilde {\cal J}}=0$ the effective action and the Schwinger-Dyson equation coincide with those in the Kourkoulou-Maldacena model \cite{Kourkoulou:2017zaj} if we identify $G_{LL}(\tau,\tau')=G_{RR}(\tau,\tau')=G_{\text{diag}}(\tau,\tau')$, $G_{LR}(\tau,\tau')=G_{\text{off}}(\tau,\tau')$.

By using the operator relations
\begin{align}
\partial_\tau (e^{\tau H}\psi_i^a e^{-\tau H}\psi_i^a)|_{\tau\rightarrow +0}=[H,\psi_i^a]\psi_i^a=qH_{\text{SYK}}^{(a)}+\mu H_{\text{int}},\quad (a=L,R)
\end{align}
together with the identification \eqref{Gabvsoperator},
% $G_{ab}(\tau,\tau')=\frac{\langle\text{tr}e^{\tau H}\psi_i^ae^{-\tau H}e^{\tau' H}\psi_i^be^{-\tau' H}e^{-\beta H}\rangle_{J^{(\alpha)}_{i_1\cdots i_q}}}{\langle \text{tr}e^{-\beta H}\rangle_{J_{i_1\cdots i_a}^{(\alpha)}}}$ ($\tau>\tau'$),
we can express the energy $E=\frac{\langle \text{tr}He^{-\beta H}\rangle_{J_{i_1\cdots i_q}^{(\alpha)}}}{\langle \text{tr}e^{-\beta H}\rangle_{J_{i_1\cdots i_q}^{(\alpha)}}}$ as
\be
\f{E}{N} = \Bigg[ \f{1}{q} \partial_{\tau} G_{LL}(\tau,0)+ \f{1}{q}\partial _{\tau}G_{RR}(\tau,0) + i \mu \Bigg( 1 - \f{2}{q}\Bigg)G_{LR}(\tau,0) \Bigg]_{\tau \to +0}.
% \f{E}{N} = \Bigg[ \f{1}{q} \partial_{\tau_1} G_{LL}+ \f{1}{q}\partial _{\tau_1}G_{RR} + i \mu \Bigg( 1 - \f{2}{q}\Bigg)G_{LR} \Bigg]_{\tau_{12} \to 0}.
\label{energy}
\ee
Using the Schwinger-Dyson equations \eqref{simplifiedSD} and the symmetry property of $G_{ab}(\tau)$ \eqref{simplifiedsymproperty} it follows
\ba
\lim_{\tau\to +0} \partial_{\tau} G_{aa}(\tau,0) = \sum_c\int d\tau \Sigma_{ac} (\tau) G_{ca}(-\tau)=-\sum_c\frac{s_{ac}{\cal J}_{ac}^2}{2q}\int d\tau (2G_{ac}(\tau))^q,
\ea
hence the energy \eqref{energy} can be further rewritten as 
\ba
\f{E}{N} &=-\sum_{a,b}\f{s_{ab}\mathcal{J}_{ab}^2}{2q^2}\int d\tau (2G_{ab}(\tau))^q + i\mu G_{LR}(0).
% \f{E}{N} &= \f{\mathcal{J}^2}{2q^2}\int d\tau \Big[ (2G_{LL}(\tau))^{q} +(2G_{RR}(\tau))^{q}  \Big] + (-1)^{\f{q}{2}}\f{\tilde{\mathcal{J}}^2}{2q^2}\int d\tau \Big[ (2G_{LR}(\tau))^{q} +(2G_{RL}(\tau))^{q}  \Big] \notag \\
% &\qquad + i\mu G_{LR}(0).
\ea

In the following sections, we study the solution of the Schwinger-Dyson equation both numerically and analytically.

\section{Finite $q$, large $N$}
\label{sec_finiteqlargeN}
In this section we study the two-coupled model \eqref{220221_HMQwithJLJRindepmodel} with $q=4$ in the large $N$ limit numerically by using the bilocal field formalism \eqref{220502ZbetainGSigma} with \eqref{220201_ZinGSigma}, \eqref{220201_EuclideanSDeq}.

\subsection{Phase diagram}
\label{sec_phasediagramq4}
\begin{figure}
\begin{center}
\includegraphics[width=8cm]{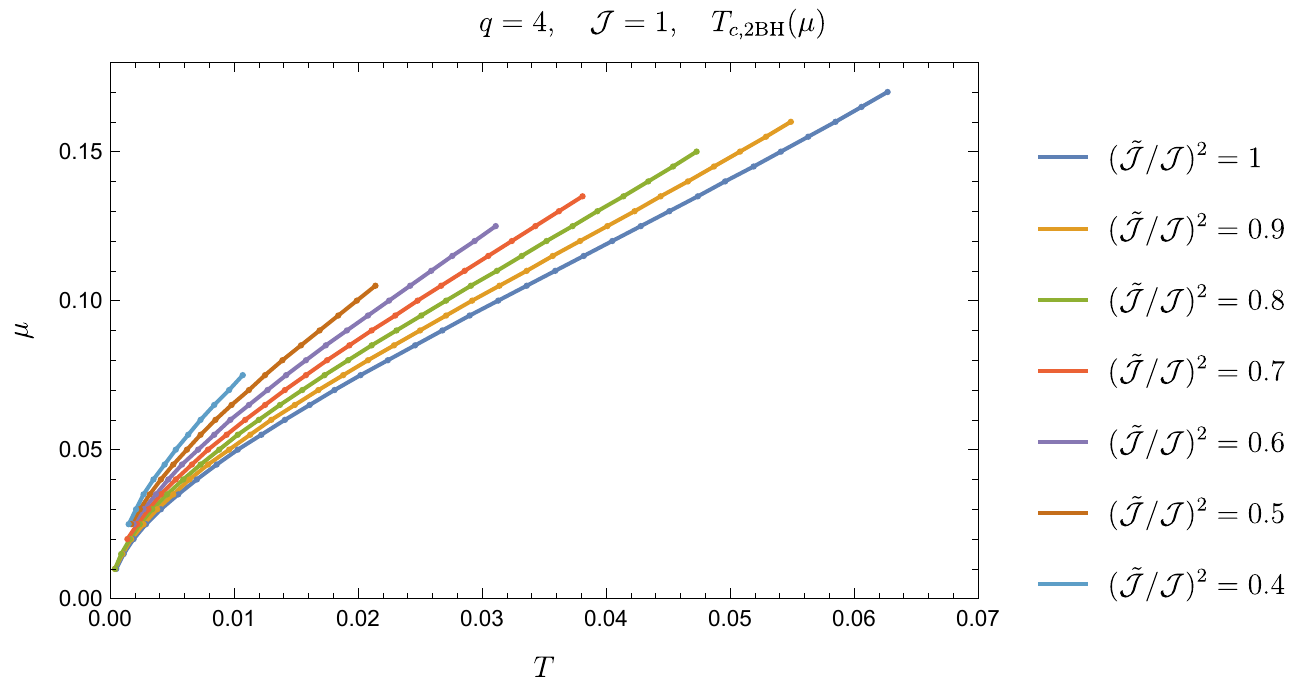}
\includegraphics[width=8cm]{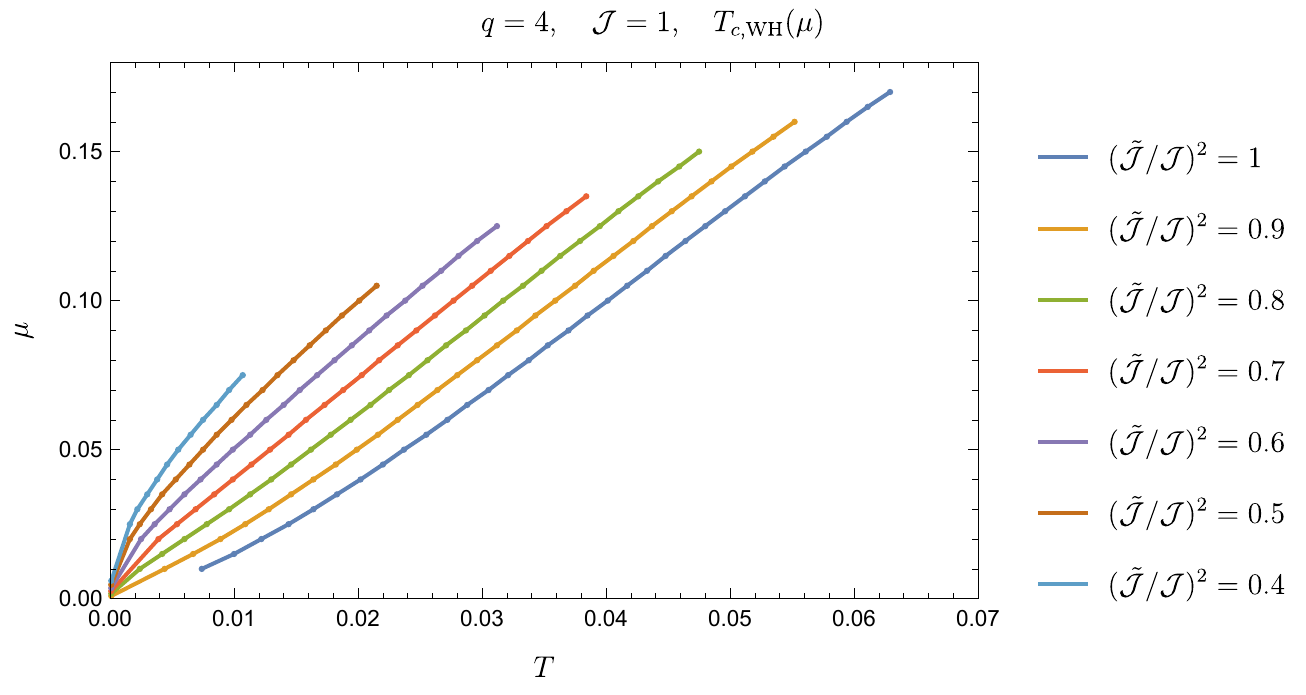}\\
\includegraphics[width=8cm]{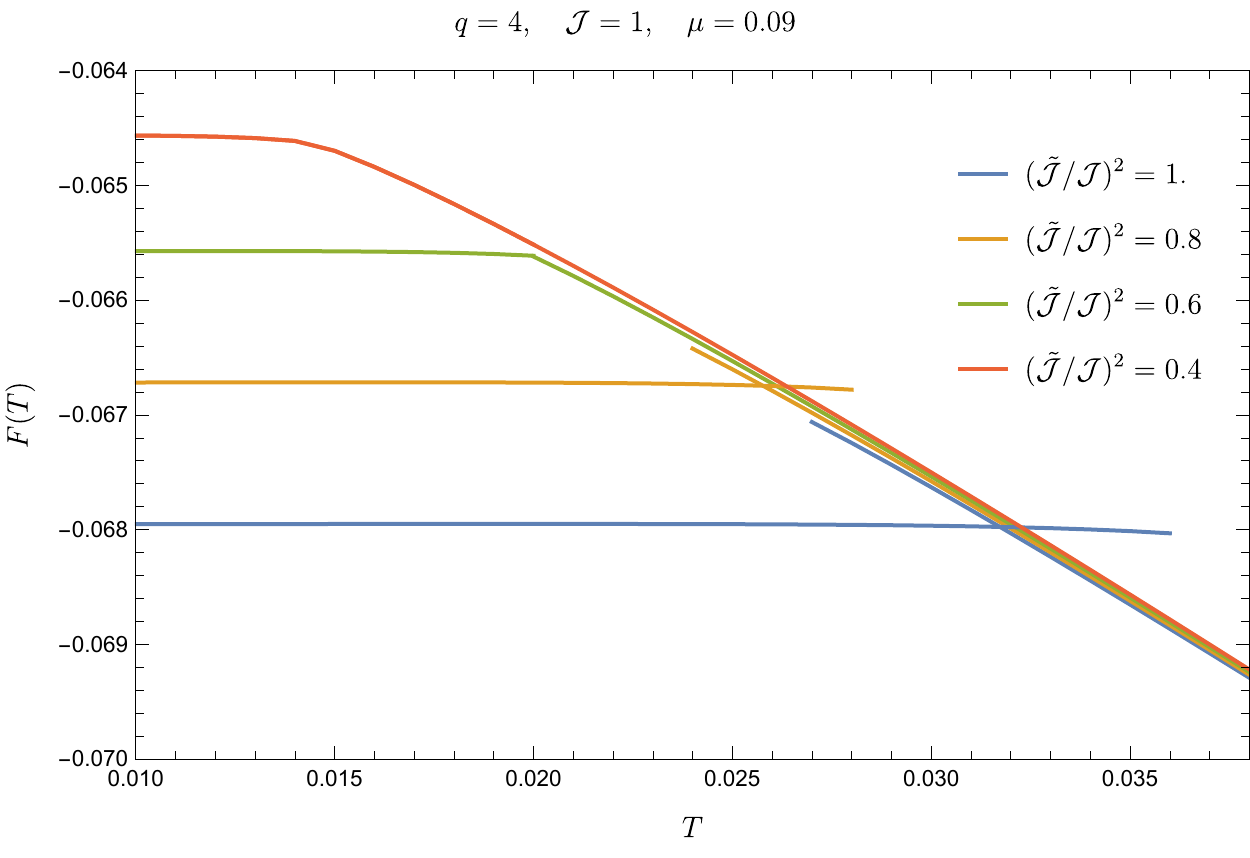}
\caption{
Top left/right: critical temperatures $T_{c,\text{2BH}}(\mu)$ and $T_{c,\text{WH}}(\mu)$ for various values of $\frac{{\tilde {\cal J}}}{{\cal J}}$.
Bottom: free energy for the two solutions for $\mu=0.09$ and various values of $\frac{{\tilde {\cal J}}}{{\cal J}}$ ($\frac{{\tilde {\cal J}}^2}{{\cal J}^2}=0.4$ the two solutions are smoothly connected with each other).
}
\label{220220_fig_Tc2BHandTcWH}
\end{center}
\end{figure}

\begin{figure}
\begin{center}
\includegraphics[width=16cm]{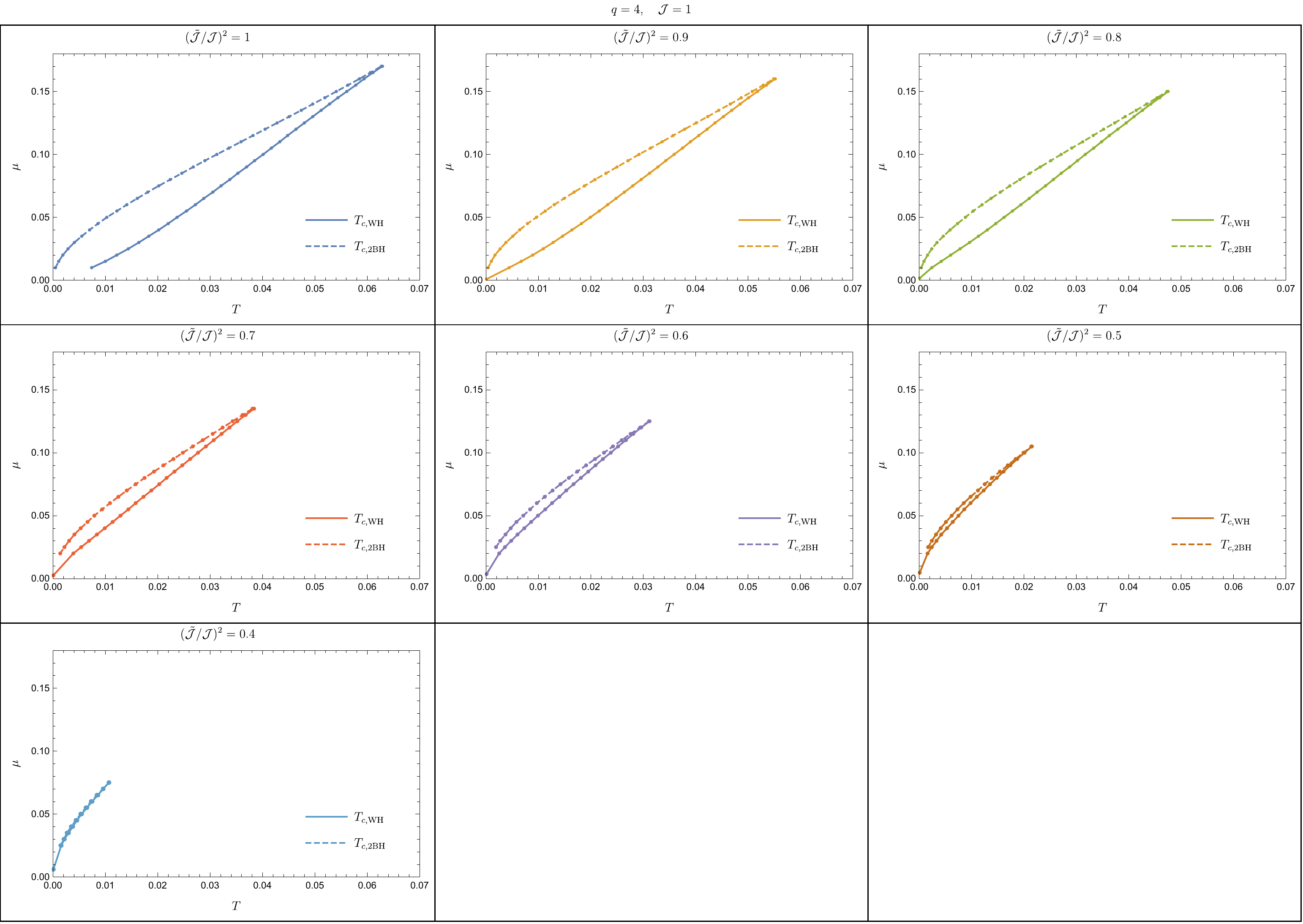}
\caption{
Phase diagram for various values of $\frac{{\tilde {\cal J}}}{{\cal J}}$.
Points connected by the solid line are $T_{c,\text{WH}}(\mu)$ such that the wormhole solution does not exists for $T>T_{c,\text{WH}}$.
Points connected by the dashed line are $T_{c,\text{2BH}}$ such that the two-black hole solution does not exist for $T<T_{c,\text{2BH}}$.
The two lines intersect at a point $(\mu_*,T_*)$ ($T_*=T_{c,\text{WH}}(\mu_*)=T_{c,\text{2BH}}(\mu_*)$) which depends on $\frac{{\tilde {\cal J}}}{{\cal J}}$.
In the regime where either $\mu>\mu_*$ or $T>T_*$ is satisfied, the wormhole solution and the two-black hole solution are smoothly connected with each other.
}
\label{220220_fig_phasediagram}
\end{center}
\end{figure}

In the large $N$ limit we can evaluate the partition function \eqref{220201_ZinGSigma} by the solution of the equations of motion \eqref{220201_EuclideanSDeq}.
% From \eqref{220201_EuclideanSDeq} it follows that
% \begin{align}
% G_{ab}(\tau,\tau')=-G_{ba}(\tau',\tau),\quad
% \Sigma_{ab}(\tau,\tau')=-\Sigma_{ba}(\tau',\tau).
% \end{align}
If we define the Fourier transformation as
% introduce the If we assume $G_{ab}(\tau,\tau')=G_{ab}(\tau-\tau')$, this implies
% \begin{align}
% {\widehat G}_{ab}(\nu)=-{\widehat G}_{ba}(-\nu),
% \end{align}
% where we have defined the Fourier transformation as
\begin{align}
f(\tau)\rightarrow {\widehat f}(\nu)=\int_0^\beta d\tau e^{i\nu\tau}f(\tau),
\end{align}
% If we further impose the following ansatz
% % \footnote{
% % The ansatz \eqref{starrelation} is a special case of the following ansatz which we will use later in section \ref{realtime}:
% % \begin{align}
% % G_{ab}(u_1,u_2)^*=-G_{ab}(-u_1^*,-u_2^*).
% % \end{align}
% % This follows if we identify $G_{ab}(u_1,u_2)$ as $G_{ab}(u_1,u_2)=N^{-1}\sum_{i=1}^N\langle {\cal T}\psi_i^a(u_1)\psi_i^b(u_2)\rangle_\beta$ \cite{Nosaka:2020nuk}.
% % }
% \begin{align}
% &G_{ab}(\tau)^*=-G_{ab}(-\tau)\quad (\Leftrightarrow {\widehat G}_{ab}(\nu)^*=-{\widehat G}_{ab}(\nu)),\label{starrelation} \\
% % G_{ab}(u_1,u_2)^*=-G_{ab}(-u_1^*,-u_2^*)
% &G_{RR}(\tau)=G_{LL}(\tau),
% \end{align}
and also impose an ansatz $G_{RR}(\tau)=G_{LL}(\tau)$, the Euclidean Schwinger-Dyson equation \eqref{220201_EuclideanSDeq} can be rewritten as
\begin{align}
&{\widehat G}_{LL}(\nu)+\frac{i\nu+{\widehat\Sigma}_{LL}(\nu)}{(i\nu+{\widehat\Sigma}_{LL}(\nu))^2+{\widehat\Sigma}_{LR}(\nu)^2}=0,\nonumber \\
&{\widehat G}_{LR}(\nu)-\frac{{\widehat \Sigma}_{LR}(\nu)}{(i\nu+{\widehat\Sigma}_{LL}(\nu))^2+{\widehat\Sigma}_{LR}(\nu)^2}=0,\nonumber \\
&\Sigma_{LL}(\tau)=\frac{{\cal J}^2}{q}(2G_{LL}(\tau))^{q-1},\quad
\Sigma_{LR}(\tau)=\frac{(-1)^{\frac{q}{2}}{\tilde {\cal J}}^2}{q}(2G_{LR}(\tau))^{q-1}+i\mu\delta(\tau),
% \nonumber \\
% &{\widehat G}_{ab}(\nu)=-{\widehat G}_{ba}(-\nu),\quad {\widehat G}_{ab}(\nu)^*=-{\widehat G}_{ab}(\nu).
\label{220201_SDeqsimplifiedwithansatz}
\end{align}
The partition function, or the free energy $F=-\frac{1}{\beta}\log Z$, can be evaluated in the large $N$ limit by the solutions of \eqref{220201_SDeqsimplifiedwithansatz} as
\begin{align}
F\approx \text{min}\Bigl\{\frac{1}{\beta}S_E[G_{ab}(\tau,\tau'),\Sigma_{ab}(\tau,\tau')]\,\Bigl|\, (G_{ab},\Sigma_{ab})\text{: solution of }\eqref{220201_SDeqsimplifiedwithansatz}\Bigr\}.
\end{align}
The set of equations \eqref{220201_SDeqsimplifiedwithansatz} can be solved numerically for each values of $(q,{\cal J},{\tilde{\cal J}},\mu)$ and the inverse temperature $\beta=T^{-1}$.
We performed the numerical analysis for $q=4, {\cal J}=1$ and various values $(\frac{{\tilde{\cal J}}}{{\cal J}},\mu,\beta)$.
In particular, as we vary $(\mu,\beta)$ we obtained the following results:
\begin{itemize}
\item [(i)] When the temperature $T$ is sufficiently large, there is a solution where $\frac{S_E}{\beta}$ is similar to the (annealed) free energy of two uncoupled SYK systems.
We shall call this solution as two-black hole solution.
\item [(ii)] As we decrease the temperature slowly (we have chosen $\Delta T=0.0001$), this solution is deformed continuously until some temperature $T=T_{c,\text{2BH}}$.
Once the temperature crosses $T_{c,\text{2BH}}$, the two-black hole solution ceases to exist and the numerical analysis detects another solution where the free energy is almost constant in $T$.
We shall call this solution as wormhole solution.
\item [(iii)] As we increase the temperature from $T<T_{c,\text{2BH}}$ the wormhole solution is deformed continuously until some temperature $T=T_{c,\text{WH}}$ which is greater than $T_{c,\text{2BH}}$.
Once $T$ exceeds $T_{c,\text{WH}}$ the wormhole solution disappears.
\item [(iv)] When $\mu$ is larger than some critical value $\mu_*$, (ii) and (iii) do not occur; the two-black hole solution and the wormhole solution merge to a single solution which exists at any value of the temperature.
\end{itemize}
See figure \ref{220220_fig_Tc2BHandTcWH} and \ref{220220_fig_phasediagram}.
\begin{figure}
\begin{center}
\includegraphics[width=15cm]{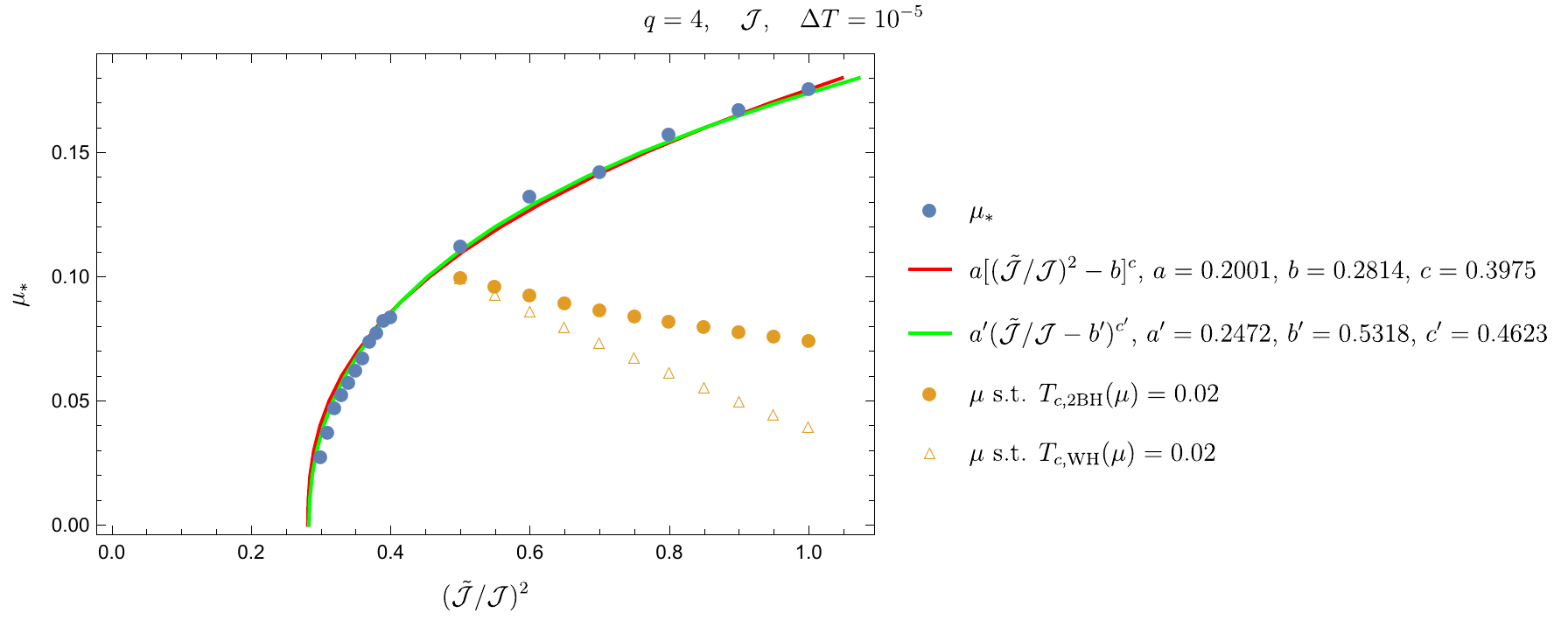}
\caption{
The critical value $\mu_*$ of $\mu$ for each $\frac{{\tilde {\cal J}}}{{\cal J}}$ such that the phase transition does not exist for $\mu>\mu_*$, with the red/green curve obtained by fitting with the ansatz $\mu_*=a((\frac{{\tilde {\cal J}}}{{\cal J}})^n-b)^c$.
For comparison, we have also displayed the points on the phase transition lines in the $(\mu,\frac{{\tilde {\cal J}}^2}{{\cal J}^2})$-plane with a fixed temperature $T=0.02$, i.e., the points where $T_{c,\text{2BH}}(\mu;\frac{{\tilde {\cal J}}^2}{{\cal J}^2})=0.02$ and $T_{c,\text{WH}}(\mu;\frac{{\tilde {\cal J}}^2}{{\cal J}^2})=0.02$.
Here we have determined $\alpha,c$ first by fitting the data of $\frac{d((\frac{{\tilde {\cal J}}}{{\cal J}})^n)}{d\mu_*}$ obtained by the numerical differentiation with the ansatz $\log\frac{d((\frac{{\tilde {\cal J}}}{{\cal J}})^n)}{d\mu_*}=(c-1)\log \mu_*+\log(\alpha c)$ and then determined $b$ by fitting again $(\frac{{\tilde {\cal J}}}{{\cal J}})^n$ with the ansatz $(\frac{{\tilde {\cal J}}}{{\cal J}})^n=(\frac{\mu_*}{a})^c+b$.
We performed the fitting for $n=1,2$ and have found almost the same value of $b^{\frac{1}{n}}$, the value of $\frac{{\tilde {\cal J}}}{{\cal J}}$ where $\mu_*$ vanishes.
}
\label{220220_fig_Mucwherephasetransitionvanishes}
\end{center}
\end{figure}
These behaviors of the solution and the free energy are qualitatively the same as those for the case with ${\tilde {\cal J}}={\cal J}$ \cite{Maldacena:2018lmt,Nosaka:2020nuk}.
In the temperature regime $T_{c,\text{2BH}}<T<T_{c,\text{WH}}$ both the two-black hole solution and the wormhole solution exists, hence the free energy is given by the smaller one of the two values of $S_E$ evaluated at these two solutions.
We observe that the two values crosses at one point $T=T_c$, where the system undergoes a phase transition.

As we further vary $\frac{{\tilde {\cal J}}}{{\cal J}}$ we observed that these behaviors change in the following way:
\begin{itemize}
\item [(v)] $T_{c,\text{WH}}$ and $T_c$ decrease as $\frac{{\tilde {\cal J}}}{{\cal J}}$ is decreased.
On the other hand, $T_{c,\text{2BH}}$ also decreases, but it is almost independent of $\frac{{\tilde {\cal J}}}{{\cal J}}$ when $\mu$ is small.
This is consistent with the fact that $T_{c,\text{2BH}}$ is determined as a property of the two-black hole solution where the off-diagonal component $G_{LR},\Sigma_{LR}$ are small and hence the correlation between $J_{i_1i_2\cdots i_q}^{(L)}$ and $J_{i_1i_2\cdots i_q}^{(R)}$ is less important.
See figure \ref{220220_fig_Tc2BHandTcWH}.
\item [(vi)] The critical value $\mu_*$ of $\mu$ where the phase transition disappears decreases as $\frac{{\tilde {\cal J}}}{{\cal J}}$ is decreased.
See figure \ref{220220_fig_Mucwherephasetransitionvanishes}.
From the results we also expect that $\mu_*$ becomes zero somewhere in the range $0.2<\frac{{\tilde {\cal J}}^2}{{\cal J}^2}<0.3$ ($b$ and $b'{}^2$ in figure \ref{220220_fig_Mucwherephasetransitionvanishes}), that is, the phase transition completely disappears as $\frac{{\tilde {\cal J}}^2}{{\cal J}^2}$ is decreased below this value.
\end{itemize}
Though the observation that $\mu_c$ depends on $\frac{{\tilde {\cal J}}}{{\cal J}}$ might be surprising, it is consistent with the fact that for $\frac{{\tilde {\cal J}}}{{\cal J}}=0$ our model \eqref{220221_HMQwithJLJRindepmodel} is equivalent in the large $N$ limit to the single-side model \cite{Kourkoulou:2017zaj} which does not exhibit phase transition at any value of $\mu$ \cite{Nosaka:2019tcx}.

Notice that our claim (vi) for the absence of the phase transition is based on the observation that $S_E$ obtained by decreasing $T$ from high temperature regime and $S_E$ obtained by increasing $T$ from low temperature regime do not deviate at discrete points spaced with $\Delta T=0.00001$, but this does not exclude the possibility that the phase transition exists with $T_{c,\text{WH}}-T_{c,\text{2BH}}<0.00001$.
In our approach it is in principle impossible to {\it prove} the absence of phase transition at $\mu>\mu_*$.
As we see in section \ref{sec_largeq}, however, we can rigorously show the absence of the phase transition in the large $q$ limit.

\subsection{Energy gap}
Next we look at the energy gap $E_{\text{gap}}$ of the two-coupled model \eqref{220221_HMQwithJLJRindepmodel} in the large $N$ limit.
In \cite{Nosaka:2019tcx} we have observed that $E_{\text{gap}}$ of the model \eqref{220221_HMQwithJLJRindepmodel} with ${\tilde {\cal J}}={\cal J}$ \cite{Maldacena:2018lmt} and $E_{\text{gap}}$ of the single sided model \cite{Kourkoulou:2017zaj} which is equivalent in the large $N$ limit to the two coupled model with with ${\tilde {\cal J}}=0$ show different power law behavior for small $\mu$: $E_{\text{gap}}({\tilde {\cal J}}={\cal J})\sim \mu^{\frac{q}{2(q-1)}}$ and $E_{\text{gap}}({\tilde {\cal J}}=0)\sim \mu^{\frac{q}{q-2}}$.
In this section we investigate how these two behaviors are interpolated as we vary ${\tilde {\cal J}}$ from $0$ to $1$.

\begin{figure}
\begin{center}
\includegraphics[width=8cm]{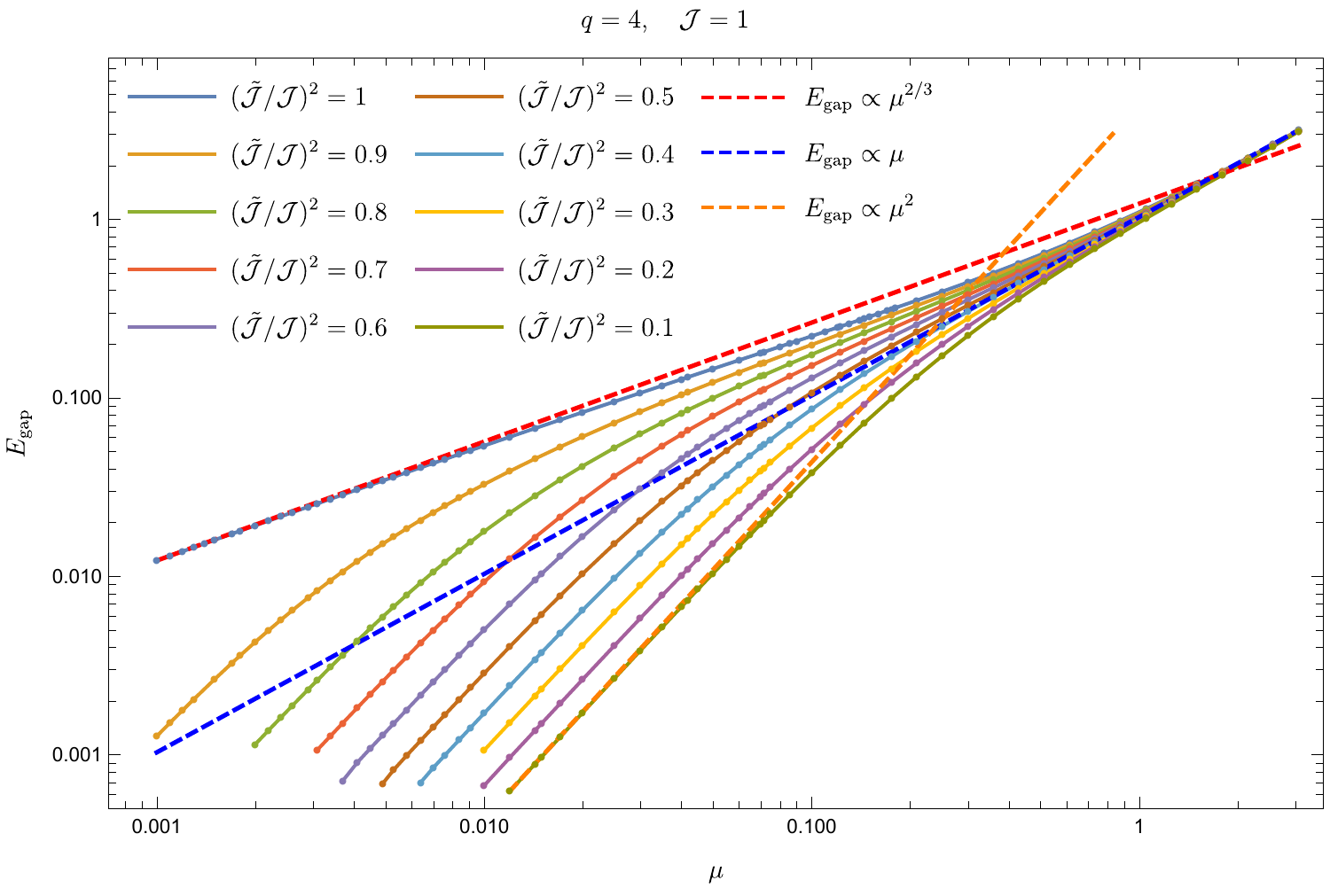}
\includegraphics[width=8cm]{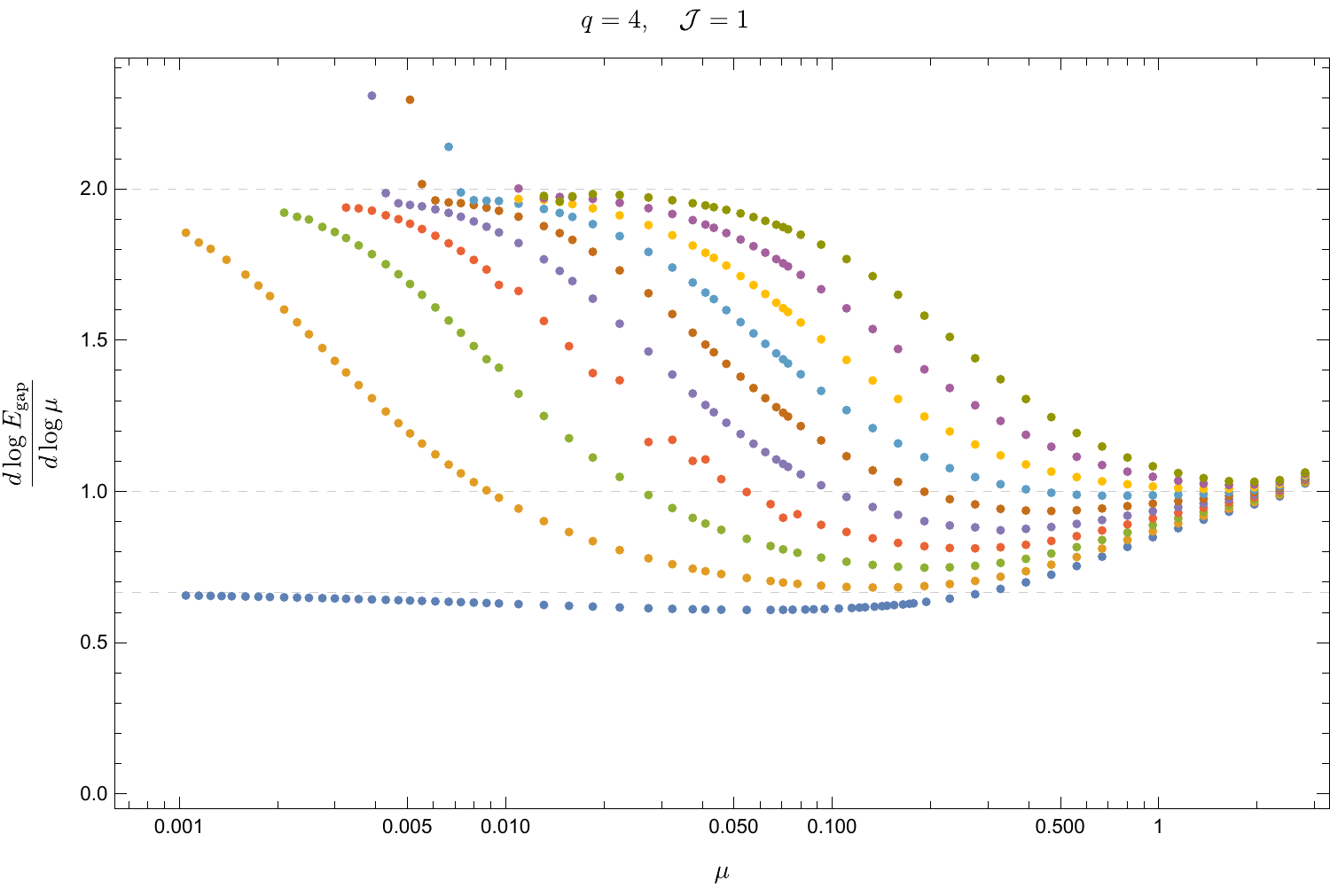}
\caption{
The energy gap $E_{gap}$ for various values of $\frac{{\tilde {\cal J}}}{{\cal J}}$ read off by fitting $G_{ab}(\tau)$ with the ansatz \eqref{220502Egapansatz}.
Here the legends of the right plot are the same as those indicated in the left plot.}
\label{220221_fig_Egap}
\end{center}
\end{figure}

The large $N$ energy gap can be read off from the Euclidean two point functions $G_{ab}(\tau)$ in the low-temperature phase as \cite{Maldacena:2018lmt}
\begin{align}
G_{LL}(\tau)\sim \cosh\Bigl[E_{\text{gap}}\Bigl(\frac{\beta}{2}-\tau\Bigr)\Bigr],\quad
G_{LR}(\tau)\sim \sinh\Bigl[E_{\text{gap}}\Bigl(\frac{\beta}{2}-\tau\Bigr)\Bigr].\quad\quad (1\ll \tau\ll \beta)
\label{220502Egapansatz}
\end{align}
By fitting $G_{ab}(\tau)$ with these ansatz we have obtained $E_{\text{gap}}$ for $0.1\le \frac{{\tilde {\cal J}}^2}{{\cal J}^2}\le 1$ as displayed in figure \ref{220221_fig_Egap}.
We observe the followings
\begin{itemize}
\item For any value of $\mu$, $E_{\text{gap}}$ increase monotonically in $\frac{{\tilde {\cal J}}^2}{{\cal J}^2}$.
\item When $\mu$ is sufficiently large, $E_{\text{gap}}$ scales as $E_{\text{gap}}\sim \mu$ regardless of the value of $\frac{{\tilde {\cal J}}^2}{{\cal J}^2}$
\item When $\mu$ is sufficiently small, $E_{\text{gap}}\sim \mu^{2/3}$ for $\frac{{\tilde {\cal J}}^2}{{\cal J}^2}=1$ while $E_{\text{gap}}\sim \mu^2$ for $\frac{{\tilde {\cal J}}^2}{{\cal J}^2}<1$.
\item When $\frac{{\tilde {\cal J}}^2}{{\cal J}^2}$ is less than $1$ but close to $1$, there is also an intermediate regime where $E_{\text{gap}}\sim \mu^{\frac{2}{3}}$.
The range of the intermediate regime becomes wider in $\mu$ as $\frac{{\tilde {\cal J}}^2}{{\cal J}^2}$ approaches $1$.
\end{itemize}

\subsection{Real time response}
\label{realtime}

Next we study real time dynamics of the two coupled model \eqref{randomcoupling}, in particular, the transmission amplitude $T_{LR}$ \cite{Plugge:2020wgc} which measures how an excitation on the right site affects on the left site at late time, and the decay of the out-of-time-ordered four point function \cite{Larkin:1969aaa} which is characeterized with the quantum chaos exponent $\lambda_L$.
For this purpose we need to continue the $G\Sigma$ formalism for real $\tau$
\eqref{220502ZbetainGSigma}-\eqref{220201_EuclideanSDeq} in section \ref{sec_JLneqJRmodel} to complex $u=\tau+it$, which can be done in the following way.\footnote{
Note the calculations in this section are completely parallel to the case with $J_{i_1\cdots i_q}^{(L)}=J_{i_1\cdots i_q}^{(R)}$ in \cite{Nosaka:2020nuk}.
Hence we shall skip the details of the derivations which can be found in section 3 in \cite{Nosaka:2020nuk}.
}
First we define the two different components of the two point functions at $\text{Re}[u]=0$, $G_{ab}^>,G_{ab}^<$, depending on how $\text{Re}[u]$ approaches $0$:
\begin{align}
G_{ab}^>(t_1,t_2)&=-iG_{ab}(it_1^-,it_2^+)=-i\lim_{\epsilon\rightarrow +0}G_{ab}(\epsilon+it_1,-\epsilon+it_2),\nonumber \\
G_{ab}^<(t_1,t_2)&=-iG_{ab}(it_1^+,it_2^-)=-i\lim_{\epsilon\rightarrow +0}G_{ab}(-\epsilon+it_1,\epsilon+it_2),
\end{align}
with which we also define the retarded/advanced components of the two point functions as
\begin{align}
G_{ab}^R(t_1,t_2)&=\theta(t_1-t_2)(G^>(t_1,t_2)-G^<(t_1,t_2)),\nonumber \\
G_{ab}^A(t_1,t_2)&=\theta(t_2-t_1)(G^<(t_1,t_2)-G^>(t_1,t_2)).
\end{align}
We also define the $>,<,R,A$ components of $\Sigma_{ab}(u)$ at $u=it$ in the same ways.

With these notations the real time continuation of the Schwinger-Dyson equations \eqref{220201_EuclideanSDeq} reduce to the followings
\begin{align}
&{\widetilde G}_{LL}^R(\omega)=\frac{-(-\omega+{\widetilde\Sigma}_{RR}^{R}(\omega))}{(-\omega+{\widetilde\Sigma}_{LL}^{R}(\omega))(-\omega+{\widetilde\Sigma}_{RR}^{R}(\omega))-({\widetilde\Sigma}_{LR}^{R}+i\mu)({\widetilde\Sigma}_{RL}^{R}-i\mu)},\nonumber \\
&{\widetilde G}_{LR}^R(\omega)=\frac{{\widetilde\Sigma}_{LR}^{R}(\omega)+i\mu}{(-\omega+{\widetilde\Sigma}_{LL}^{R}(\omega))(-\omega+{\widetilde\Sigma}_{RR}^{R}(\omega))-({\widetilde\Sigma}_{LR}^{R}+i\mu)({\widetilde\Sigma}_{RL}^{R}-i\mu)},\nonumber \\
&{\widetilde G}_{RL}^R(\omega)=\frac{{\widetilde\Sigma}_{RL}^{R}(\omega)-i\mu}{(-\omega+{\widetilde\Sigma}_{LL}^{R}(\omega))(-\omega+{\widetilde\Sigma}_{RR}^{R}(\omega))-({\widetilde\Sigma}_{LR}^{R}+i\mu)({\widetilde\Sigma}_{RL}^{R}-i\mu)},\nonumber \\
&{\widetilde G}_{RR}^R(\omega)=\frac{-(-\omega+{\widetilde\Sigma}_{LL}^{R}(\omega))}{(-\omega+{\widetilde\Sigma}_{LL}^{R}(\omega))(-\omega+{\widetilde\Sigma}_{RR}^{R}(\omega))-({\widetilde\Sigma}_{LR}^{R}+i\mu)({\widetilde\Sigma}_{RL}^{R}-i\mu)},\nonumber \\
% &-i\partial_{t_1}G_{ab}^R(t_1,t_2)-\sum_c\Bigl(-i\mu \rho_{ac}G_{cb}^R(t_1,t_2)-\int dt_3\Sigma_{ac}^{R}(t_1,t_3)G_{cb}^R(t_3,t_2)\Bigr)=-\delta_{ab}\delta(t_1-t_2), \label{manipulation3} \\
&\Sigma_{ab}^{>}(t_1,t_2)=-\frac{i^q{\cal J}_{ab}^2}{q}s_{ab}G^>_{ab}(t_1,t_2)^{q-1},\label{MQrealtimeSDeqsecondline}\\
&\Sigma_{ab}^{R}(t_1,t_2)=\theta(t_1-t_2)(\Sigma_{ab}^{>}(t_1,t_2)+\Sigma_{ba}^{>}(t_2,t_1)),\nonumber \\
&{\widetilde G}_{ab}^>(\omega)=\frac{{\widetilde G}_{ab}^R(\omega)-({\widetilde G}_{ba}^R(\omega))^*}{1+e^{-\beta \omega}},
\end{align}
where we have defined the Fourier transform in real time $t$ as
\begin{align}
{\widetilde f}^X(\omega)&=\int_{-\infty}^{\infty}dt e^{i\omega t}f^X(t),\quad
f^X(t)=\int_{-\infty}^{\infty}\frac{d\omega}{2\pi} e^{-i\omega t}{\widetilde f}^X(t),\quad (f=G_{ab},\Sigma_{ab},\quad X=>,<,R,A).
\end{align}

Note that we can obtain $G_{ab}(u)$ for general $u\in\mathbb{C}$ from ${\widetilde G}_{ab}^R(\omega)$ as
\begin{align}
G_{ab}(u)=iG_{ab}^>(t=-iu)=i\int \frac{d\omega}{2\pi}e^{-\omega u}\frac{{\widetilde G}_{ab}^R(\omega)-({\widetilde G}_{ba}^R(\omega))^*}{1+e^{-\beta\omega}},
\label{fixed200806}
\end{align}
which we use to compute the chaos exponent in section \ref{sec_chaosexponent_MQ}.

\subsubsection{Transmission amplitude in low temperature regime}
\begin{figure}
\begin{center}
\includegraphics[width=12cm]{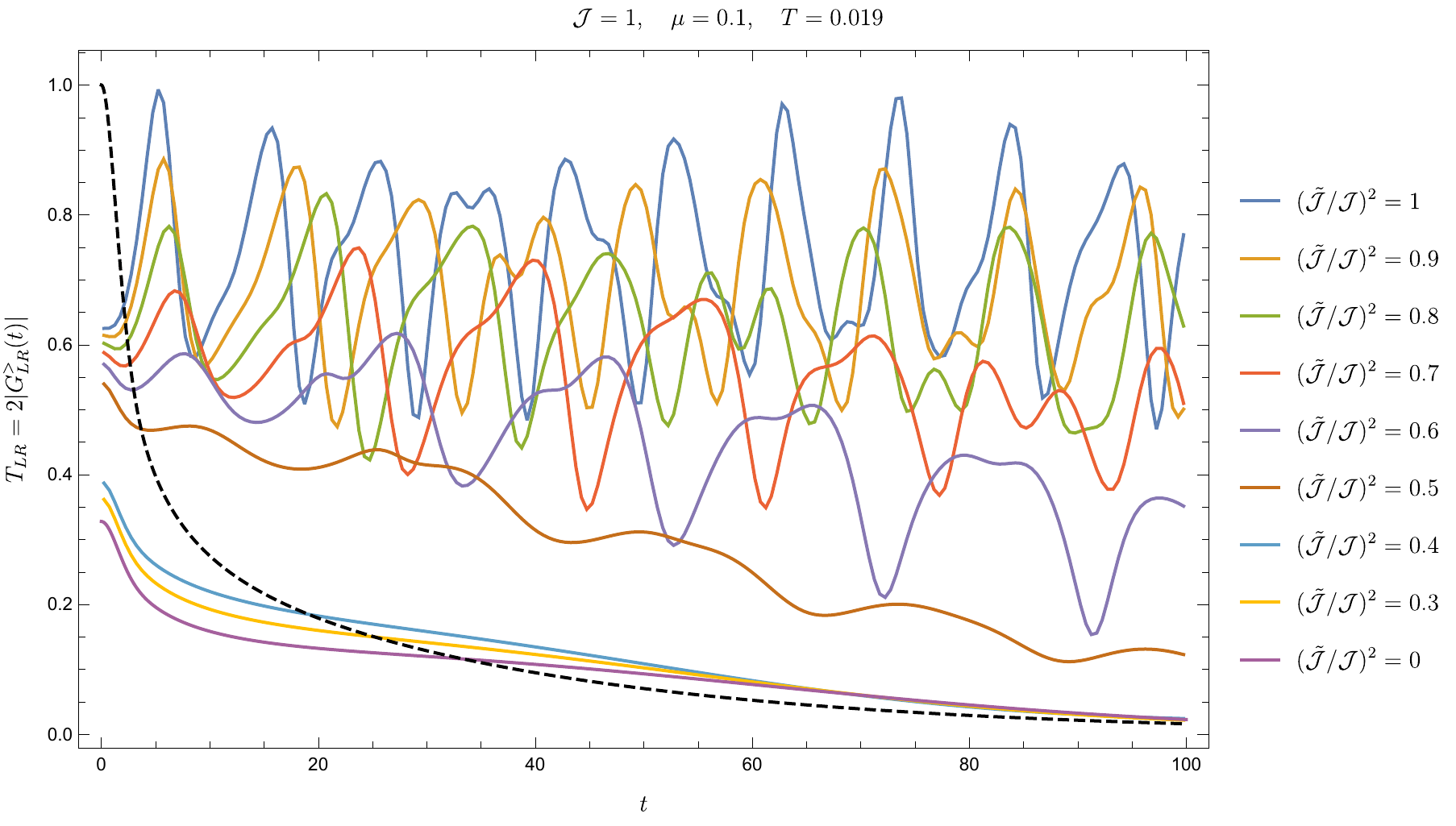}
\caption{
Transmission amplitude $T_{LR}(t)=2|G_{LR}^>(t)|$ for various values of $\frac{{\tilde {\cal J}}^2}{{\cal J}^2}$ in the low-temperature phase.
Note that for $\frac{{\tilde {\cal J}}^2}{{\cal J}^2}=0.4$ there are no phase transition for $\mu=0.1$ (see figure \ref{220220_fig_phasediagram}), hence $(\mu,T)=(0.1,0.019)$ belongs to the supercritical regime.
For comparison we have also plotted $2|G^>(t)|$ for a single SYK model (dashed black line).
}
\label{220318_fig_TLR}
\end{center}
\end{figure}

\begin{figure}
\begin{center}
\includegraphics[width=12cm]{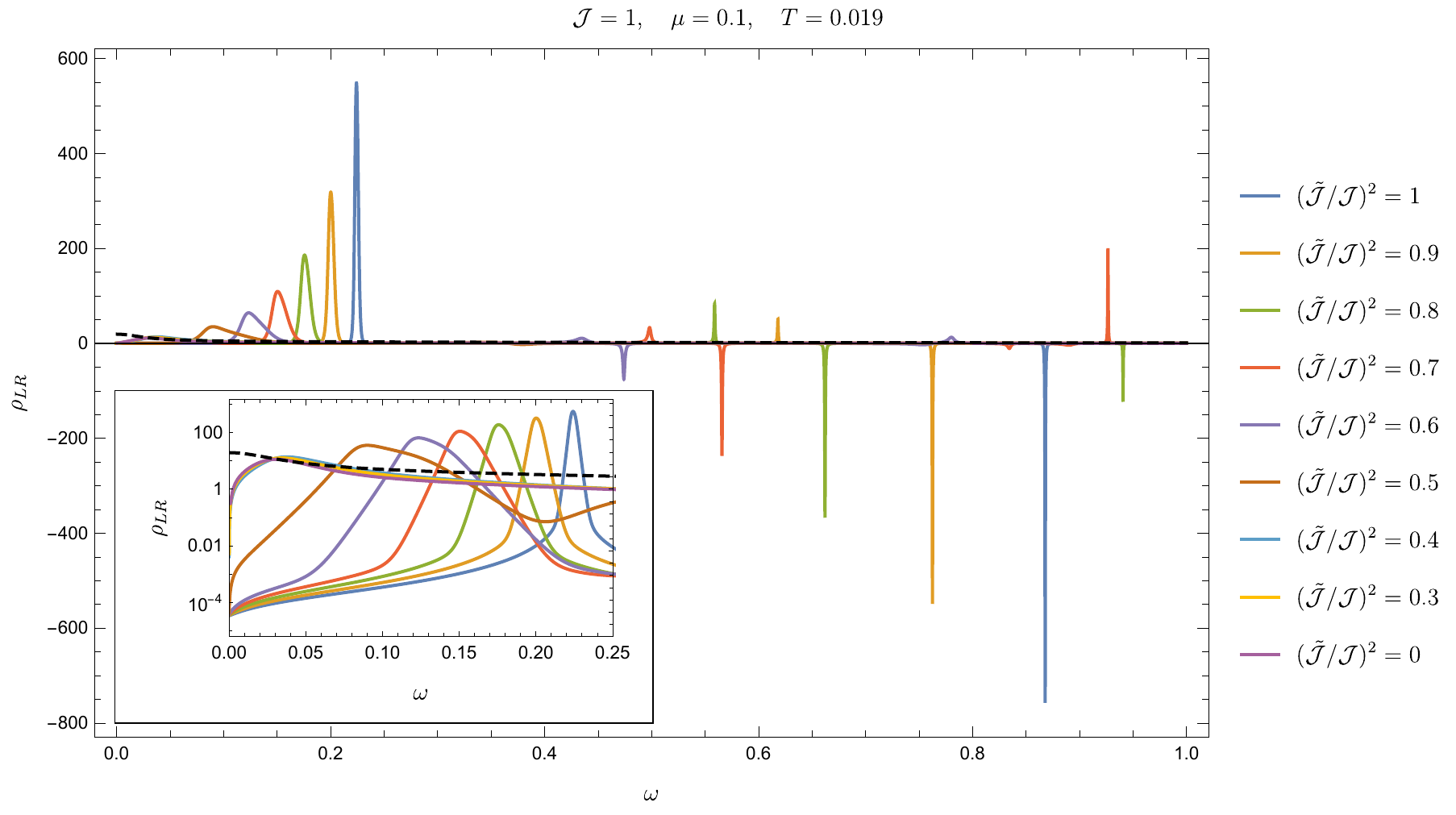}
\caption{
Spectral function $\rho_{LR}(\omega)=-2\text{Re}[{\widetilde G}_{LR}^R(\omega)]$.
}
\label{220318_fig_spectralfunctionLR}
\end{center}
\end{figure}

Let us define the transmisson amplitude $T_{LR}(t)$ as $T_{LR}(t)=2|G^>_{LR}(t)|$, which measures the probability to find the excitation of $\psi_i^L$ at time $t$ after the insertion of $\psi_i^R$ at time $t=0$ \cite{Plugge:2020wgc}.
We have displayed $T_{LR}(t)$ for $q=4, {\cal J}=1, \mu=0.1, T=0.019$ and various values of $\frac{{\tilde {\cal J}}^2}{{\cal J}^2}$ in figure \ref{220318_fig_TLR}.
We find that the transmisson is reduced by decreasing LR correlation $\frac{{\tilde {\cal J}}^2}{{\cal J}^2}$.
Note that when the temperature is sufficiently small, $G_{LR}^>(t)$ is well approximated by a single quasi-particle with the speed $\omega_1$ and a finite life time $\Gamma_1$: $G_{LR}^>(t)\sim e^{-i\omega_1 t-\Gamma_1 t}$.
Hence the suppression of $T_{LR}(t)$ can be explained by the decrease of $\omega_1$ and increase of $\Gamma_1$, both of which are encoded in the first peak of the spectral function $\rho_{LR}(\omega)=-2\text{Re}[{\widetilde G}_{LR}(\omega)]$, as $\frac{{\tilde {\cal J}}^2}{{\cal J}^2}$ is decreased.
See figure \ref{220318_fig_spectralfunctionLR}.

\subsubsection{Chaos exponent}
\label{sec_chaosexponent_MQ}
To compute the chaos exponent, we consider the four point function
\begin{align}
\frac{1}{N^2}\sum_{i,j}^{N}\langle
\psi_i^a(u_1)
\psi_i^b(u_2)
\psi_j^c(u_3)
\psi_j^d(u_4)
\rangle=
\frac{1}{Z}\int {\cal D}G_{ab}{\cal D}\Sigma_{ab}G_{ab}(u_1,u_2)G_{cd}(u_1,u_2)e^{-NS}.
\end{align}
with $u_1=\frac{3\beta}{4}+it_1$, $u_2=\frac{\beta}{4}+it_2$, $u_3=\frac{\beta}{2}$, $u_4=0$.
The chaos exponent $\lambda_L$ is given by the following late-time behavior of the four point function:
\begin{align}
&\frac{1}{N^2}\sum_{i,j}^{N}\langle
\psi_i^a(u_1)
\psi_i^b(u_2)
\psi_j^c(u_3)
\psi_j^d(u_4)
\rangle=G_{ab}(u_1,u_2)G_{cd}(u_3,u_4)+\frac{1}{N}{\cal F}_{abcd}(u_1,u_2,u_3,u_4),\nonumber \\
&{\cal F}_{abcd}(u_1,u_2,u_3,u_4)\sim e^{\frac{\lambda_L(t_1+t_2)}{2}}\quad (t_1,t_2\gg 1).
\end{align}
In the large $N$ limit and at late time we find that ${\cal F}_{abcd}(t_1,t_2)$ obeys the following equation \cite{Nosaka:2020nuk}
\begin{align}
{\cal F}_{abcd}(t_1,t_2)&\approx \sum_{ef}\int dt dt' {\cal K}^R_{abef}(t_1,t_2,t,t'){\cal F}_{efcd}(t,t'),\nonumber \\
{\cal K}^R_{abcd}(t_1,t_2,t_3,t_4)&=
-\frac{\mathcal{J}_{cd}^2 2^{q-1}(q-1)}{q}G_{ac}^{R}(t_1-t_3)G_{bd}^{R}(t_2-t_4)s_{cd}G_{cd}\Bigl(\frac{\beta}{2}+i(t_3-t_4)\Bigr)^{q-2}.
% -\frac{\mathcal{J}^2 2^{q-1}(q-1)}{q}G_{ac}^{R}(t_1-t_3)G_{bd}^{R}(t_2-t_4)s_{cd}G_{cd}\Bigl(\frac{\beta}{2}+i(t_3-t_4)\Bigr)^{q-2}.
\label{MQladderFKF}
\end{align}
If we substitute the ansatz ${\cal F}_{abcd}(t_1,t_2)=e^{\frac{\lambda_L(t_1+t_2)}{2}}f_{abcd}(t_{12})$ this equation can be rewritten as an eigenequation of $f_{ab\cdot\cdot}(t)$ with eingenvalue $1$ \cite{Nosaka:2020nuk}:
\begin{align}
&\sum_{e,f}\int dt'{\cal M}_{abef}(\lambda_L;t,t')f_{efcd}(t')\approx f_{abcd}(t)\nonumber \\
&{\cal M}_{abcd}(\lambda_L;t,t')=-\frac{\mathcal{J}_{cd}^2 2^{q-1}(q-1)}{q}e^{-\frac{\lambda_L(t-t')}{2}}\biggl[\int dt'' G_{ac}^{R}(t-t'-t'')G_{bd}^{R}(-t'')e^{\lambda_Lt''}\biggr]
% &{\cal M}_{abcd}(\lambda_L;t,t')=-\frac{\mathcal{J}^2 2^{q-1}(q-1)}{q}e^{-\frac{\lambda_L(t-t')}{2}}\biggl[\int dt'' G_{ac}^{R}(t-t'-t'')G_{bd}^{R}(-t'')e^{\lambda_Lt''}\biggr]
s_{cd}G_{cd}\Bigl(\frac{\beta}{2}+it'\Bigr)^{q-2}.
\label{ladderfinalMQ}
\end{align}
The quantum chaos exponent $\lambda_L$ can be determined so that the largest eigenvalue of the $\lambda_L$-dependent kernel ${\cal M}_{abcd}(\lambda_L;t,t')$ crosses $1$.

Notice that \eqref{ladderfinalMQ} can be decomposed into the following two equations with $\sigma=\pm 1$ \cite{Nosaka:2020nuk}
\begin{align}
&\begin{pmatrix}
f_{2,LL}(t)+\sigma f_{2,RR}(t)\\
f_{2,LR}(t)-\sigma f_{2,RL}(t)
\end{pmatrix}_a\nonumber \\
&=
\sum_b\int dt'{\cal M}_{ab}(\sigma,\lambda_L;t,t')
\begin{pmatrix}
f_{2,LL}(t')+\sigma f_{2,RR}(t')\\
f_{2,LR}(t')-\sigma f_{2,RL}(t')
\end{pmatrix}_{b},
\end{align}
where $a,b=1,2$ and
\begin{align}
&{\cal M}_{ab}(\sigma,\lambda_L;t,t')\nonumber \\
&=\begin{pmatrix}
(M_{1,LLLL}(t-t')+\sigma M_{1,LRLR}(t-t'))M_{2,LL}(t')&(M_{1,LLLR}(t-t')-\sigma M_{1,LRLL}(t-t'))M_{2,LL}(t')\\
-(M_{1,LLLR}(t-t')-\sigma M_{1,LRLL}(t-t'))M_{2,LR}(t')&(M_{1,LLLL}(t-t')+\sigma M_{1,LRLR}(t-t'))M_{2,LR}(t')
\end{pmatrix}
\label{ladderMQfinalsimplified}
\end{align}
with
\begin{align}
f_{2,ab}(t_{12})&=e^{\frac{\lambda_Lt_{12}}{2}}f_{ab}(t_{12}),\nonumber \\
M_{1,abcd}(t)&=\int dt'G_{ab}^{R}(t-t')G^{R}_{cd}(-t')e^{\lambda_Lt'},\nonumber \\
M_{2,ab}(t)&=
-\frac{\mathcal{J}_{ab}^2 2^{q-1}(q-1)}{q}s_{ab}G_{ab}\Bigl(\frac{\beta}{2}+it\Bigr)^{q-2}.
% -\frac{\mathcal{J}^2 2^{q-1}(q-1)}{q}s_{ab}G_{ab}\Bigl(\frac{\beta}{2}+it\Bigr)^{q-2}.
\end{align}
% Note that $M_{1,abcd}$ itself can also be written as a convolution: $M_{1,abcd}=G^{(0)R}_{ab}\circ ({\widehat G}^{(0)R}_{cd}e^{\lambda_Lt})$ with ${\widehat G}^{(0)R}_{ab}(t)=G^{(0)}_{ab}(-t)$.
Hence we can compute the chaos exponent $\lambda_L(\sigma)$ for each sector rather than the chaos exponent of the full system $\lambda_L=\text{max}\{\lambda_L(1),\lambda_L(-1)\}$.

\begin{figure}
\begin{center}
\includegraphics[width=16cm]{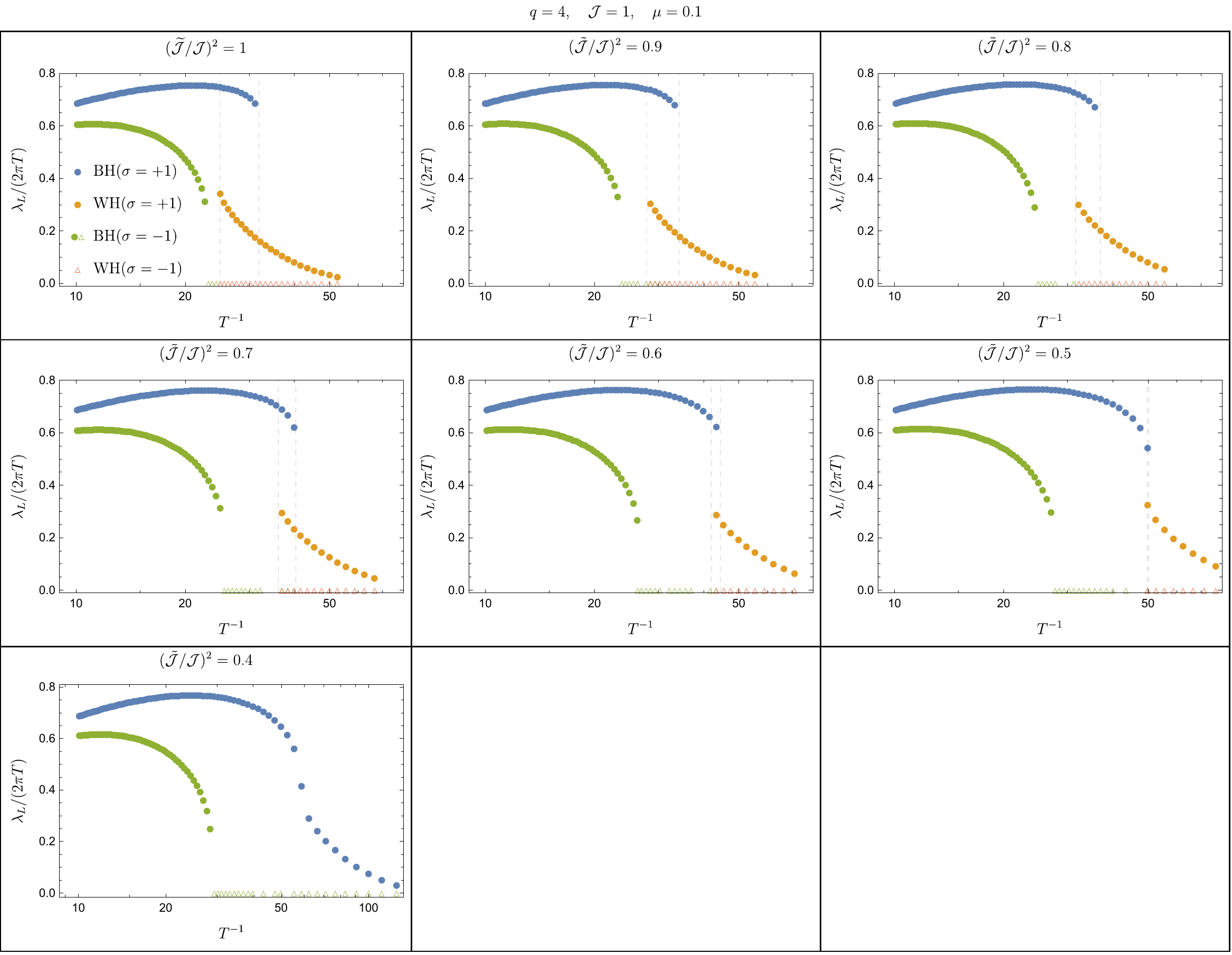}
\caption{
The chaos exponent $\lambda_L$ for the two sectors $\sigma=\pm 1$.
Here we have displayed the data points where the largest absolute value of the eigenvalues of ${\cal M}_{ab}(-1,\lambda_L;t,t')$ does not cross $1$ in $0<\lambda_L\le 2\pi T$ with empty triangle markers.
}
\label{220415_Lyapunov}
\end{center}
\end{figure}

By performing a binary search for the value of $\lambda_L$ in the range $0<\lambda_L\le 2\pi T$ such that the largest eigenvalue of ${\cal M}_{ab}(\sigma,\lambda_L;t,t')$ is $1$, we obtained the chaos exponent for $\sigma=\pm 1$ and various values of $\frac{{\tilde {\cal J}}^2}{{\cal J}^2}$, as displayed in figure \ref{220415_Lyapunov}.
We found that as $\frac{{\tilde {\cal J}}^2}{{\cal J}^2}$ is decreased the chaos exponent of each sector increases while their temperature dependence remains qualitatively the same.
This result may be interpreted in the following way.
Let us assume that the chaos exponent is associated with the operator growth over a single side (say Left).
In the two-coupled system the operator growth in a single side should be reduced due to the leakage of the operator to the operators supported on the other side.
As we have seen in the previous section, as $\frac{{\tilde {\cal J}}^2}{{\cal J}^2}$ is decreased the $LR$ transmission becomes suppressed.
Hence the chaos exponent is expected to become larger as $\frac{{\tilde {\cal J}}^2}{{\cal J}^2}$ is decreased, which is consistent with the results in figure \ref{220415_Lyapunov}.

\begin{figure}
\begin{center}
\includegraphics[width=8cm]{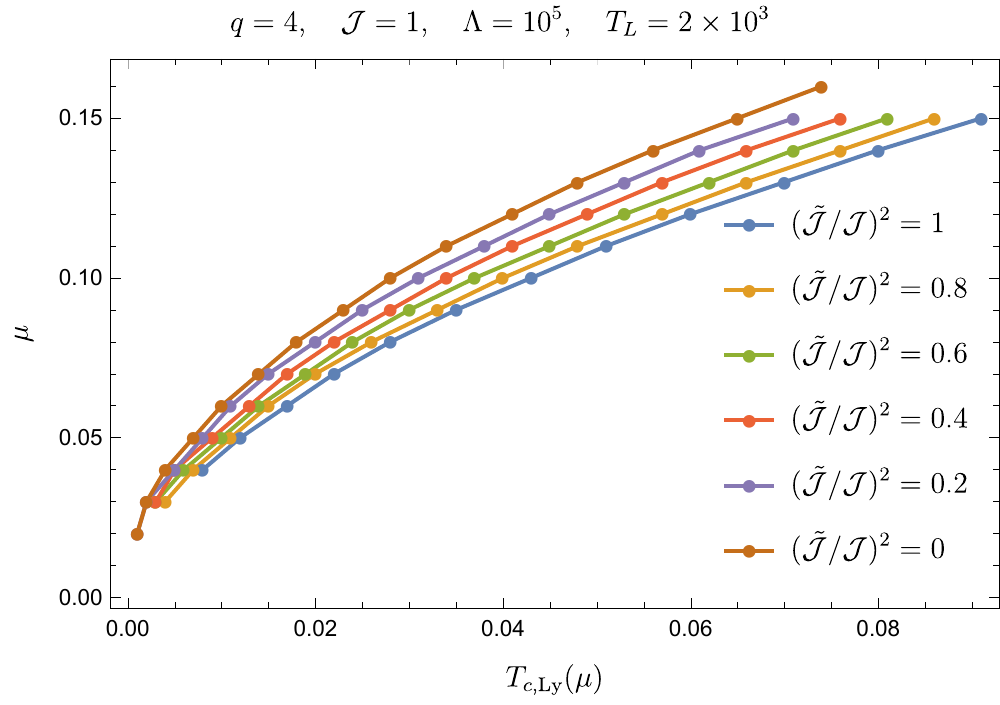}
\caption{
The temperature $T_{c,\text{Ly}}$ where the binary search to determine $\lambda_L(-1)$ fails.
}
\label{fig_TcLy}
\end{center}
\end{figure}

Interestingly we also found that when the temperature is lower than some critical value $T_{c,\text{Ly}}(\frac{{\tilde {\cal J}}}{{\cal J}},\mu)$ (which is larger than $T_{c,\text{2BH}}$), the absolute values of the eigenvalues of ${\cal M}_{ab}(-1,\lambda_L;t,t')$ are smaller than $1$ for any value of $\lambda_L$ in $0< \lambda_L\le 2\pi T$.
This would imply that there are no exponentially growing modes in $\sigma=-1$ sector for $T< T_{c,\text{Ly}}$.
In figure \ref{fig_TcLy} we display the observed values of $T_{c,\text{Ly}}$ for various ${\tilde {\cal J}}/{\cal J}$ and $\mu$.
% We discuss more in section \ref{sec_discuss} the possible interpretations for this observation.

\section{large $q$ limit}

\label{sec_largeq}

In the large $q$ limit, we can study the $J_{i_1\cdots i_q}^{(L)}\neq J_{i_1\cdots i_q}^{(R)}$ model analytically.
In this section we study this limit and  compare with the former section of the numerical analysis at finite $q$ .

\subsection{large $q$ limit at zero temperature}
In the large $q$ limit, the $G\Sigma$ action reduces to the Liouville action:
\ba
\f{S_E}{N} &= \f{1}{8q^2} \int d\tau_1 \int d\tau_2 \Big( \partial_{\tau_1} g_{LL}(\tau_1,\tau_2) \partial_{\tau_2}g_{LL}(\tau_1,\tau_2) - \partial _{\tau_1 }g_{LR}(\tau_1,\tau_2)\partial_ {\tau_2 }g_{LR}(\tau_1,\tau_2) \Big) \notag \\
& -\f{\mathcal{J}^2}{2q^2} \int d\tau_1 \int d \tau_2  e^{g_{LL}(\tau_1,\tau_2)}- \f{\tilde{\mathcal{J}}^2}{2q^2} \int d\tau_1 \int d \tau_2  e^{g_{LR}(\tau_1,\tau_2)}  - \f{\hat{\mu}}{ 2 q^2}\int d\tau g_{LR}(\tau,\tau),
\label{largeqLiouville}
\ea
with the following ansatz for the large $q$ expansion 
\ba
G_{LL}(\tau) &= G_{RR}(\tau)=\f{1}{2} \text{sgn}(\tau)\Big( 1 + \f{1}{q}g_{LL}(\tau) + \cdots \Big) ,\notag \\
G_{LR}(\tau)&= -G_{RL}(\tau)=\f{i}{2} \Big( 1 + \f{1}{q} g_{LR}(\tau) + \cdots \Big) ,
\label{GLLGLRlargeq}
\ea
and we also scale $\mu$ so that $\hat{\mu} = \mu q$ is kept finite in the large $q$ limit.
%The derivation is shown in the appendix \ref{appendix_ZinGandSigma}.
At small temperature and the late time of order $\tau \sim q$, this approximation is not valid because of the exponential decay of the correlation functions.
In this case, we also consider the solution in $\tau \gg q $ regime and impose the matching condition between $\tau \ll q$ and $\tau \gg q$ solutions, as we discuss in  section \ref{sec:largeqfiniteT}.
The Schwinger-Dyson equation reduces to the following two equation:
\ba
\partial_{\tau}^2 g_{LL}(\tau) &= 2 \mathcal{J}^2e^{g_{LL}(\tau)}, \qquad (\text{for} \ \ \ \tau > 0) \notag \\
\partial_{\tau}^2 g_{LR}(\tau) &= -2 \tilde{\mathcal{J}}^2e^{g_{LR}(\tau)}-2 \hat{\mu} \delta(\tau), \label{eq:liouville1}
\ea
with the boundary conditions 
\ba
&g_{LL}(0) = 0, \qquad  \partial _{\tau} g_{LR}(0^+) = - \hat{\mu}, \notag \\
&g_{LL}(\tau)- g_{LR}(\tau)\to 0 , \qquad  \text{as} \ \ \  \tau \to \infty. \label{eq:bdycond1}
\ea
The general solutions of the equations (\ref{eq:liouville1}) are
\ba
e^{g_{LL}(\tau)} &= \f{\alpha^2 }{\mathcal{J}^2 \sinh^2(\alpha |\tau| + \gamma)}, \notag \\
e^{g_{LR}(\tau)} &= \f{\tilde{\alpha}^2 }{\tilde{\mathcal{J}}^2 \cosh^2(\tilde{\alpha} |\tau| + \tilde{\gamma})}, \label{eq:earlytau}
\ea
with constants of the integration $\alpha, \tilde{\alpha}, \gamma, \tilde{\gamma}$. 
These parameters are determined by the boundary conditions which depend on the temperature.

Each of the boundary conditions (\ref{eq:bdycond1}) fixes the constants of integration in a following way.
First, the boundary conditions at $\tau = 0$ give the relations
\ba
&g_{LL}(0) = 0 \qquad \Rightarrow \qquad  \f{\alpha}{\mathcal{J} \sinh\gamma } = 1 \notag \\
&\partial_{\tau}g_{LR}(0^+) = -\hat{\mu} \qquad  \Rightarrow \qquad   \ 2\tilde{\alpha}\tanh\tilde{\gamma} =  \hat{\mu} ,
\label{bcattau0}
\ea
whereas the boundary condition at $\tau = \infty$ gives
\be
g_{LL}(\tau)- g_{LR}(\tau)\to 0 , \qquad  \text{as} \ \  \tau \to \infty \qquad  \Rightarrow \qquad  \tilde{\gamma} = \gamma + s, \qquad \alpha = \tilde{\alpha}.
\ee
Here $s = \log \f{\mathcal{J}}{\tilde{\mathcal{J}}}$ is a positive parameter and vanish when $\tilde{\mathcal{J}} = \mathcal{J}$.
%In terms of $\cos 2 c$, the parameter is given as 
%\be
%s = -\f{1}{2}\log(\cos 2c). 
%\ee
Finally, $\gamma$ satisfies the equation 
\be
 \sinh \gamma \tanh (\gamma + s) = \f{\hat{\mu}}{2\mathcal{J}}, \label{eq:gammarelation}
\ee
and other parameters are determined through $\gamma$.
The physical gap $E_{gap}$ is given by
\be
E_{gap} %= 2 \alpha \Delta 
= \f{2 \alpha}{q}.
\ee
For small $j \equiv \mathcal{J}-\tilde{\mathcal{J}}$ limit, we can separate the scale of Maldacena-Qi behavior and Kourkoulou-Maldacena behavior.
When $j \ll \mu $, we can ignore the parameter $s$ and we obtain 
\be
\sinh \gamma \tanh \gamma =  \f{\hat{\mu}}{2\mathcal{J}},
\ee
which is exactly the same equation with the Maldacena-Qi model.
$\gamma$ is given by 
\be
\tanh^2 \gamma =  \f{\epsilon}{2} (\s{4 + \epsilon^2} - \epsilon) , \qquad \epsilon = \f{\hat{\mu}}{2\mathcal{J}}.
\ee
In the range $j \ll \hat{\mu} \ll \mathcal{J}$ 
%\textcolor{red}{[25 Mar 2022 Nosaka, $j\ll\hat{\mu}\ll\mathcal{J}$?]}
, we can expand as 
\be
\gamma \approx \s{\epsilon}= \s{\f{\hat{\mu}}{2\mathcal{J}}}, \qquad \alpha = \mathcal{J}\sinh \gamma  \approx \s{\f{\hat{\mu}\mathcal{J}}{2}}
\ee
and the gap is given by
\be
E_{gap}=\f{2\alpha }{q} \approx  \f{\s{2\hat{\mu}\mathcal{J}}}{q}.
\ee 
This parameter regime exists only when $j \ll \mathcal{J}$.
In the regime $\hat{\mu}\ll j$, we can ignore the $\gamma$ from the  term $\tanh(\gamma + s)$.
Therefore we obtain the equation 
\be
\sinh \gamma \tanh s \approx \f{\hat{\mu}}{2\mathcal{J}}.
\ee
We can evaluate $\tanh s$ as 
\be
\tanh s = \f{e^s - e^{-s}}{e^s + e^{-s}} = \f{\mathcal{J}^2 - \tilde{\mathcal{J}}^2}{\mathcal{J}^2 + \tilde{\mathcal{J}}^2} 
%\approx \f{j}{\mathcal{J}}.
\ee
Then, $\gamma$ is evaluated as 
\be
\sinh \gamma \approx \f{\hat{\mu}}{2\mathcal{J}\tanh s}.
\ee
When $\tilde{\mathcal{J}} = 0$ this reduces to the relation of Kourkoulou-Maldacena model.
This parameter regime always exists but we need to take $\hat{\mu}$ to be smaller than $j$.
When $\tilde{\mathcal{J}} = \mathcal{J}$, i.e., the perfect correlation between left and right SYK model, this regime vanish which occurs in the Maldacena-Qi model.
The parameter $\alpha$ becomes 
%\be
% \alpha = \mathcal{J} \sinh \gamma \approx  \f{\hat{\mu}}{2\tanh s}
$\alpha \approx  \frac{\hat{\mu}}{2\tanh s}$.
%\ee
The mass gap in this limit becomes 
\be
E_{gap}=\f{2\alpha }{q} \approx  \f{\hat{\mu} }{q \tanh s}.
\ee
For $j={\cal J}$, i.e., when there are no correlations between $J_{i_1\cdots i_q}^{(L)}$ and $J_{i_1\cdots i_q}^{(R)}$, we have $E_{gap} = \f{\hat{\mu}}{q}$, which is the same with that of the Kourkoulou-Maldacena model.

\begin{figure}[ht]
\begin{center}
\includegraphics[width=11.0cm]{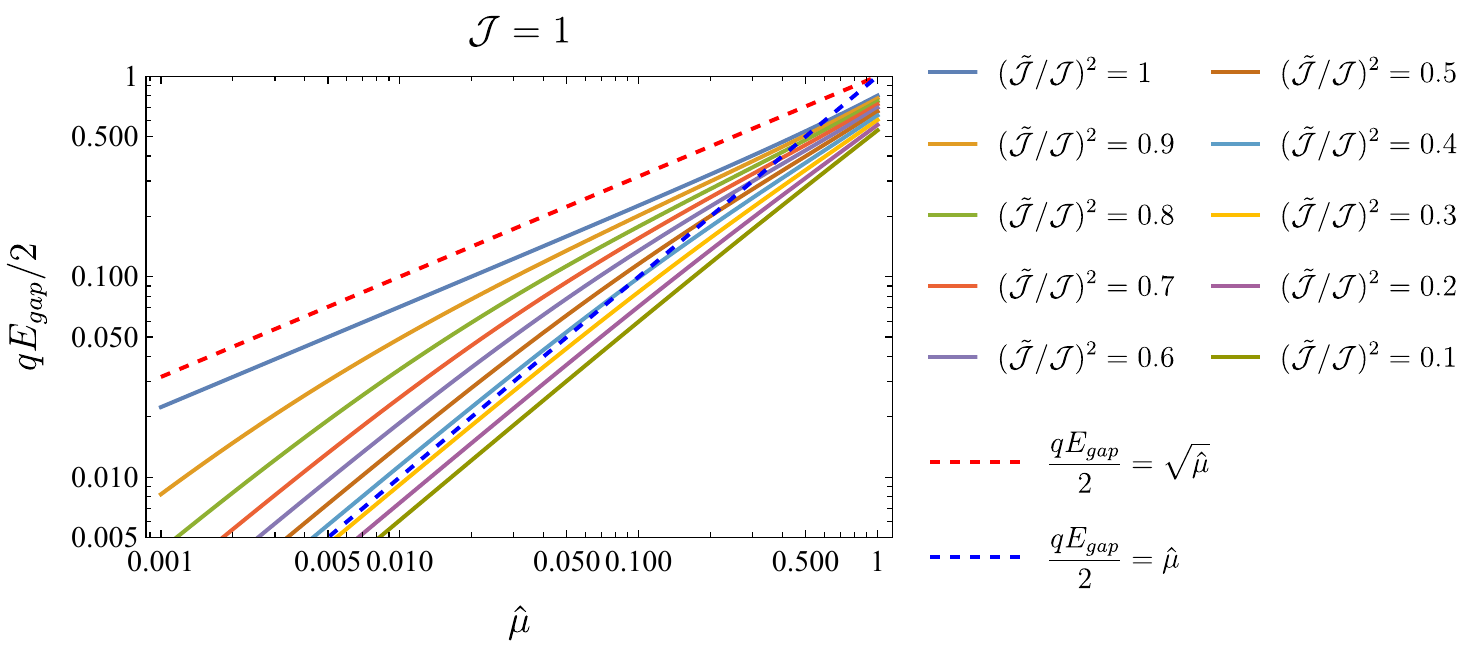}
\caption{Plot of $E_{gap}$ as a function of $\hat{\mu}$.
}
\label{fig:EgapMuPlotq96}
\end{center}
\end{figure}

\begin{figure}[ht]
\begin{center}
\includegraphics[width=7.0cm]{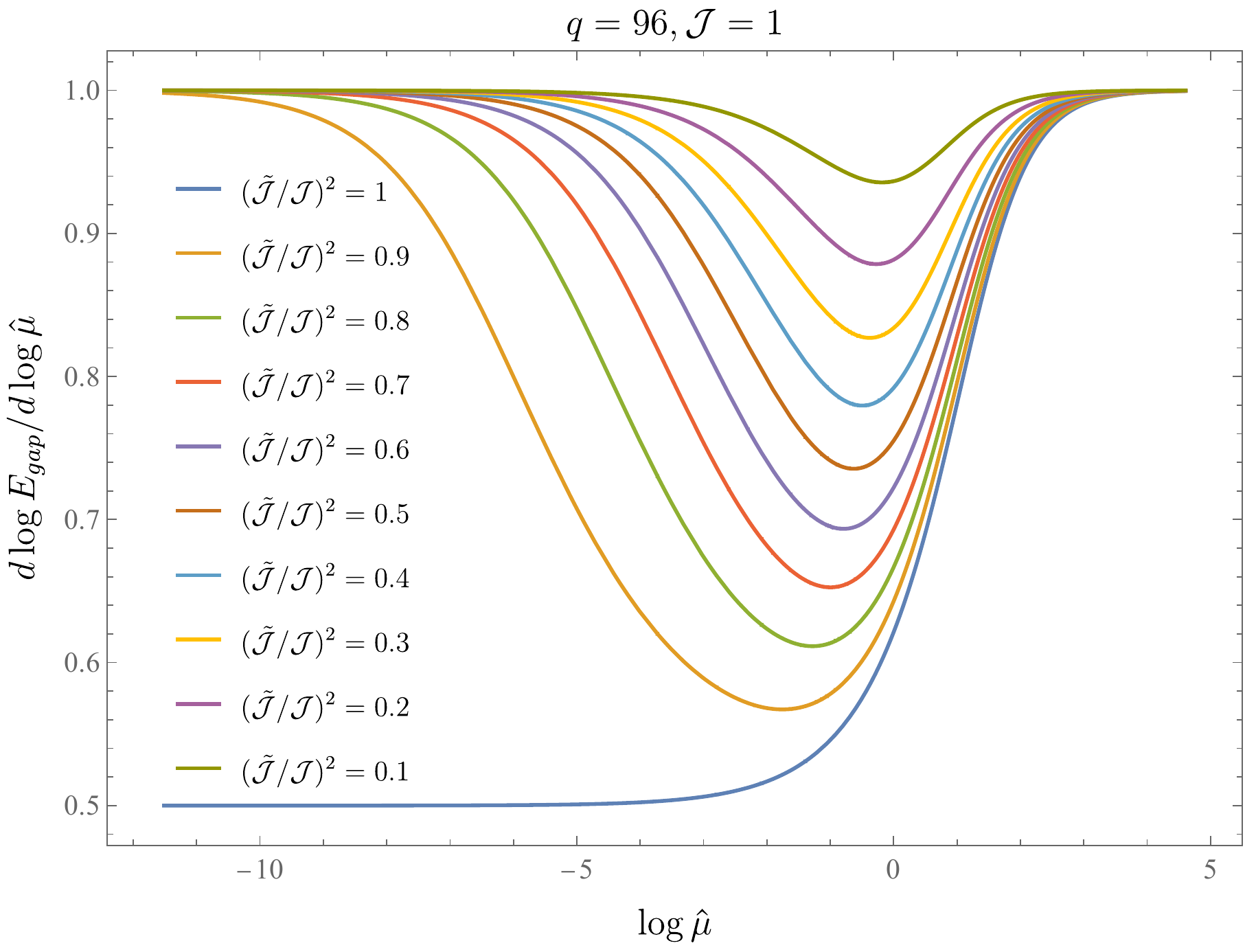}
\includegraphics[width=7.0cm]{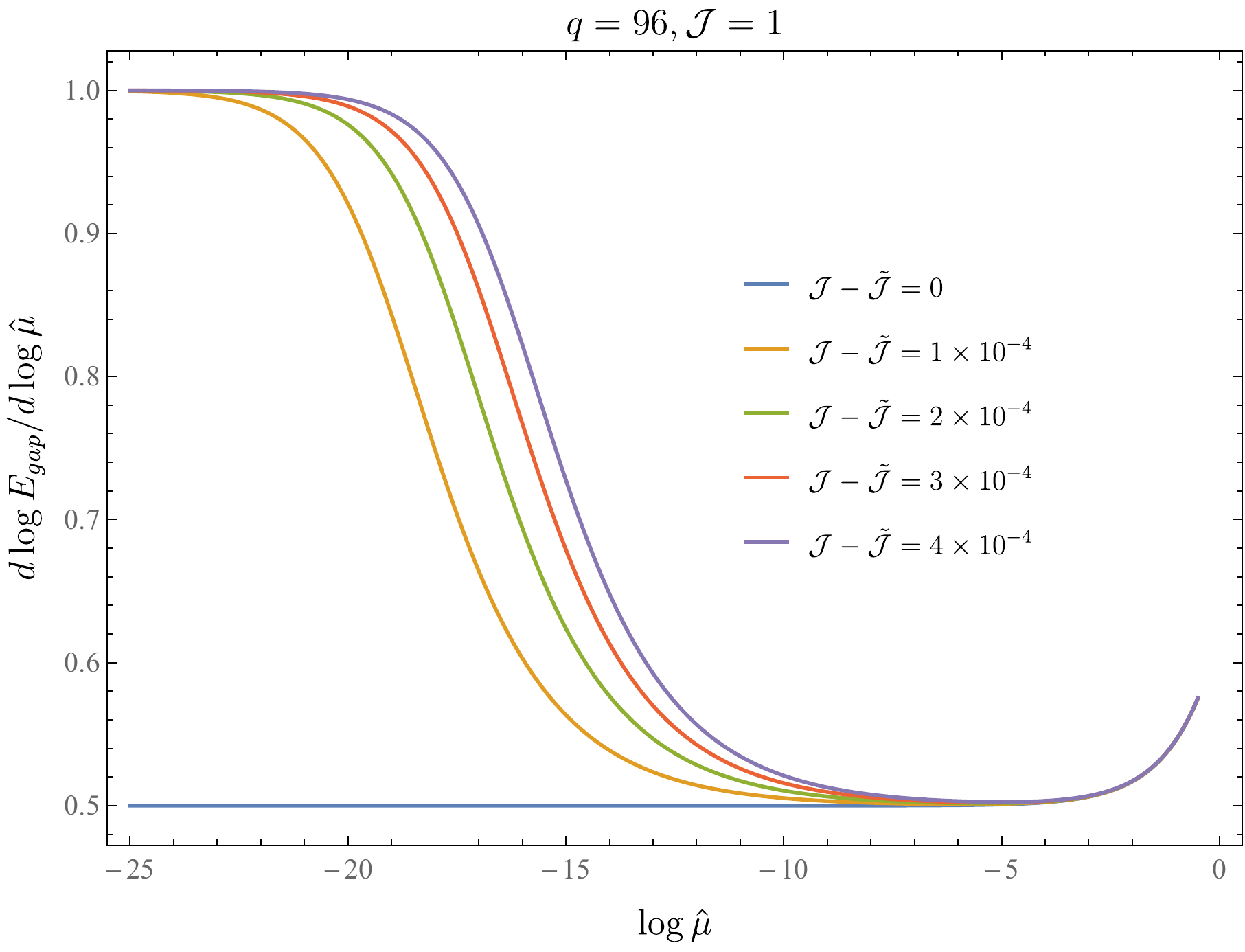}
\caption{Plots of the power of $E_{gap}$ in $\hat{\mu}$.
}
\label{fig:LargeQgapscale}
\end{center}
\end{figure}

%It is interesting to study the power of $E_{gap}$ in $\mu$ in the large $q$ limit.
Let us study how the behavior of $E_{gap}$ in $\mu$ changes when we decrease $\mu$.
The plot of $E$ as a function of $\mu$ is shown in figure \ref{fig:EgapMuPlotq96}.
The power of the gap in $\mu$ is defined as $\f{d \log E_{gap}}{d \log \mu} $. 
Here we take the derivative with respect to $\mu$ while we fix $\mathcal{J}$ and $\mathcal{\tilde{J}}$.
Then $s$ is also fixed and this is equivalent to take the derivative with respect to $\gamma$ while fixing $s$.
This becomes 
\ba
\f{d \log E_{gap}}{d \log \mu} &= \f{\hat{\mu}}{\alpha} \f{\partial \gamma}{\partial \hat{\mu}} \f{\partial \alpha}{\partial \gamma} = \f{\sinh (\gamma + s)\cosh (\gamma + s)}{\sinh (\gamma + s)\cosh(\gamma + s) + \tanh \gamma}.  \label{eq:LargeQgapscale}
\ea
The plot of \eqref{eq:LargeQgapscale} is shown in figure \ref{fig:LargeQgapscale}.
We can see that for very small $j$, there is a region where the power in $\mu$ is almost $1/2$, which is the behavior in the Maldacena-Qi model.
On the other hand, even for small $j$, the power in $\mu$ approaches to $1$ for sufficiently small $\mu$ as we expect.

The real time correlation function is obtained by analytically continuing to the Lorentzian time.
At low temperature of order $\beta = O(q\log q)$ we see that there are no decay and the return amplitude just oscillate.
This is because the decay rate is of order $e^{-(\f{q}{2}-1)E_{gap}}$, which is non-perturbatively small in $q$ \cite{Nosaka:2020nuk,Qi:2020ian} at large $q$ limit, and we did not take into account this decay rate at large $q$.
Since the decay rate is also suppressed by the energy gap $E_{gap}$, the decay rate will increase as we decrease the correlation of the random couplings between the two sides.

\subsection{Finite temperature \label{sec:largeqfiniteT}}
For $\tau \ll q$ we can still use the solution (\ref{eq:earlytau}) together with the boundary conditions at $\tau\rightarrow 0$ \eqref{bcattau0}.
For $\tau \gg q$, we can make a different approximation for the Schwinger-Dyson equation \eqref{simplifiedSD} which goes as follows.
First, from the second equation in \eqref{220201_EuclideanSDeq} we can approximate the self energy $\Sigma_{ab}(\tau)$ at the time scale of $G_{ab}(\tau)$ by the delta function configurations (see section 5.4 in \cite{Maldacena:2018lmt})
% For $\tau \gg q$, we can approximate the self energy $\Sigma_{LR}(\tau)$ by the delta functional configuration
\be
\Sigma_{LL}(\tau),\Sigma_{RR}(\tau)\approx \rho\partial_\tau\delta(\tau),\quad
\Sigma_{LR}(\tau)=-\Sigma_{RL}(\tau) \approx -i \nu \delta(\tau), \qquad \nu = i \int _0^\beta \Sigma_{LR}(\tau) d\tau = \f{2\tilde{\alpha}}{q},
% \Sigma_{LR}(\tau)=-\Sigma_{RL}(\tau) \approx -i \nu \delta(\tau), \qquad \nu = i \int _{-\infty}^{\infty} \Sigma_{LR}(\tau) d\tau = \f{2\tilde{\alpha}}{q},
\label{SigmaLRdelta}
\ee
where $\rho$ is a constant of order ${\cal O}(q^{-1})$.
We have evaluated the last integration by using $G_{LR}(\tau)$ in the small $\tau$ regime \eqref{GLLGLRlargeq},\eqref{eq:earlytau} which gives $i\Sigma_{LR}(\tau)\approx \frac{{\tilde\alpha}^2}{q\cosh^2({\tilde\alpha}|\tau|+{\tilde\gamma})}+\mu\delta(\tau)$, and replacing the domain of integration with $(-\infty,\infty)$.
With \eqref{SigmaLRdelta}, the other Schwinger-Dyson equation is simplified as
% Then, the equation of motion is simplified as 
\ba
&(1-\rho)\partial_{\tau}G_{LL}(\tau)-i\nu G_{LR}(\tau)=0,\quad \\
&(1-\rho)\partial_{\tau}G_{LR}(\tau)+i\nu G_{LL}(\tau)=0,
% &\partial _{\tau}G_{LL}(\tau)  + i\nu G_{RL}(\tau) = 0, \notag \\
% &\partial _{\tau}G_{RL}(\tau)  + i\nu G_{RR}(\tau) = 0
\ea
We can ignore $\rho$ in each equation since it gives a sub-leading correction in the large $q$ limit.
Solving these equations we have
\be
G_{LL} (\tau) = A \cosh \Big[\nu \Big(\f{\beta}{2} -\tau \Big) \Big], \qquad G_{LR} (\tau) =i A \sinh \Big[\nu \Big(\f{\beta}{2} -\tau \Big) \Big],
\ee
where $A$ is an integration constant.
The other integration constant (translation in $\tau$) is fixed by the conditions $G_{LL}(\tau)=G_{LL}(\beta-\tau), G_{LR}(\tau)=-G_{LR}(\beta-\tau)$ which follows from \eqref{simplifiedsymproperty}.
\begin{figure}[ht]
\begin{center}
\includegraphics[width=8cm]{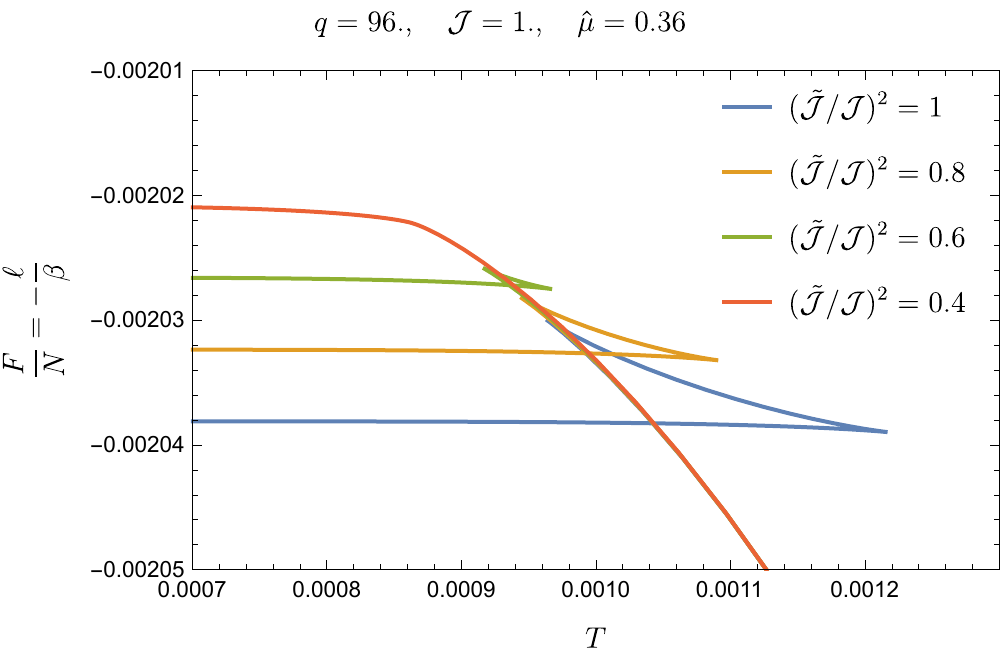}
\caption{Plots of the free energy as a function of the temperature for representative $\tilde{\mathcal{J}}/\mathcal{J}$.
}
\label{fig:FandTplotq96}
\end{center}
\end{figure}

Let us define
% defined
$\sigma = q e^{-\beta \nu}$ as a new parameter corresponding to the temperature, and assume that $\sigma$ is of order ${\cal O}(q^0)$ (i.e., $\beta={\cal O}(q\log q)$).
Matching the two solutions in the overlapping regime, i.e., the large $\tau$ expansion of the solution for $\tau\ll q$ with the small $\tau$ expansion of the solution for $\tau \gg q$ as
\begin{align}
G_{LL}(\tau)\approx \frac{1}{2}+\frac{1}{q}\log\frac{2\alpha}{{\cal J}}-\frac{\gamma}{q}-\frac{\alpha\tau}{q}\approx A\Bigl(\cosh\frac{\nu\beta}{2}-\nu\tau\sinh\frac{\nu\beta}{2}\Bigr),\nonumber \\
-iG_{LR}(\tau)\approx \frac{1}{2}+\frac{1}{q}\log\frac{2{\tilde\alpha}}{{\tilde {\cal J}}}-\frac{{\tilde\gamma}}{q}-\frac{{\tilde\alpha}\tau}{q}\approx
A\Bigl(\sinh\frac{\nu\beta}{2}-\nu\tau\cosh\frac{\nu\beta}{2}\Bigr),
\end{align}
we can find the integration constants as
% (up to ${\cal O}(q^{-1})$ corrections)
\be
\tilde{\alpha} = \alpha ,\qquad \tilde{\gamma} = \gamma + s + \sigma, \qquad \alpha = \mathcal{J}\sinh \gamma, \qquad  \hat{\mu} = 2\tilde{\alpha} \tanh \tilde{\gamma}.
\label{intconstrelations2}
\ee

\begin{figure}[ht]
\begin{center}
\includegraphics[width=17.0cm]{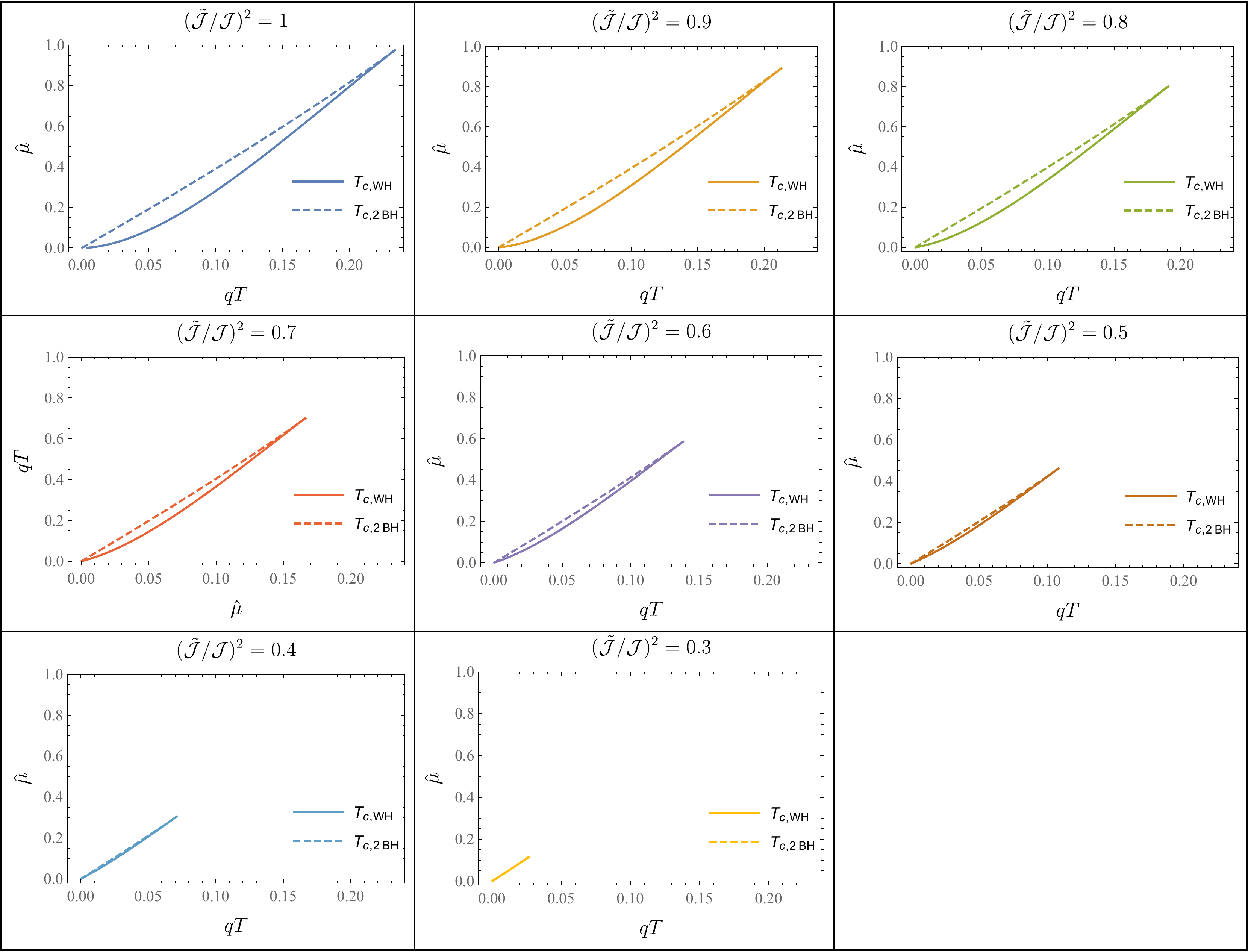}
\caption{The plots of phase diagrams for representative $\tilde{\mathcal{J}}$.}  
\label{fig:LargeQphases}
\end{center}
\end{figure}
\begin{figure}[ht]
\begin{center}
\includegraphics[width=8.0cm]{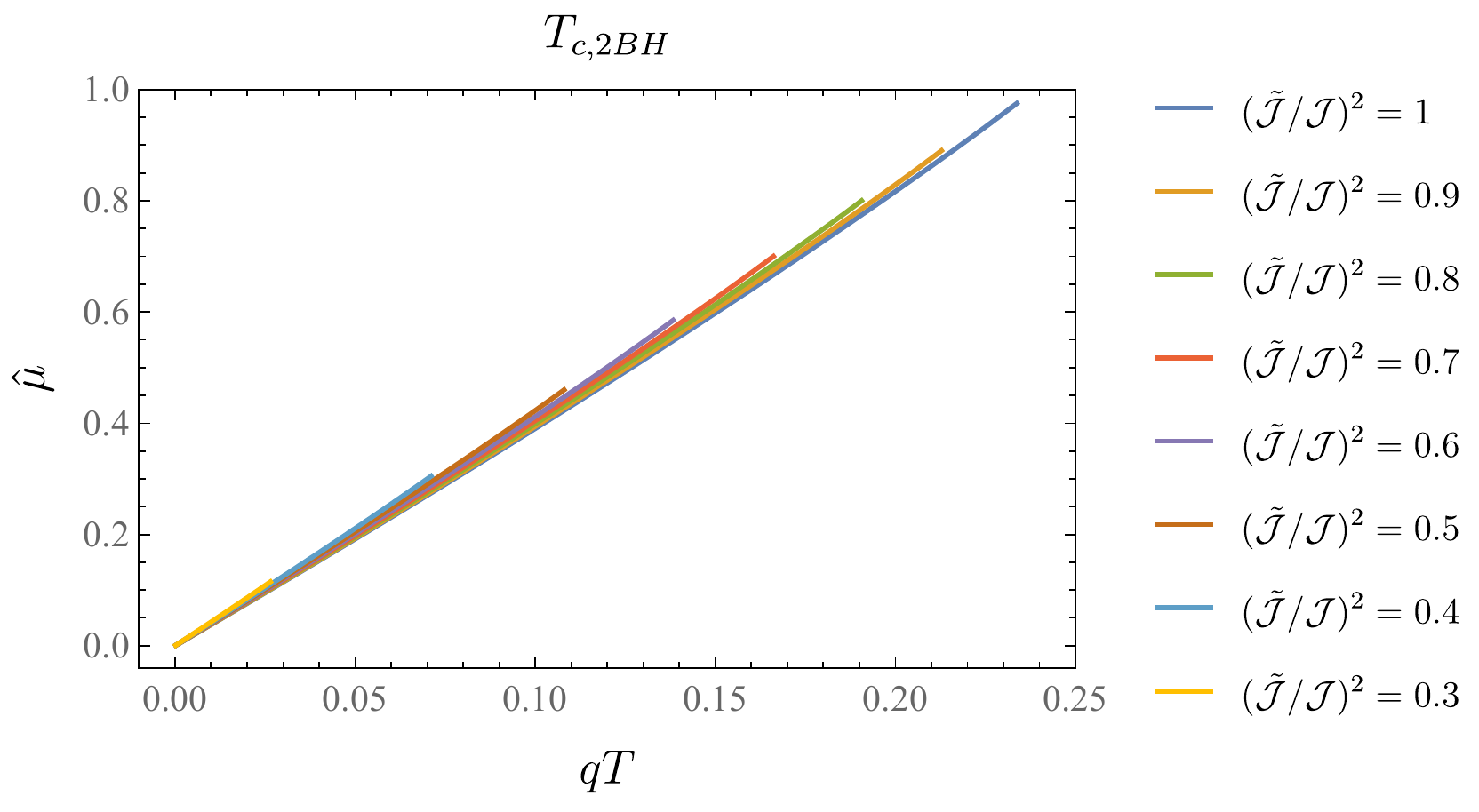}
\includegraphics[width=8.0cm]{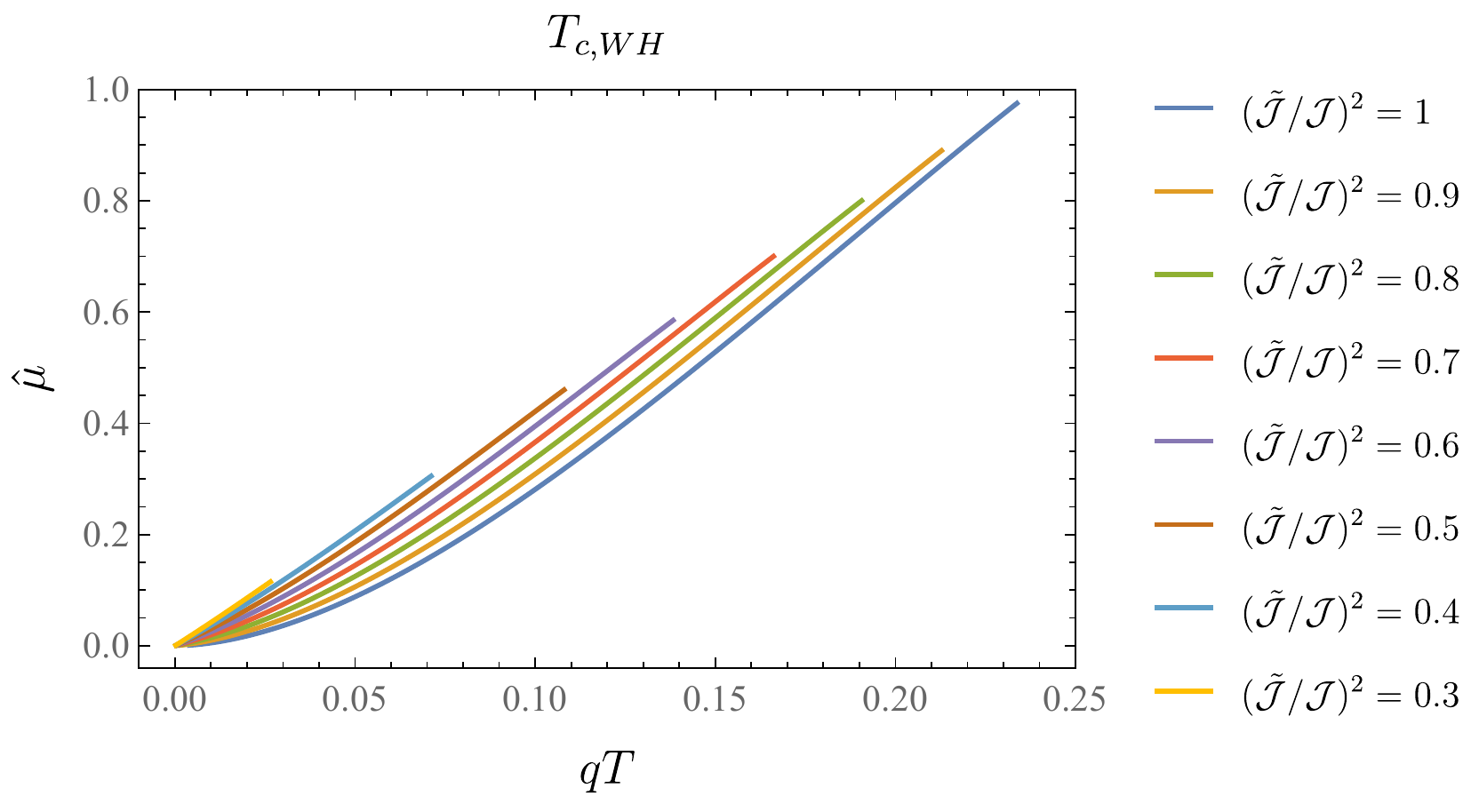}
\caption{Plots of $T_{c,2BH}$ and $T_{c,WH}$.
The parameters are taken to be $\mathcal{J} = 1$, $q = 96$.}  
\label{fig:LargeQTcWHTcBH}
\end{center}
\end{figure}

Using the relation between the correlation functions and the energy \eqref{energy} we obtain
\ba
\f{E}{N}% &= \Bigg[ \f{1}{q} \partial_{\tau_1}G_{LL} + \f{1}{q}\partial _{\tau_1}G_{RR} + i \mu \Bigg( 1 - \f{2}{q}\Bigg)G_{LR} \Bigg]_{\tau_{12} \to 0} \notag \\
&=  \f{1}{2q} \partial_{\tau}g_{LL}(0) +  \f{1}{2q} \partial_{\tau}g_{RR}(0) +  i \mu \Big( 1 - \f{2}{q}\Big) \f{i}{2} \Big( 1 + \f{1}{q} g_{LR}(0)\Big) \notag \\
&= -\f{2\mathcal{J}}{q^2} \cosh \gamma - \f{\hat {\mu}}{2 q}   + \f{\hat{\mu}}{q^2 } \Big(1 +  \log \f{e^s \sinh \gamma}{\cosh \tilde{\gamma}} \Big). \label{eq:energylargeq}
\ea
The effective action $\ell = \f{1}{N}\log Z$ is then (see appendix \ref{app_largeqderivation})
\be
\ell(\sigma,\gamma) = \f{\tanh \tilde{\gamma} \log \f{q}{\sigma}}{q} \Big(\f{q}{2} -1 + \f{1}{\tanh \gamma \tanh \tilde{\gamma}} + \log \f{\sinh \gamma}{\cosh \tilde{\gamma}} + \f{\sigma}{\tanh \tilde{\gamma}} +s\Big) + \f{\sigma}{q}. \label{eq:LargeqPF}
\ee
Now we study the free energy $\frac{F}{N}=-\frac{\ell}{\beta}$ for representative $\tilde{\mathcal{J}}$.
% Now we study the partition function for representative $\tilde{\mathcal{J}}$.
This can be worked out in the following way.
First we choose $q,{\cal J},{\tilde {\cal J}},{\hat \mu}$ to a particular set of values.
Then, by using the relations ${\hat \mu}=2{\cal J}\sinh\gamma\tanh{\tilde\gamma}$ \eqref{intconstrelations2} and $\beta=-\frac{1}{2{\cal J}\sinh\gamma}\log\frac{\sigma}{q}$ we can compute $\gamma(\sigma)$ and $T(\sigma)=\beta(\sigma)^{-1}$ as functions of $\sigma$, with which the data points  $(T(\sigma),-\frac{\ell(\sigma,\gamma(\sigma))}{\beta(\sigma)})$ for the free energy can be generated.
% As this approach suggests it is more natural to use $\sigma$ as a fundamental parameter rather than the temperature $T$.
% Indeed, $T(\sigma)$ is not a monotonic function when $({\tilde {\cal J}}/{\cal J})^2$ is sufficiently large and ${\hat \mu}$ is sufficiently small, hence a single point in $T$-$\mu$ plane may correspond to several different values of $\sigma$.
% On such a point, different phases corresponding to each value of $\sigma$ coexist together.
First, the plot of the free energy as a function of $T$ for representative $(\tilde{\mathcal{J}}/\mathcal{J})^2$ is shown in figure \ref{fig:FandTplotq96}.
In figure \ref{fig:LargeQphases}, we plot the phase diagram of for
%$(\tilde{\mathcal{J}}/\mathcal{J})^2 = 1, 0.9, 0.8, 0.7, 0.6, 0.5, 0.4, 0.3$ and $q = 96$.
representative $(\tilde{\mathcal{J}}/\mathcal{J})^2$ with $q = 96$.
Clearly, we can see that the phase transition exists for smaller $\hat{\mu}$ and $T$ for sufficiently large  $\tilde{\mathcal{J}}$.
However, as we decrease $\tilde{\mathcal{J}}$ finally the phase transition disappears for any $\hat{\mu}$ and $\mathcal{J}$
(in the example of $q=96, \hat{\mu} = 0.36$ and $\mathcal{J}=1$ in figure \ref{fig:FandTplotq96} the phase transition does not exist for $(\tilde{\mathcal{J}}/\mathcal{J})^2 = 0.4$)
.
In figure \ref{fig:LargeQTcWHTcBH}, we show the phase boundaries $T_{c,2BH}$ and $T_{c,WH}$ for the same $\tilde{\mathcal{J}}$ simultaneously.
We see that the power of $T_{c,2BH}$ is not changed much but $T_{c,WH}$ becomes more straight for smaller $\tilde{\mathcal{J}}$.

\begin{figure}[ht]
\begin{center}
\includegraphics[width=5.5cm]{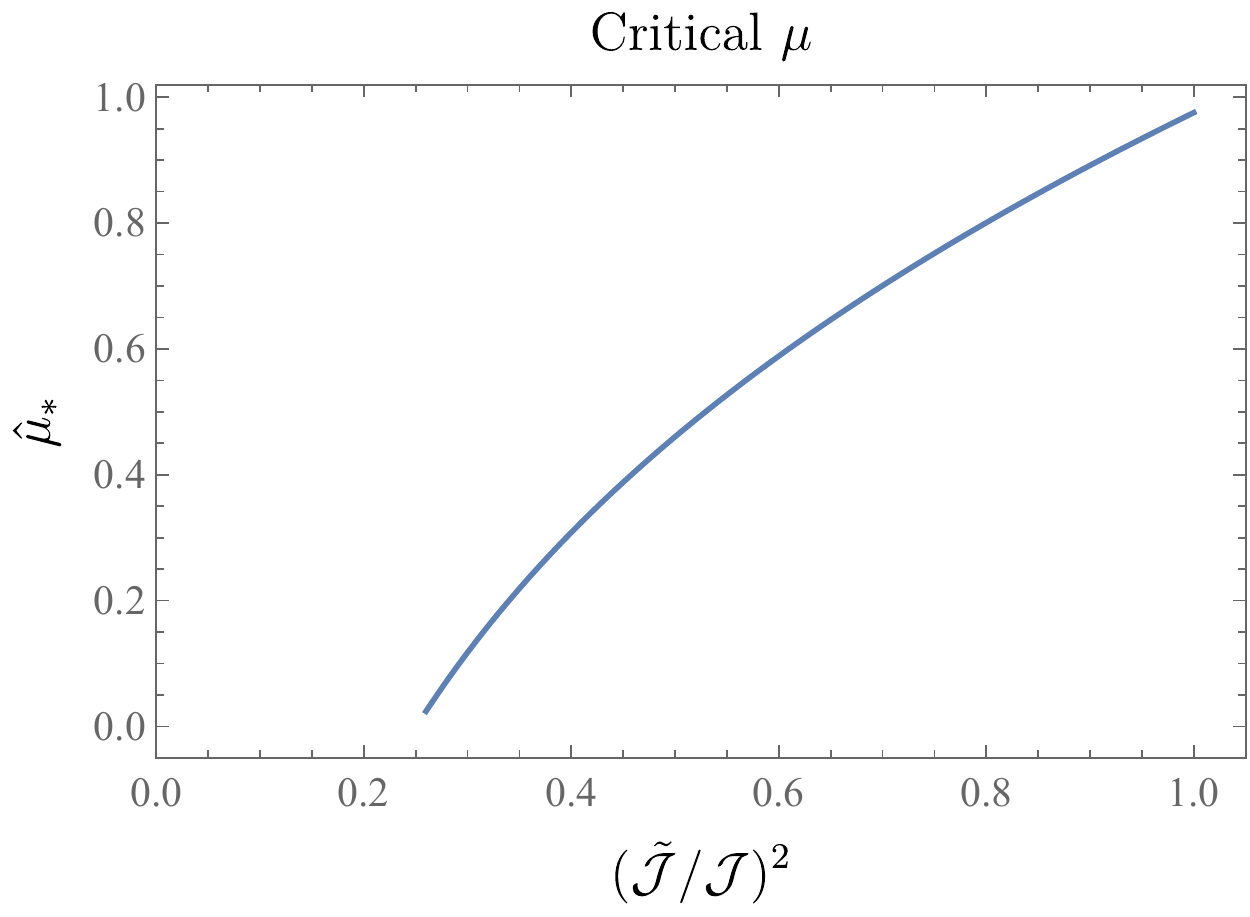}
\includegraphics[width=5.5cm]{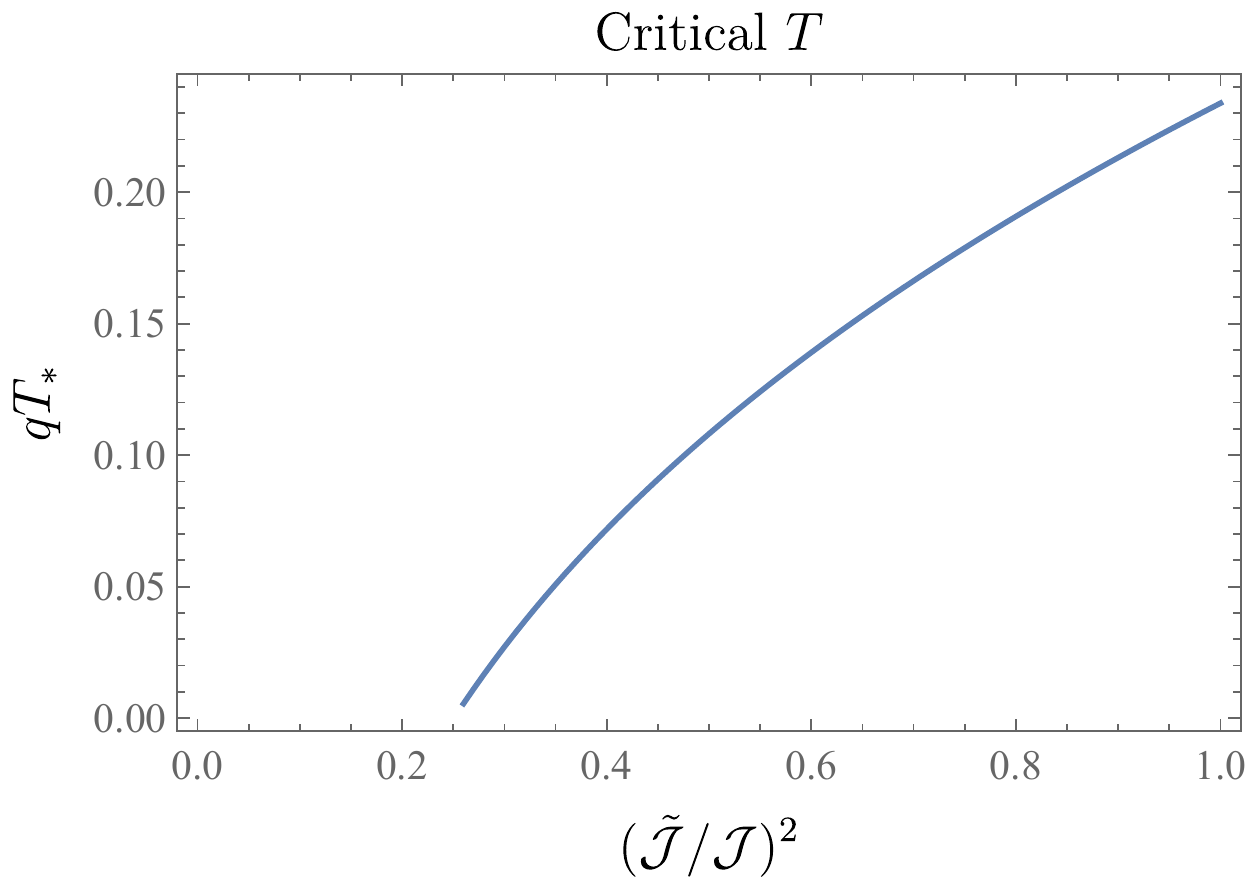}
\includegraphics[width=5.5cm]{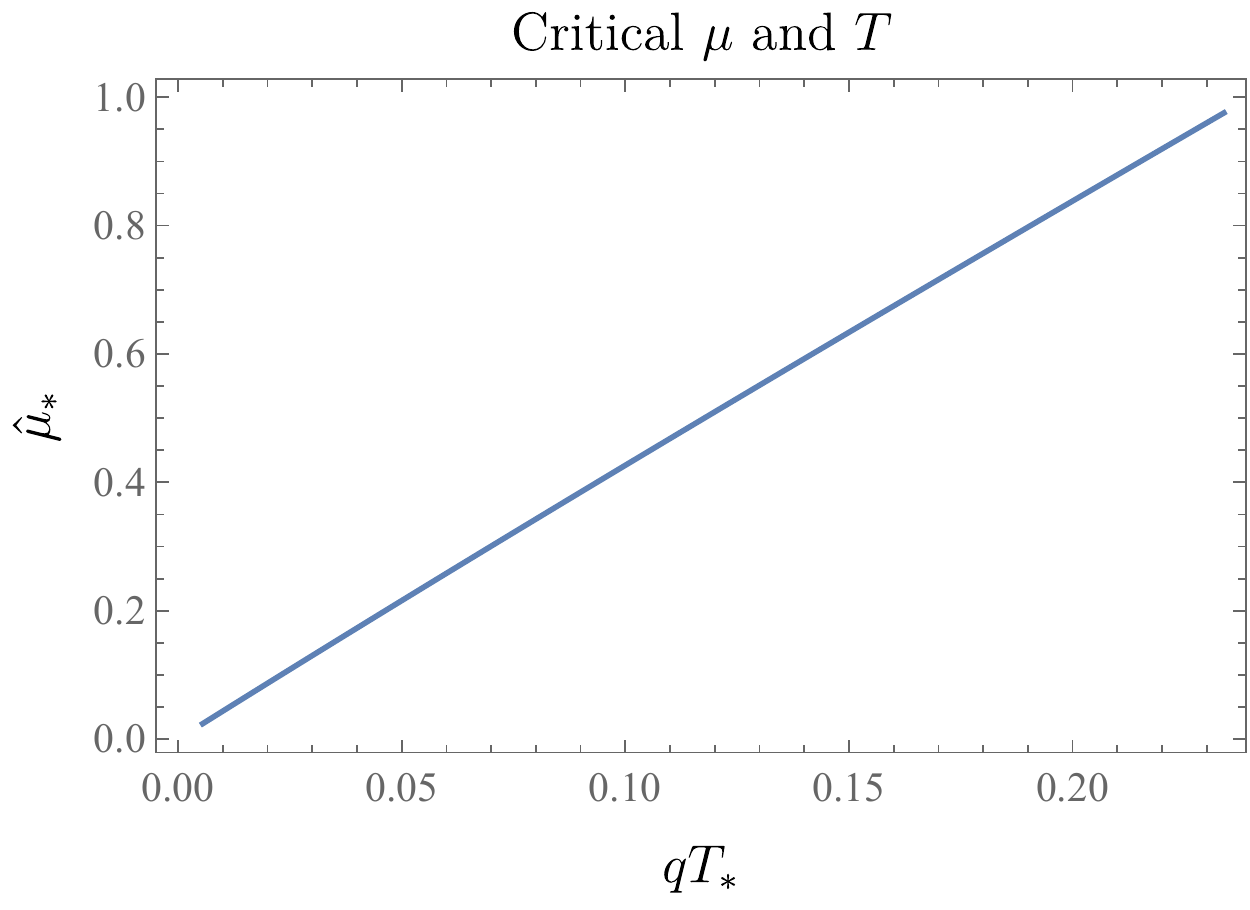}
\caption{Plots of $T_*$ and $\mu_*$.
The parameters are taken to be $\mathcal{J} = 1$, $q = 96$.
{\bf Left:} Plot of critical $\mu$ as a function of $\tilde{\mathcal{J}}/\mathcal{J}$.
{\bf Middle:} Plot of critical $T$ as a function of $\tilde{\mathcal{J}}/\mathcal{J}$.
{\bf Right:} Parametric plot of  $T_*(\tilde{\mathcal{J}}/\mathcal{J})$ and $\mu_*(\tilde{\mathcal{J}}/\mathcal{J})$.
}
\label{fig:LargeQcrit}
\end{center}
\end{figure}

The phase transition is replaced by  crossover at $\mu = \mu_*$ where $T_{c,2BH}$ and $T_{c,WH}$ meet in the phase diagrams.
We call this temperature $T_*$.
These are functions of $\tilde{\mathcal{J}}/\mathcal{J}$. 
We plot $\mu_*$ and $T_*$ as a function in $\tilde{\mathcal{J}}/\mathcal{J}$ in figure \ref{fig:LargeQcrit}.
In particular, $\mu_* = 0$, or phase transition itself entirely disappears when $\tilde{\mathcal{J}} \approx 0.50 \mathcal{J}$ for $q=96$.
For smaller $\tilde{\mathcal{J}}$ the phase transition do not exists for all $\mu$ and we only see the crossover.
In particular, even at large $q$ limit the phase transition disappears at finite $\tilde{\mathcal{J}}/\mathcal{J}$.

\subsection{Absence of phase transitions for small $\mathcal{\tilde{J}}/\mathcal{J}$.}

Let us regard $({\cal J},{\tilde {\cal J}},{\hat\mu},\sigma)$ as fundamental variables instead of $({\cal J},{\tilde {\cal J}},{\hat\mu},T)$ and express $T$ as a function of $\sigma$ (and ${\cal J},{\tilde {\cal J}},{\hat \mu}$) as explained above.
% This can be worked out in the following way.
% First we choose $q,{\cal J},{\tilde {\cal J}},{\hat \mu}$ to a particular set of values.
% Then, by using the relations ${\hat \mu}=2{\cal J}\sinh\gamma\tanh{\tilde\gamma}$ \eqref{intconstrelations2} and $\beta=-\frac{1}{2{\cal J}\sinh\gamma}\log\frac{\sigma}{q}$ we can compute $\gamma(\sigma)$ and $T(\sigma)=\beta(\sigma)^{-1}$ as functions of $\sigma$, with which the data points  $(T(\sigma),-\frac{\ell(\sigma,\gamma(\sigma))}{\beta(\sigma)})$ for the free energy can be generated.
%
% As this approach suggests it is more natural to use $\sigma$ as a fundamental parameter rather than the temperature $T$.
When $({\tilde {\cal J}}/{\cal J})^2$ is sufficiently large and ${\hat \mu}$ is sufficiently small, $T(\sigma)$ is not a monotonic function of $\sigma$, hence a single point in $T$-$\mu$ plane may correspond to several different values of $\sigma$.
On such a point, different phases corresponding to each value of $\sigma$ coexist together.
On the other hand, when $T(\sigma)$ is monotonic, there are no phase transitions since $\ell(\sigma)$ is a smooth function and hence the free energy $F$ is a smooth function of the temperature $T$.

\begin{figure}
\begin{center}
\includegraphics[width=8cm]{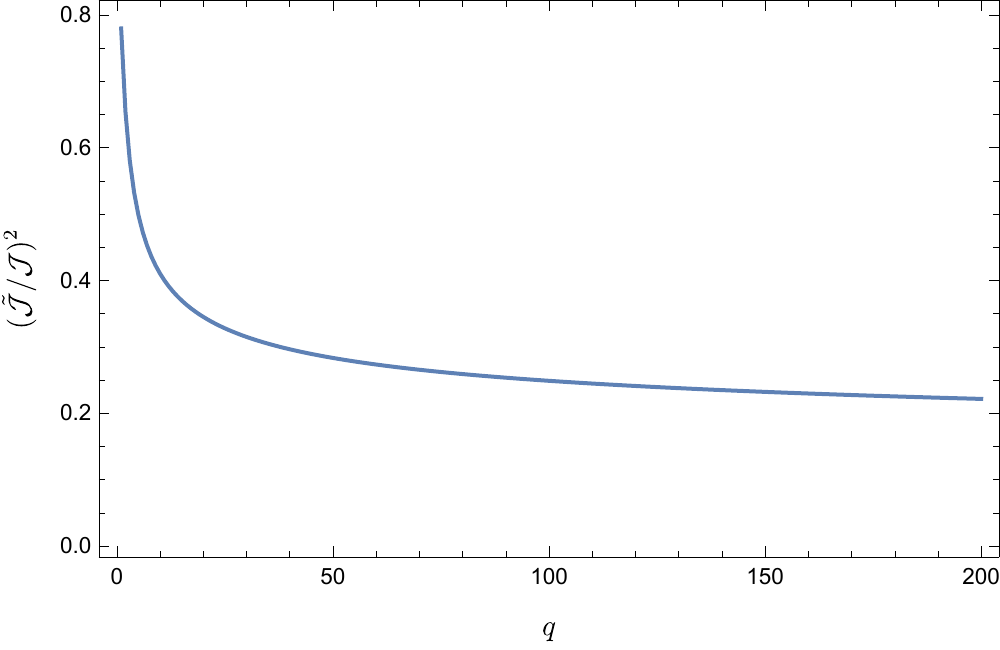}
\caption{
The critical value of $({\tilde {\cal J}}/{\cal J})^2$ in the large $q$ approximation where the phase transition becomes a smooth crossover in the entire $({\hat \mu},T)$-plane.
}
\label{221014_criticalcalJtilde_largeq_qdependence}
\end{center}
\end{figure}

Now we study the monotonicity of $\beta(\sigma)$.
The derivative becomes 
\be
\f{d\beta }{d \sigma} \equiv \f{d \beta(\gamma(\sigma),\sigma)}{d\sigma}  = \f{q}{\hat{\mu}} \tanh \tilde{\gamma} \Big( \f{\log \f{q}{\sigma}}{\sinh \tilde{\gamma}\cosh \tilde{\gamma} + \tanh \gamma} - \f{1}{\sigma} \Big). \label{eq:dbetadsigma}
\ee
Here we used the chain rule 
$\f{d\beta}{d\sigma} = \f{d \gamma}{d\sigma} \f{\partial \beta}{\partial \gamma} + \f{\partial \beta}{\partial \sigma}$
, together with $\frac{d\gamma(\sigma)}{d\sigma}=-\frac{\tanh\gamma}{\sinh{\tilde\gamma}\cosh{\tilde\gamma}+\tanh\gamma}$
which follows from the relation ${\hat \mu}=2{\cal J}\sinh\gamma\tanh{\tilde\gamma}$ \eqref{intconstrelations2}.
Here we take the $\sigma$ derivative while keeping $\hat{\mu}$.
For $\tilde{\mathcal{J}} = \mathcal{J}$, which is the equal coupling in left and right, \eqref{eq:dbetadsigma} is not monotonic for sufficiently small $\hat{\mu}$  and show the first order phase transition.
In figure \ref{fig:LargeQphases}, we have observed that as ${\tilde {\cal J}}/{\cal J}$ is decreased, $\mu_*$, the critical value of $\mu$ where the phase transition disappears becomes smaller.
When ${\tilde {\cal J}}/{\cal J}$ crosses some critical value, $\mu_*$ finally reaches zero, where the phase transition completely disappears on the $(\mu,T)$-plane.
To determine this critical value of ${\tilde {\cal J}}/{\cal J}$ as a function of $q$, let us consider the limit ${\hat \mu}\rightarrow 0$, which gives us
\begin{align}
\gamma(\sigma)\approx \frac{{\hat \mu}}{2{\cal J}\tanh(s+\sigma)},\quad
\frac{d\beta}{d\sigma}\approx \frac{q}{2{\hat\mu}\sigma\cosh^2(s+\sigma)}\Bigl(\frac{2\sigma\log\frac{q}{\sigma}}{1+\frac{{\hat \mu}}{\sinh^2(s+\sigma)}}-\sinh(2s+2\sigma)\Bigr).
\end{align}
We are interested only in whether $\frac{d\beta}{d\sigma}$ flips its sign or not.
Writing $\frac{d\beta}{d\sigma}$ in the following form
\begin{align}
\frac{d\beta}{d\sigma}=(positive)\times (2\sigma\log\frac{q}{\sigma}-\sinh(2s+2\sigma)),
\end{align}
we find that the last factor is negative at $\sigma\rightarrow 0,\infty$ and gain its maximum at some finite $\sigma_*$ where $\frac{d}{d\sigma}(\frac{d\beta}{d\sigma})|_{\sigma_*}=0$, which is given by
\begin{align}
2\log\frac{q}{\sigma_*}-2-2\cosh(2s+2\sigma_*)=0.
\label{criticalconditionmuhat0}
% ,\quad
% \rightarrow \frac{d\beta}{d\sigma}\biggr|_{\sigma_*}=(positive)\times \Bigl(2\sigma_*\log\frac{q}{\sigma_*}-\sinh(2s+2\sigma_*)\Bigr).
\end{align}
Hence the critical value of $({\tilde {\cal J}}/{\cal J})^2$ is determined by the condition $\frac{d\beta}{d\sigma}|_{\sigma_*}=0$.
Solving this condition numerically we obtain figure \ref{221014_criticalcalJtilde_largeq_qdependence}.
In particular, when $q=96$ we obtain ${\tilde {\cal J}}/{\cal J}=0.500577$, which is consistent with the critical value we have observed in the previous section.

If we further consider the case ${\tilde {\cal J}}\ll {\cal J}$, i.e., $s\gg 1$, the condition \eqref{criticalconditionmuhat0} and for $\sigma_*$ reduces to $\log\frac{q}{\sigma_*}\approx \frac{1}{2}e^{2s}$, and the condition $\frac{d\beta}{d\sigma}|_{\sigma_*}=0$ gives the critical value of $({\tilde {\cal J}}/{\cal J})^2$ as
\begin{align}
\Bigl(\frac{{\tilde {\cal J}}}{{\cal J}}\Bigr)^2\approx \frac{1}{2\log(2q)}.
\end{align}
This result also suggests that the phase transition remains at  $q = \infty$ limit as far as the $J^{(L)}_{i_1\cdots i_q}$ and $J^{(R)}_{i_1\cdots i_q}$ are correlated even slightly.

The derivative of $\ell$ is 
\ba
\f{d \ell}{d\sigma} &= \Big(\f{\log \f{q}{\sigma}}{\sinh \tilde{\gamma} \cosh \tilde{\gamma} + \tanh \gamma} - \f{1}{\sigma } \Big)\f{1}{\log \f{q}{\sigma}} \beta \f{\partial \ell}{\partial \beta} \notag \\ 
&= \f{q}{\hat{\mu} \tanh \tilde{\gamma} \log \f{q}{\sigma}} \f{d\beta }{d\sigma} \beta \f{\partial \ell}{\partial \beta} .
\ea
Since $\f{\partial \ell}{\partial \beta} = - E/N$ is always positive from the expression \eqref{eq:energylargeq},  $\ell(\sigma)$ is also a monotonic function when $\beta (\sigma)$ is monotonic. 
Therefore, when $\beta(\sigma)$ is monotonic the free energy $F$ is also a monotonic function of the temperature $T$.

\subsection{Inverse temperature of order $\beta \sim q$ and beyond}

First we consider the order of $\beta \sim q$.
In this regime, $G_{LR}$ is smaller than $1/2$ even for $\tau \ll q$. 
Therefore we can put $G_{LR}^q = 0$ for $\tau \gg q$ and the effective decay rate $\nu$ becomes the naive one $\nu = \mu$.
Matching with the solution with $\tau \ll q$ regime we obtain
\be
\alpha = \mathcal{J} \sinh \gamma = \f{\hat{\mu}}{2} \tanh \f{\beta \mu}{2}
\ee
The ratio $\tilde{\mathcal{J}}/\mathcal{J}$ do not enter in the correlation function and $\tilde{\mathcal{J}}$ dependence disappears at the order of $\beta \sim q$.
Indeed the correlation function and the partition function take the same form both in Maldacena-Qi model \cite{Maldacena:2018lmt} and Kourkoulou-Maldacena model \cite{Nosaka:2019tcx}. 
Since the behaviors are exactly the same for any $\tilde{\mathcal{J}}$, we only draw the results from \cite{Maldacena:2018lmt,Nosaka:2019tcx} in this paper.

The free energy now becomes 
\be
-\f{\beta F}{N} = \log \Big( 2 \cosh \f{\beta \mu}{2} \Big) + \f{\beta \mu}{q} \tanh \f{\beta \mu}{2} \Big[ \log (2\sinh \gamma) + \f{1}{\tanh \gamma} - \gamma -1  \Big]
\ee
This is independent from the ratio $\mathcal{\tilde{J}}/\mathcal{J}$ and does not depend on the incompleteness of the random couplings.

At the order of $\beta \sim \s{q}$, the chaos exponent increases from very small value and finally saturates the chaos bound \cite{Maldacena:2018lmt,Nosaka:2019tcx}.
The correlation function $G_{LR}$ at this order  becomes
\be
G_{LR}(\tau) = \f{i}{2}\mu  \Big(\f{\beta}{2} -\tau \Big),
\ee
which is of order $1/\s{q}$.
The free energy is 
\be
-\f{\beta F}{N} = \log 2 + \f{(\beta \mu)^2}{8} +\f{2 \beta \mathcal{J}}{q^2} + \f{(\beta \mu)^2}{2q} \log (\beta \mathcal{J}) + \f{h[q(\beta \mu)^2]}{q^2}.
\ee
where $h$ is a function that we have not determined.

At the order of $\beta \sim 1$, we can set $\mu = 0$ to compute $g_{LL}, g_{RR}$ and we recover the decoupled SYK models at large $q$ limit.
The free energy is 
\be
\f{F}{N} = \f{2F_{SYK}}{N} - \f{\beta \mu^2 }{8},
\ee
where $F_{SYK}$ is the free energy of the SYK.

\subsubsection{Comments on subleading Lyapunov exponents}
\label{sec_Lyapunovlargeq}
At finite $q$, we find that there is a subleading Lyapunov exponents in the $\sigma = -1$ sector.
In the large $q$ perspective, the problem to find  Lyapunov exponents reduces to studying the bound states of a Schr\"odinger equation, and we can understand the subleading Lyapunov exponents using that language as follows.
At $\mu = 0$ we have two copies of the SYK and the Lyapunov exponents are degenerate.
After introducing $\mu$, two degeneracies are resolved and we will get two different exponents, which leads to the subleading Lyapunov exponents.

However, at leading order of $1/q$ expansion, the degeneracy is not resolved   because $\sigma$ dependent term is actually of order $1/q$ at large $q$ limit.
By taking the large $q$ limit of \eqref{ladderMQfinalsimplified}, we obtain
\ba
&M_{1,LLLL} (t) \propto (G_{LL}^R)^2 = O(1), \qquad M_{1,LLLR} (t) \propto G_{LL}^RG_{LR}^R = O(1/\s{q}),\notag \\
& M_{1,LRLR} (t) \propto (G_{LR}^R)^2 = O(1/q), \qquad 
M_{2,LL} (t) = - 2\mathcal{J}^2 e^{g_{LL}(\beta/2 + it)}, \qquad M_{2,LR} (t) = 0 .
\ea
where we have used the fact that $G_{LR}$ is of order $1/{\s{q}}$. 
Therefore the only surviving term at large $q$ is $M_{1,LLLL} (t)$ and $M_{2,LL} (t)$ and we find that $\sigma$ dependent terms drop at the leading of $1/q$ expansion.
This means that what we get is the degenerate Lyapunov exponents at any temperature at large $q$.
We will pose the calculation of the leading correction to see the resolution of the degeneracy.

\section{Structure of ground state for imperfectly correlated disorders}
\label{sec_structureofgs}
In this section we investigate the structure of the ground state of the coupled model \eqref{220221_HMQwithJLJRindepmodel}.
Let us consider the following state $|I(\beta)\rangle$
\begin{align}
|I(\beta)\rangle=\frac{1}{{\cal N}}e^{-\frac{\beta}{4}(H_{SYK}^{(L)}+H_{SYK}^{(R)})}|I\rangle,
\label{Ibeta}
\end{align}
where $|I\rangle$ is the ground state of $H_{int}=i\sum_{i=1}^N\psi_i^L\psi_i^R$ and ${\cal N}=\sqrt{\langle I|e^{-\frac{\beta}{2}(H_{SYK}^{(L)}+H_{SYK}^{(R)})}|I\rangle}$ is the normalization factor.
When the correlation of $J^{(L)}_{i_1\cdots i_q}$ and $J^{(R)}_{i_1\cdots i_q}$ is perfect, this state is the thermofield double state.
Therefore the state $\ket{I(\beta)}$ is a generalization of the  thermofield double state.
In the limit $\mu\rightarrow\infty$, the ground state of $H$ approaches $|I(\beta)\rangle$ with $\beta=0$.
Also in the limit $\beta\rightarrow\infty$, $|I(\beta)\rangle$ is approximated with the ground state of $H_{SYK}^{(L)}+H_{SYK}^{(R)}$.
If we assume that the ground state of $H_{SYK}^{(L)}+H_{SYK}^{(R)}$ is non-degenerate, it coincides with the ground state of the coupled model in the limit $\mu\rightarrow 0$.
Therefore $|I(\beta)\rangle$ should be a good one-parameter ansatz to approximate the ground state of the two-coupled system \eqref{220221_HMQwithJLJRindepmodel} at least in the limit $\mu\rightarrow \infty$ and $\mu\rightarrow 0$.
When the two random couplings are perfectly correlated, $|I(\beta)\rangle$ was found to be a good approximation of the ground state also for finite $\mu$ \cite{Maldacena:2018lmt,Garcia-Garcia:2019poj,Alet:2020ehp}.
In this appendix we provide some pieces of evidence that $|I(\beta)\rangle$ approximate the ground state for finite $\mu$ well even when the correlation between $J^{(L)}_{i_1\cdots i_q}$ and $J^{(R)}_{i_1\cdots i_q}$ is imperfect.

\subsection{Variational approximation in large $q$ limit}

To understand the ground state of imperfect left-right coupling model, here we study the variational   approximation by the generalized thermofield double state
\be
\ket{I(\beta)} =\f{1}{\s{Z_{LR}}} e^{-\f{\beta}{4}H_L}e^{-\f{\beta}{4}H_R}\ket{I}.
\ee
Here the $\ket{I}$ is the maximally entangled state defined by 
\be
\psi_L^i\ket{I} = -i \psi_R^i \ket{I}.
\ee
This is the ground state of the coupling Hamiltonian $H_{int} = -i\mu \sum_{i=1}^N \psi_L^i \psi_R^i$.
$Z_{LR}$ is the normalization factor defined by 
\be
Z_{LR} = \bra{I} e^{-\f{\beta}{2}H_L}e^{-\f{\beta}{2}H_R} \ket{I} = \Tr( e^{-\f{\beta}{2}H_L}e^{-\f{\beta}{2}H_R}).
\ee
In Maldacena-Qi model, choosing the apprpriate $\beta$, $\ket{I(\beta)}$ is a very good approximation for the exact ground state in the sense that the leading term of the overlap between them becomes $1$ in small $\mu$ or in the large $q$ limit.
In Kourkoulou-Maldacena model, this state is still a good approximation in the sense that the leading overlap is $1$ at large $q$.
Therefore, we expect that the state $\ket{I(\beta)}$ is a good approximation for the exact ground state even when the correlation of left right coupling is imperfect.

\begin{figure}
\begin{center}
\includegraphics[width=15.0cm]{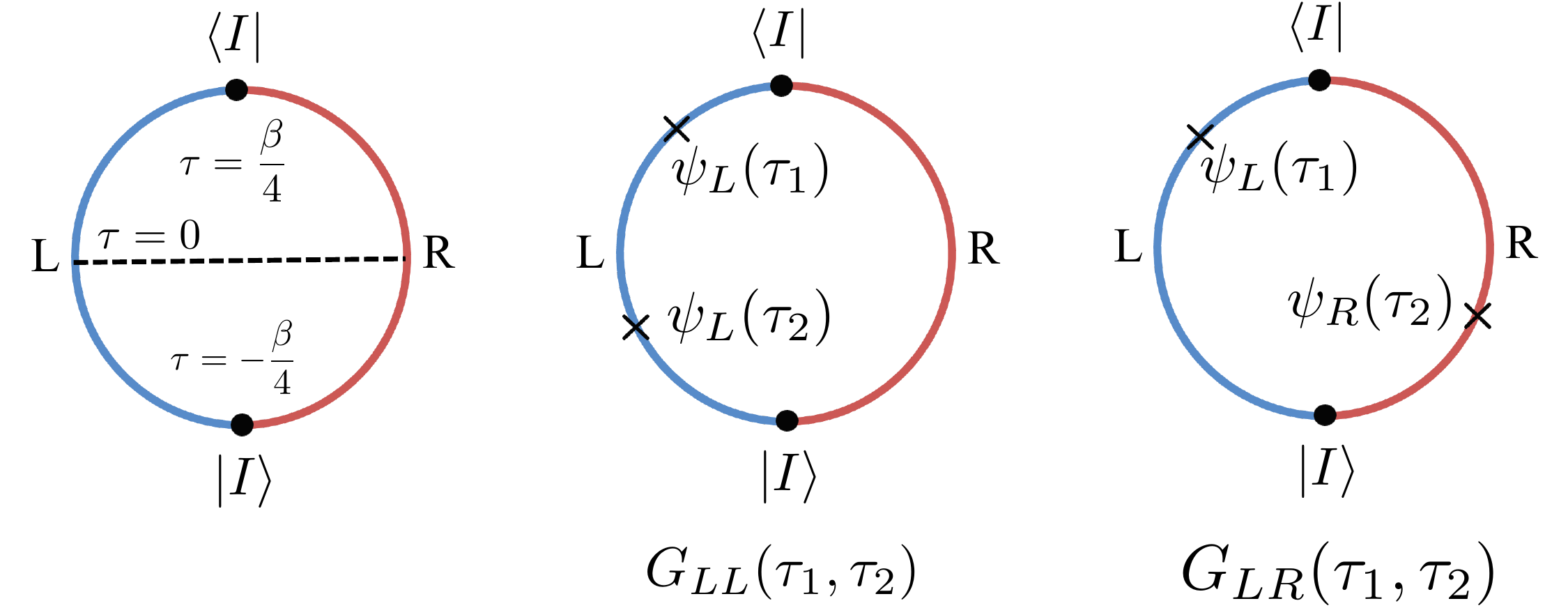}
\caption{
The schematic pictures of the correlation functions.
{\bf Left:} The path integral representation for the partition function $Z_{LR}$.
{\bf Middle:} The correlation function $G_{LL}$.
{\bf Right:} The correlation function $G_{LR}$.
}
\label{fig:IbetaCor}
\end{center}
\end{figure}

We study the variational approximation by $\ket{I(\beta)}$ at large $q$.
To do that, we minimize the trial energy 
\be
E_{\text{trial}}(\beta) = \braket{I(\beta)|H_L + H_R + H_{int}|I(\beta)}. 
\ee
In terms of the Euclidean correlation functions on the thermal circle with interfaces which are schematically depicted in figure \ref{fig:IbetaCor}, 
the trial energy is 
\be
\braket{H_L + H_R + H_{int}}_{\ket{I(\beta)}} = \f{1}{q}\partial _{\tau} G_{LL}(\tau,0)\Big | _{\tau \to 0_+} +\f{1}{q}\partial _{\tau} G_{LR}(\tau,0)\Big | _{\tau \to 0_+} + i \mu G_{LR}(0,0).
\ee
At large $q$ limit, the correlation function becomes $G_{\alpha\beta}(\tau_1,\tau_2) = G_{0,\alpha\beta}\Big( 1 + \f{1}{q}g_{\alpha\beta}(\tau_1,\tau_2) \Big) $ with \cite{Almheiri:2021jwq,Streicher:2019wek}
\ba
e^{g_{LL}(\tau_1,\tau_2)} &= e^{g_{RR}(\tau_1,\tau_2)} = \bigg( \f{\check{\alpha}}{\mathcal{J} \sin (\check{\alpha} |\tau_1-\tau_2| + \check{\gamma})}\bigg)^2 , \ \  \text{for} \qquad -\f{\beta}{4} < \tau_1,\tau_2 < \f{\beta}{4},  \notag \\
e^{g_{LR}(\tau_1,\tau_2)} &= \bigg( \f{\check{\alpha}^2 /\mathcal{J}^2}{-\lambda^2 \sin(\check{\alpha} (\tau_1 + \f{\beta}{4}))\sin(\check{\alpha} (\tau_2 - \f{\beta}{4}))+ \sin (\check{\alpha} (\f{\beta}{4}-\tau_1  ) + \check{\gamma})\sin ( \check{\alpha} ( \f{\beta}{4}-\tau_2) + \check{\gamma})    }\bigg)^2, \notag \\ 
& \qquad \qquad \qquad \qquad \qquad \qquad \qquad \qquad \qquad \qquad \qquad \text{for} \qquad -\f{\beta}{4}< \tau_1, \tau_2 < \f{\beta}{4}.
\ea
Here we introduced a parameter $\lambda = \tilde{ \mathcal{J}}/\mathcal{J}$.
The parameters $\check{\alpha}$ and $\check{\gamma}$ satisfy
\be
\check{\alpha}= \mathcal{J}\sin \check{\gamma}, \qquad \sin \Big(\f{\check{\alpha} \beta}{2}  + 2\check{\gamma} \Big) = \lambda^2 \sin \Big( \f{\check{\alpha} \beta}{2} \Big).
\ee
Then, $E_{\text{trial}}(\beta)$ becomes 
\be
E_{\text{trial}}(\beta)  = - \f{2\mathcal{J}}{q^2} \cos \check{\gamma} - \f{\hat{\mu}}{2q} -  \f{\hat{\mu}}{q^2} \log \bigg[ \f{\sin^2 \check{\gamma}}{ (1-\lambda^2) + \s{(1-\lambda^2)^2 + 4\lambda^2 \sin^2 \check{\gamma}} } \bigg].
\ee
Here we used the relation 
\ba
&\f{1}{q}\partial _{\tau} G_{LL}(\tau,0)\Big | _{\tau \to 0_+} =  -\f{\mathcal{J}}{q^2} \cos \check{ \gamma}, \notag \\
&e^{g_{LR}(0,0)} =  \bigg( \f{\check{\alpha}^2/\mathcal{J}^2 }{- \lambda^2 \sin^2 \f{\check{\alpha} \beta}{4} + \sin^2( \f{\check{\alpha} \beta}{4} + \check{\gamma})} \bigg)^2 = \bigg( \f{2 \sin ^2\check{\gamma} }{(1-\lambda^2) + \s{(1-\lambda^2)^2 + 4\lambda^2 \sin ^2 \check{\gamma} }} \bigg)^2
\ea
Because of the chain rule, we can instead take the derivative w.r.t.~$\check{\gamma}$ to minimize the trial energy:
\be
q^2 \f{\partial E_{\text{trial}}}{\partial \check{\gamma}} = 2\mathcal{J}\sin \check{\gamma}  - \f{\hat{\mu}}{\tan \check{\gamma}} \f{(1-\lambda^2) + \s{(1-\lambda^2)^2 + 4\lambda^2 \sin ^2 \check{\gamma} }}{\s{(1-\lambda^2)^2 + 4\lambda^2 \sin ^2 \check{\gamma} }}  = 0.
\ee
This equation is solved as 
\be
\cos \check{\gamma} = \cosh \gamma - \sinh \gamma \tanh \tilde{\gamma},
\ee
where $\gamma$ is the solution of the equation \eqref{eq:gammarelation}.
This determines the variational parameter  $\beta$ as a function of $\mu$.

We find that the SYK energy and the expectation value of the interaction Hamiltonian completely agree:
\ba
&\braket{H_L + H_R}_{\ket{G(\mu)}} = \braket{H_L + H_R}_{\ket{I(\beta(\mu))}} = -\f{2\mathcal{J}}{q^2} \cos \check{\gamma} \notag \\
&  \braket{\psi_L\psi_R}_{\ket{G(\mu)}} = \braket{\psi_L\psi_R}_{\ket{I(\beta(\mu))}}  =  \f{i}{2} \bigg[ 1 +
% \f{2}{q}\biggl(s+\log  \f{\sinh \gamma}{\cosh\tilde{\gamma}}\biggr)
\f{1}{q} \log  \Big( \f{e^s \sinh \gamma}{ \cosh\tilde{\gamma}} \Big) ^2
\bigg]
\ea
Therefore, the variational energy actually is equal to the true energy \eqref{eq:energylargeq}
\be
E_{\text{trial}}(\beta(\mu)) = E_{g}.
\ee
This means that the $\ket{I(\beta(\mu))} = \ket{G(\mu)}$ in the large $q$ limit.

\subsection{Overlap between ground state and $|I(\beta)\rangle$ for finite $q$}
In the previous section we have found that $|I(\beta)\rangle$ \eqref{Ibeta}, with $\beta(\mu)$ chosen appropriately, approximate the energy of the ground state well when both $q$ and $N$ are large.
This is a strong evidence that $|I(\beta)\rangle$ approximate well the true ground state $|gs\rangle$ of the two-coupled Hamiltonian \eqref{220221_HMQwithJLJRindepmodel}.
In this appendix we study the validity of this approximation for finite $q$ by comparing the two states directly for finite $N$.

Note that there are several subtleties.
First, since the full Hamiltonian as well as $i\sum_{i=1}^N\psi_i^L\psi_i^R$ commute with the fermion number parity $\Gamma_c$ \eqref{Gammac}, both $|I(\beta)\rangle$ and the ground state of $H$ are eigenstates of $\Gamma_c$.
When $\mu$ is not sufficiently large, the parity of the ground state of $H$ depends on the realization of $J_{i_1\cdots i_q}^{(a)}$ and hence not always the same as the parity of $|I(\beta)\rangle$.
Hence to make the comparison reasonable we should compare $|I(\beta)\rangle$ with $|gs,(-1)^{\frac{N(N-1)}{2}}\rangle$, the eigenstate of $H$ with the lowest energy in the same parity sector as $|I(\beta)\rangle$, $\Gamma_c=(-1)^{\frac{N(N-1)}{2}}$, rather than the true ground state of $H$.

Second, although $|I(\beta)\rangle$ approximate well the ground state at $\mu\approx 0$ when the ground state of $H_{SYK}^{(L)}+H_{SYK}^{(R)}$ is non-degenerate, the spectrum of single SYK Hamiltonian has the following degeneracy depending of the value of $q$ and $N$ mod 8 \cite{Kanazawa:2017dpd}:
\begin{align}
\begin{tabular}{|c|c|}
\hline
$q,N$                                      &degeneracy of single SYK spectrum\\ \hline
$q=0\text{ mod 4},\quad N=2,6\text{ mod }8$&2 (between different parity sectors)\\ \hline
$q=0\text{ mod 4},\quad N=4\text{ mod }8$  &2 (in each parity sector)\\ \hline
$q=0\text{ mod 4},\quad N=0\text{ mod }8$  &non-degenerate\\ \hline
$q=2\text{ mod 4}$                         &non-degenerate\\ \hline
\end{tabular},
\end{align}
which implies that the ground state of $H_{SYK}^{(L)}+H_{SYK}^{(R)}$ is also degenerate in the cases of the first two rows.
On the other hand, $|I\rangle$ contains only the certain linear combination of them (which can be identified explicitly for $J_{i_1\cdots i_q}^{(L)}=J_{i_1\cdots i_q}^{(R)}$ case, as summarized in appendix \ref{app_IineigenstatesofHSYK}).
Note that this degeneracy cannot be removed completely by the total fermion number parity $(-1)^F$.
When $\mu$ is small but non-zero, the degeneracy is removed and the true ground state is approximately a certain linear combination of the degenerate ground states at $\mu=0$.
This linear combination varies depending on the realization of $J_{i_1\cdots i_q}^{(a)}$ and is not necessarily the same as the linear combination contained in $|I\rangle$.

\begin{figure}
\begin{center}
\includegraphics[width=8cm]{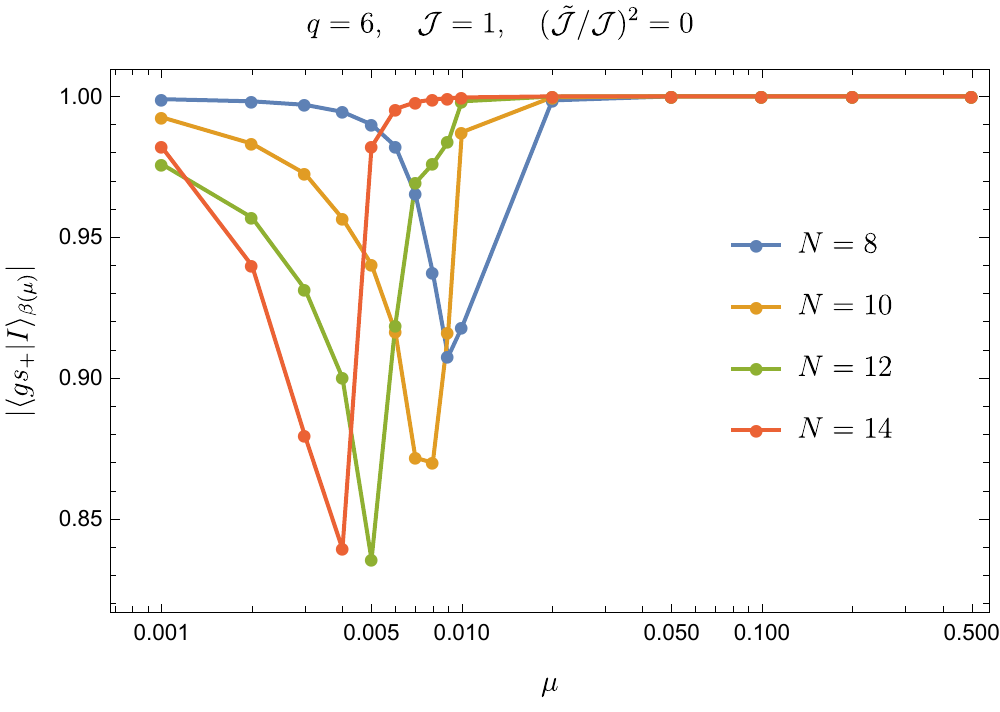}
\includegraphics[width=8cm]{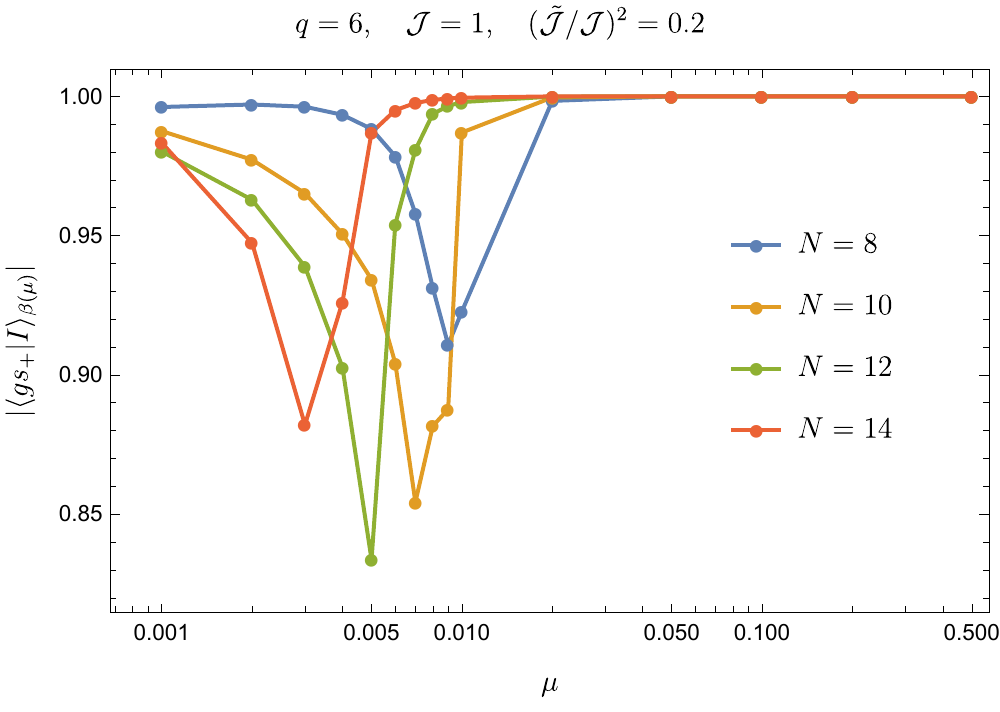}
\includegraphics[width=8cm]{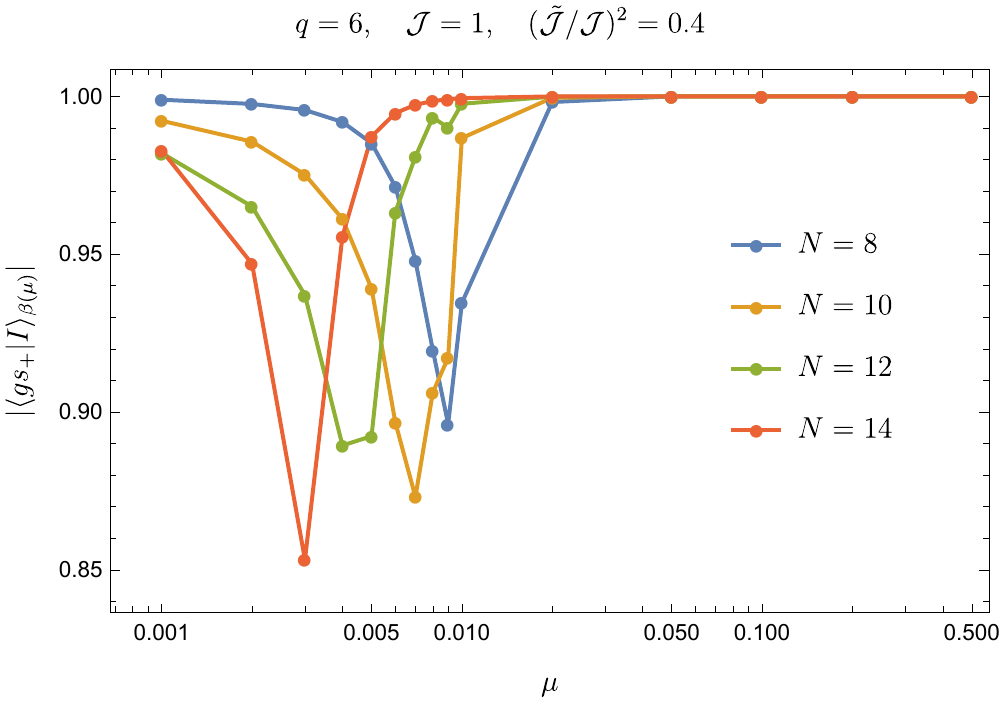}
\includegraphics[width=8cm]{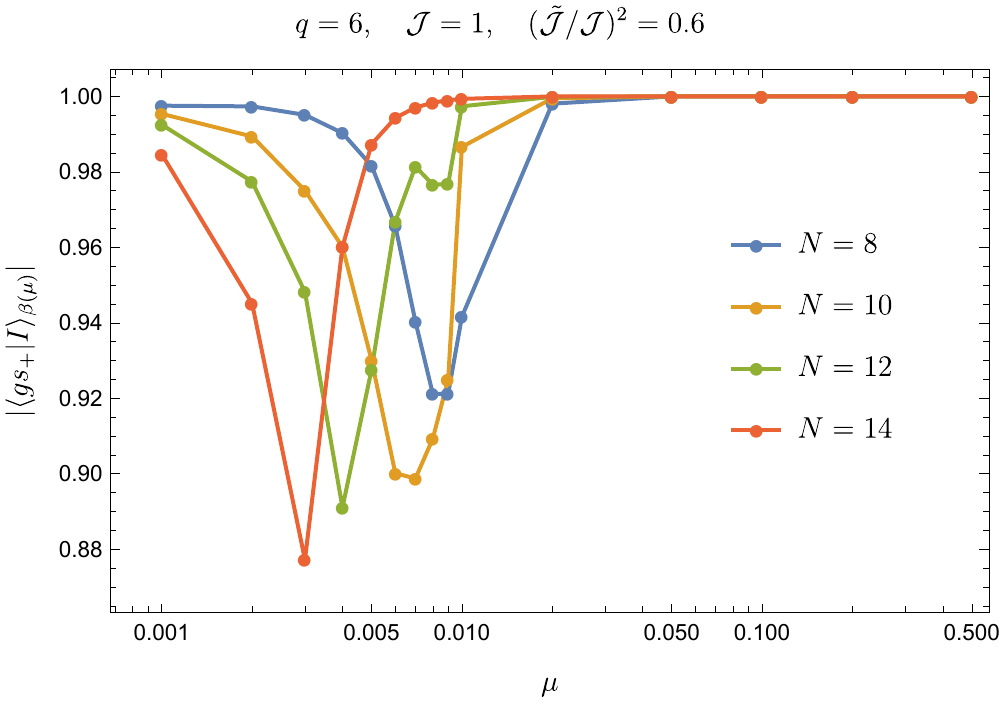}
\includegraphics[width=8cm]{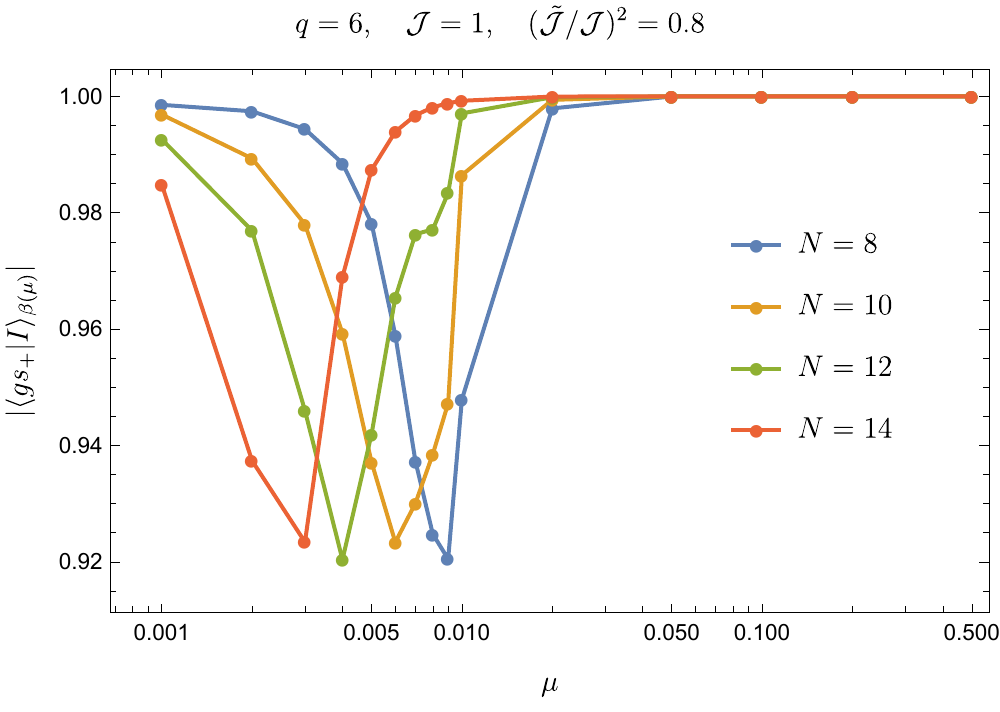}
\includegraphics[width=8cm]{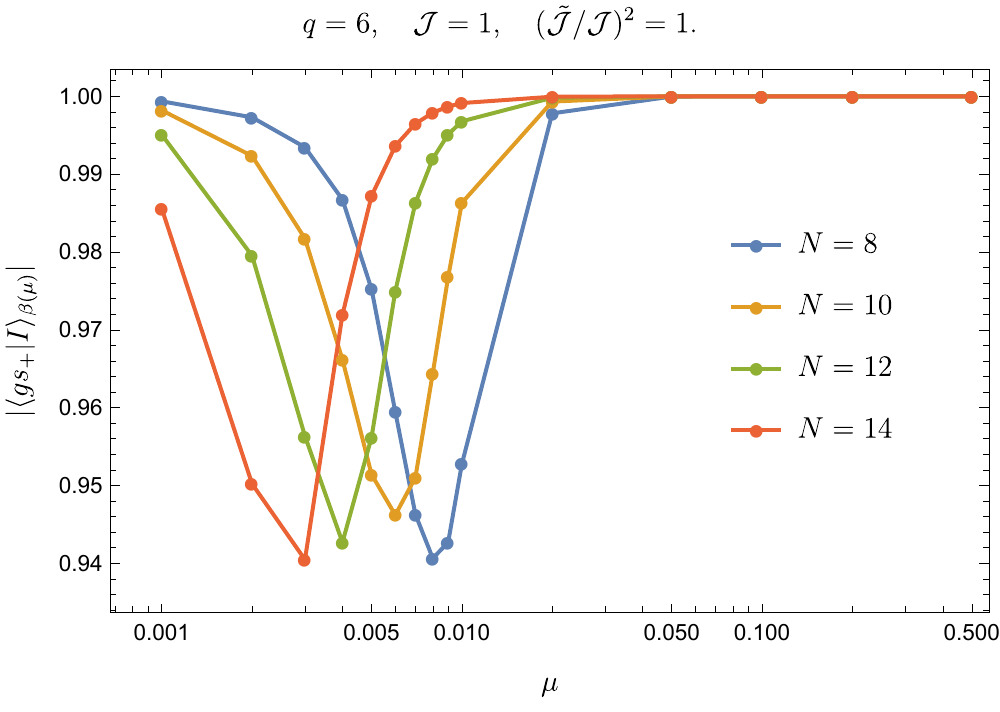}
\caption{
The overlap $|\langle I_\beta|gs,(-1)^{\frac{N(N-1)}{2}}\rangle|$ maximized with respect to $\beta$.
}
\label{fig_q6finiteNoverlap}
\end{center}
\end{figure}

To avoid these subtleties, here we choose $q=6$
where the ground state in $\Gamma_c=(-1)^{\frac{N(N-1)}{2}}$ sector at $\mu=0$ is non-degenerate for any $N$, and consider the overlap between $|I(\beta)\rangle$ and $|gs,(-1)^{\frac{N(N-1)}{2}}\rangle$, maximized with respect to $\beta$.
As a result we obtain figure \ref{fig_q6finiteNoverlap}.
The results suggest that $|I(\beta)\rangle$ is indeed a good approximation to $|gs,(-1)^{\frac{N(N-1)}{2}}\rangle$ also for finite $\mu$.

\section{Discussion and Future works}
\label{sec_discuss}
In this paper we have studied the thermodynamic and chaotic properties of the two-coupled SYK model where the two random couplings are not completely the same.
As a result we have found that the phase transition temperature becomes smaller as the correlation of the random couplings is reduced.
This is consistent with the intuition that the correlation of between the random couplings make it easier for the wormhole to form between the two sites.

Then, we studied the ground state properties of the coupled SYK with imperfect correlations of the random couplings.
As we change the left right correlation, the behavior of the energy gap also changes.
When  $\tilde{\mathcal{J}}/\mathcal{J}$ is close to $1$, the SYK interaction still helps to make the gap larger than the naive one $\mu$.
However as we decrease $\mu$, finally the effect of imperfect left right correlation wins and the energy gap becomes smaller than the naive one, which we expect when we have no correlation between left and right.
For $\tilde{\mathcal{J}}$ where the thermal phase transition disappears, the energy gap is close to that without left and right correlation as far as we checked.

We have also found that the transmission amplitude between two SYK sites in the wormhole phase for fixed temperature $T$ and the direct LR coupling $\mu$ becomes smaller as the correlation of the random couplings is reduced.
On the other hand, as the correlation is reduced the largest chaos exponent becomes larger.
% The first two results are consistent with the intuition that the correlation of between the random couplings make it easier for the wormhole to form between the two sites.
Assuming that the largest chaos exponent is associated with the information spreading within each site rather than between the two sites, this behavior of the chaos exponent is also consistent with the same intuition.
Interestingly, we have also found that the phase transition completely disappears when the correlation between the random couplings is smaller than some non-zero finite value which is around $\langle J^{(L)}_{i_1\cdots i_q}J^{(R)}_{i_1\cdots i_q}\rangle\approx 0.3\langle (J_{i_1\cdots i_q}^{(L)})^2\rangle$ for $q=4$ and $\langle J^{(L)}_{i_1\cdots i_q}J^{(R)}_{i_1\cdots i_q}\rangle\sim (\log q)^{-1}\langle (J_{i_1\cdots i_q}^{(L)})^2\rangle$ for large $q$.

Finally, to understand the properties of the ground state, we studied how it is close to the generalized thermofield double state
This genralized theromfield double state has a jump of couplings in Euclidean time and studied in the context of black holes \cite{Bak:2007jm,Nakaguchi:2014eiu,Bak:2011ga,Bak:2007qw}.
It turns out that the ground state coincides with the generalized thermofield double state state at large $q$. 
In that sense, the coupled Hamiltonian prepare an ``SYK Janus black holes'' as its ground state.
In the case of SYK Janus black holes, we have an expanded interiors.
Intuitively, this expanded interior makes the length between two mouths of the wormholes longer and it takes much time to traverse the wormhole.
This is an intuitive explanation of why the $E_{gap}$, which is roughly the time to traverse the wormhole,  becomes larger for smaller $\tilde{\mathcal{J}}/\mathcal{J}$.

In section \ref{sec_chaosexponent_MQ} we have found that the ladder kernel for the four point functions can be block-diagonalized into two sectors labelled by $\sigma=\pm 1$, and have observed that the chaos exponent in the $\sigma=-1$ sector vanishes at some temperature $T=T_{c,\text{Ly}}(\mu,{\tilde {\cal J}}/{\cal J})$ which is larger than $T_{c,\text{WH}}$ (the vanishing chaos exponent was also observed in \cite{Nosaka:2020nuk} for $J_{i_1\cdots i_q}^{(L)}=J_{i_1\cdots i_q}^{(R)}$).
It would be interesting to investigate the physical interpretation of this phenomenon.
For this purpose it would be important to reproduce the same phenomenon analytically in the large $q$ limit.
As commented in section \ref{sec_Lyapunovlargeq} this requires the analysis of the sub-leading correction in $1/q$.

In this paper we have studied the effect of the correlation between $J^{(L)}_{i_1\cdots i_q}$ and $J^{(R)}_{i_1\cdots i_q}$ in the two-coupled model without modifying the probability distributions of each $J_{i_1\cdots i_q}^{(a)}$ themselves.
One may also consider different modifications of the distribution of the random couplings such as an imbalanced rescaling $J_{i_1\cdots i_q}^{(R)}=cJ_{i_1\cdots i_q}^{(L)}$ with $c\neq 1$ \cite{Haenel:2021fye} or the sparse couplings \cite{Garcia-Garcia:2020cdo,Xu:2020shn,Caceres:2021nsa} instead of full $J^{(a)}_{i_1\cdots i_q}$.
It would also be interesting to study how the traversability and other properties of these models change as the correlation between the random couplings on two sides is varied.

One may also adopt different kinds of LR interactions instead of the one $i\sum_i\psi^L_i\psi^R_i$ we have used.
For example, the LR interaction can also be turned on by considering $H=f(H_{SYK}^{(L)}+H_{SYK}^{(R)})$ with any non-linear function $f$.
Such a transformation of the Hamiltonian naturally arises in the context of $T\bar{T}$ deformation \cite{Zamolodchikov:2004ce,Smirnov:2016lqw,Cavaglia:2016oda,LeFloch:2019wlf,He:2019vzf,Gross:2019uxi,Jorjadze:2020ili}, whose quench protocol was studied in \cite{He:2021dhr}.
It would be interesting to investigate the properties of such models further.

\section*{Acknowledgement}
We thank Kanato Goto, Cheng Peng, Dario Rosa and Yingyu Yang for valuable discussions and useful comments.
The numerical analyses in this paper were performed on sushiki server in Yukawa Institute Compute Facility and on Ulysses cluster v2 in SISSA.
TN is supported by  MEXT KAKENHI Grant-in-Aid for Transformative Research Areas A ``Extreme Universe'' Grant Number 22H05248.

%%%%%%%%%%%%%%%%%%%%
%%%%%%%%%%%%%%%%%%%%
\appendix

\section{A derivation of the large $q$ partition function}
\label{app_largeqderivation}
The solution for $\tau \ll q$ is 
\ba
e^{g_{LL}(\tau)} &= \f{\alpha^2 }{\mathcal{J}^2 \sinh^2(\alpha |\tau| + \gamma)}, \notag \\
e^{g_{LR}(\tau)} &= \f{\tilde{\alpha}^2 }{\tilde{\mathcal{J}}^2 \cosh^2(\tilde{\alpha} |\tau| + \tilde{\gamma})},
%\label{eq:earlytau}
\ea
and for $\tau \gg q$ 
\ba
G_{LL}(\tau) &= A \cosh \Big[ \nu \Big(\f{\beta}{2} - \tau \Big) \Big], \notag \\
G_{LR}(\tau) &=  i A \sinh \Big[ \nu \Big(\f{\beta}{2} - \tau \Big)\Big].
\ea
where 
\be
\nu = i \int_{-\infty}^{\infty} \Sigma_{LR} (\tau) d\tau   = \f{2\tilde{\alpha}}{q} = \f{\hat{\mu}}{q \tanh \tilde{\gamma}}.
\ee
In the matching region, $G_{LL}$ and $G_{LR}$ is expanded as 
\ba
G_{LL}  &\sim \f{1}{2} - \f{1}{q} \Big( \log \f{\mathcal{J}}{\alpha} + \gamma + \alpha \tau \Big) 
 \sim A\Big[ \cosh  \f{\nu\beta}{2} - \tau \nu \sinh \f{\nu\beta}{2} \Big] \notag  \\
-iG_{LR} & \sim \f{1}{2} - \f{1}{q} \Big( \log \f{\tilde{\mathcal{J}}}{\tilde{\alpha}} + \tilde{\gamma} + \tilde{\alpha} \tau \Big)  \sim  A\Big[ \sinh \f{\nu  \beta}{2}    - \tau \nu \cosh \f{\nu\beta}{2}  \Big].
\ea
This determines the parameters to be 
\be
\alpha = \tilde{\alpha},\qquad    \tilde{\gamma} = \gamma + \sigma + s,  \label{eq:matchingcond}
\ee
Here we defined $\sigma$ by 
\be
\sigma  = q e^{-\beta \nu}, \label{eq:sigmadef}
\ee
which is of order $O(1)$.
The boundary conditions at $\tau = 0$ gives 
\ba
\alpha = \mathcal{J} \sinh \gamma, \qquad \hat{\mu} = 2\tilde{\alpha }\tanh \tilde{\gamma}. \label{eq:bdycondition}
\ea
The conditions \eqref{eq:matchingcond}, \eqref{eq:sigmadef} and \eqref{eq:bdycondition} determines the relation between the physical parameters $\hat{\mu}, \mathcal{J}, \tilde{\mathcal{J}}, \beta$ and $\sigma, \gamma, s$.
The energy is 
\ba
\f{E}{N} = - \f{1}{N} \partial_{\beta} \log Z
&=  \f{1}{2q} \partial_{\tau}g_{LL}(0) +  \f{1}{2q} \partial_{\tau}g_{RR}(0) +  i \mu \Big( 1 - \f{2}{q}\Big) \f{i}{2} \Big( 1 + \f{1}{q} g_{LR}(0)\Big) \notag \\
&= -\f{2\mathcal{J}}{q^2} \cosh \gamma - \f{\hat {\mu}}{2 q}   + \f{\hat{\mu}}{q^2 } \Big(1 +  \log \f{e^s \sinh \gamma}{\cosh \tilde{\gamma}} \Big).
\ea

In the following we choose the fundamental variables as either $({\hat \mu},{\cal J},s,\beta)$ or $(\sigma,\gamma,s,\beta)$ interchangeably.
Using the effective action we can write\footnote{Here we take the $\mathcal{J}$ derivative with fixed $s$ and $\partial_{\mathcal{J}}$ also acts on $\tilde{\mathcal{J}}$ terms.}
\ba
\mathcal{J} \f{\partial \ell}{\partial \mathcal{J}}\biggr|_{{\hat \mu},s,\beta} &= \beta \int _0^{\infty} (\mathcal{J}^2 e^{g_{LL}} + \tilde{\mathcal{J}}^2 e^{g_{LR}} ) d\tau = \f{\beta \hat{\mu}}{q^2} \bigg[ \f{1}{\tanh \gamma \tanh \tilde{\gamma}} -1 \bigg]. \notag \\
\mu  \f{\partial \ell}{\partial {\hat \mu}}\biggr|_{{\cal J},s,\beta} & = - i \beta \mu G_{LR}(0) = \f{\beta \hat{\mu}}{q^2} \bigg[\f{q}{2} + \log \Big( \f{\sinh \gamma}{\cosh \tilde{\gamma}} 
\Big)
+s
\bigg]
\ea
By changes of variables, partial derivatives are given by
\ba
&\f{1}{\hat{\mu}} \f{\partial \hat{\mu}}{\partial \gamma}\biggr|_{\sigma,s,\beta} - \f{1}{\mathcal{J}}\f{\partial \mathcal{J}}{\partial \gamma}\biggr|_{\sigma,s,\beta} = \f{1}{\tanh \gamma} + \f{1}{\sinh \tilde{\gamma} \cosh \tilde{\gamma}}, \qquad \f{1}{\beta \hat{\mu}} \f{\partial (\beta \hat{\mu})}{\partial \gamma}\biggr|_{\sigma,s,\beta} = \f{1}{\sinh \tilde{\gamma} \cosh \tilde{\gamma}}, \notag \\
&\f{1}{\hat{\mu}} \f{\partial \hat{\mu}}{\partial \sigma}\biggr|_{\gamma,s,\beta} - \f{1}{\mathcal{J}}\f{\partial \mathcal{J}}{\partial \sigma}\biggr|_{\gamma,s,\beta} =  \f{1}{\sinh \tilde{\gamma} \cosh \tilde{\gamma}},
\qquad \f{1}{\beta \hat{\mu}} \f{\partial (\beta \hat{\mu})}{\partial \sigma}\biggr|_{\gamma,s,\beta} = \f{1}{\sinh \tilde{\gamma} \cosh \tilde{\gamma}} - \f{1}{\sigma \log  \f{q}{\sigma}} \label{eq:changevar}
\ea
Then, the derivative of $\ell$ in $\gamma, \sigma$ is determined through 
\ba
&\f{\partial \ell}{\partial \gamma}\biggr|_{\sigma,s,\beta} = \f{1}{\beta \hat{\mu}} \f{\partial (\beta \hat{\mu})}{\partial \gamma}\biggr|_{\sigma,s,\beta} \beta \f{\partial \ell}{\partial \beta }\biggr|_{{\hat \mu},{\cal J},s} - \Big(\f{1}{\hat{\mu}} \f{\partial \hat{\mu}}{\partial \gamma}\biggr|_{\sigma,s,\beta} - \f{1}{\mathcal{J}}\f{\partial \mathcal{J}}{\partial \gamma}\biggr|_{\sigma,s,\beta} \Big)\mathcal{J} \f{\partial \ell}{\partial \mathcal{J}}\biggr|_{{\hat \mu},s,\beta} \notag \\
&\f{\partial \ell}{\partial \sigma}\biggr|_{\gamma,s,\beta} = \f{1}{\beta \hat{\mu}} \f{\partial (\beta \hat{\mu})}{\partial \sigma}\biggr|_{\gamma,s,\beta} \beta \f{\partial \ell}{\partial \beta }\biggr|_{{\hat \mu},{\cal J},s} - \Big(\f{1}{\hat{\mu}} \f{\partial \hat{\mu}}{\partial \sigma}\biggr|_{\gamma,\beta,s} - \f{1}{\mathcal{J}}\f{\partial \mathcal{J}}{\partial \sigma}\biggr|_{\gamma,\beta,s} \Big)\mathcal{J} \f{\partial \ell}{\partial \mathcal{J}}\biggr|_{{\hat \mu},s,\beta}.
\label{eq:ellderivative}
\ea
Here we used the relation $(\mu \partial_{\mu} + \mathcal{J}\partial_{\mathcal{J}} - \beta \partial_{\beta}) \ell = 0$ since $\ell$ is a function of dimensionless parameters $\ell(\beta \hat{\mu},\mathcal{J} \beta, s )$.
Integrating \eqref{eq:ellderivative}, we obtain
\be
\ell(\sigma,\gamma) = \f{\tanh \tilde{\gamma} \log \f{q}{\sigma}}{q} \Big(\f{q}{2} -1 + \f{1}{\tanh \gamma \tanh \tilde{\gamma}} + \log \f{\sinh \gamma}{\cosh \tilde{\gamma}} + s + \f{\sigma}{\tanh \tilde{\gamma}}  \Big) + \f{\sigma}{q},
\ee
which is the result in \eqref{eq:LargeqPF}.
Interestingly, the effect of incomplete correlation $\mathcal{J}\neq \tilde{\mathcal{J}}$ is  included in $\tilde{\gamma}$ and the partition function takes the same form with the completely correlated random couplings $\mathcal{J} = \tilde{\mathcal{J}}$.

\section{Relation between ground state of $H_{int}$ and eigenstates of $H_{SYK}^{(a)}$}
\label{app_IineigenstatesofHSYK}
In this section we display the explicit relation between $|I\rangle$, the ground state of $H_{int}$ \eqref{Hint}, and the eigenstates of $H_{SYK}^{(L)}+H_{SYK}^{(R)}$ for $J^{(L)}_{i_1\cdots i_q}=J^{(R)}_{i_1\cdots i_q}$.
For $q=4$ the results are also written in \cite{Garcia-Garcia:2019poj}.

\subsection{Gamma matrices and charge conjugation matrix}
Let us first fix the covnention for the gamma matrices, and also introduce the charge conjugation operator $C$ for single site, which play a crucial role in fixing the ambiguities of the overall phases of the eigenstates of $H_{SYK}^{(L)}+H_{SYK}^{(R)}$.

We choose the representation of the single-site gamma matrices $\gamma_i$ and single-site fermion number parity matrix $\gamma_c$ as
\begin{align}
&\gamma_1=X\otimes 1\otimes 1\otimes 1\otimes \cdots\otimes 1\otimes 1,\nonumber \\
&\gamma_2=Y\otimes 1\otimes 1\otimes 1\otimes \cdots\otimes 1\otimes 1,\nonumber \\
&\gamma_3=Z\otimes X\otimes 1\otimes 1\otimes \cdots\otimes 1\otimes 1,\nonumber \\
&\gamma_4=Z\otimes Y\otimes 1\otimes 1\otimes \cdots\otimes 1\otimes 1,\nonumber \\
&\gamma_5=Z\otimes Z\otimes X\otimes 1\otimes \cdots\otimes 1\otimes 1,\nonumber \\
&\quad \vdots\nonumber \\
&\gamma_N=Z\otimes Z\otimes Z\otimes Z\otimes \cdots\otimes Z\otimes Y,\nonumber \\
&\gamma_c=(-i)^{\frac{N}{2}}\gamma_1\gamma_2\cdots\gamma_N=Z^{\otimes \frac{N}{2}},
\end{align}
where
\begin{align}
X=\begin{pmatrix}
0&1\\
1&0
\end{pmatrix},\quad
Y=\begin{pmatrix}
0&-i\\
i&0
\end{pmatrix},\quad
Z=\begin{pmatrix}
1&0\\
0&-1
\end{pmatrix}.
\label{XYZinspinbasis}
\end{align}
With these $\gamma_i$ and $\gamma_c$, we define the gamma matrices for the two-coupled system $\Gamma_i^{(a)}=\sqrt{2}\psi_i^{(a)}$ ($a=L,R$) as
\begin{align}
\Gamma_i^{(L)}=\gamma_i\otimes 1,\quad
\Gamma_i^{(R)}=\gamma_c\otimes \gamma_i,
\label{GammaiLGammaiR}
\end{align}
and define the fermion number parity matrix $\Gamma_c$ for a two-coupled system as\footnote{
We have chosen the convention of $\Gamma_c$ so that the fermion number of the two-coupled system always coincides with the sum of the fermion number of each site for any $N\in 2\mathbb{N}$.
Note that our choice is different from the one $\Gamma_c^{\text{(another)}}=(-4)^{\frac{N}{2}}\psi_1^{(L)}\psi_1^{(R)}\psi_2^{(L)}\psi_2^{(R)}\cdots \psi_N^{(L)}\psi_N^{(R)}$ with which the fermion parity of $|I\rangle$ is independent of $N$: $\Gamma_c^{\text{(another)}}|I\rangle=|I\rangle$.
}
\begin{align}
\Gamma_c=(-i)^N
\Gamma^{(L)}_1
\Gamma^{(L)}_2
\cdots
\Gamma^{(L)}_N
\Gamma^{(R)}_1
\Gamma^{(R)}_2
\cdots
\Gamma^{(R)}_N=\gamma_c\otimes \gamma_c.\label{Gammac}
\end{align}
Since $|I\rangle$, the ground state of $H_{int}$ \eqref{Hint}, satisfies $i\Gamma_i^{(L)}\Gamma_i^{(R)}|I\rangle=-|I\rangle$ for all $i=1,2,\cdots,N$, we find that the fermion number parity of $|I\rangle$ is $\Gamma_c=(-1)^{\frac{N(N-1)}{2}}$.

The charge conjugation operator $C$ of single site can be defined as
\begin{align}
C=\gamma_2\gamma_4\cdots\gamma_NK,
\end{align}
where $K$ is the complex conjugation in the basis which define the matrix element of $X,Y,Z$ as \eqref{XYZinspinbasis}.
Note that $CK=\gamma_2\gamma_4\cdots\gamma_N$ is a unitary operator but $C$ is not a unitary operator.
Let us list several important properties of $C$:
\begin{align}
&C\gamma_i=(-1)^{\frac{N}{2}}\gamma_iC,\quad
C\gamma_c=(-1)^{\frac{N}{2}}\gamma_cC,\quad
C^2=\begin{cases}
1&\quad (N=0,6\text{ mod }8)\\
-1&\quad (N=2,4\text{ mod }8)
\end{cases},\label{Cgammaigammac}\\
&\langle\phi|\gamma_i|\psi\rangle=(-1)^{\frac{N}{2}}(C|\psi\rangle)^{\dagger}\gamma_iC|\phi\rangle,\label{Cformula}
\end{align}
which we use in the rest of this section.

\subsection{$q=0\text{ mod }4$}
For $q=0\text{ mod }4$, $H_{SYK}^{(L)}$ and $H_{SYK}^{(R)}$ with $J^{(L)}_{i_1\cdots i_q}=J^{(R)}_{i_1\cdots i_q}$ are written in the basis \eqref{GammaiLGammaiR} as
\begin{align}
H_{SYK}^{(L)}=H_{SYK}\otimes 1,\quad
H_{SYK}^{(R)}=1\otimes H_{SYK},
\end{align}
with
\begin{align}
H_{SYK}=\frac{1}{2^{\frac{q}{2}}}i^{\frac{q}{2}}\sum_{i_1<\cdots <i_q}J^{(L)}_{i_1\cdots i_q}\gamma_{i_1}\cdots \gamma_{i_q}.
\end{align}
Since $H_{SYK}$ commutes with $\gamma_c$, we can choose the eigenstates of $H_{SYK}$ as simultaneous eigenstates of $\gamma_c$.
Also, since $H_{SYK}$ commutes with $C$, we can classify these eigenstates by using $C$.
As the relations \eqref{Cgammaigammac} suggest, the classification, as well as the consequent expression of $|I\rangle$, depends on $N\text{ mod }8$.

\subsubsection{$q=0\text{ mod }4$, $N=0\text{ mod }8$}
First let us consider the case $q=0\text{ mod }4$ and $N=0\text{ mod }8$.
In this case $\gamma_c$ and $C$ also commute with each other.
Therefore, if $|n,\sigma\rangle$ is an eigenstate of $H_{SYK}$ with $H_{SYK}=E_{n,\sigma}$ and $\gamma_c=\sigma$, $C|n,\sigma\rangle$ is also an eigenstate of $H_{SYK},\gamma_c$ with $H_{SYK}=E_{n,\sigma}$ and $\gamma_c=\sigma$.
Moreover, since $C^2=1$, we can choose the eigenstates $|n,\sigma\rangle$ such that $C|n,\sigma\rangle=|n,\sigma\rangle$: if $C|n,\sigma\rangle\neq |n,\sigma\rangle$ we can redefine $(|n,\sigma\rangle+C|n,\sigma\rangle)\times (\text{real number})$ and/or $i(|n,\sigma\rangle-C|n,\sigma\rangle)\times (\text{real number})$ as $|n,\sigma\rangle$.
It turns out that there are no degeneracy with generic $J^{(L)}_{i_1\cdots i_q}$, hence the eigenstates of $H_{SYK}$ are summarized as
\begin{align}
\begin{tabular}{|c|c|c|}
\hline
eigenstates of $H_{SYK}$                                               &$\gamma_c$&$H_{SYK}$\\ \hline
$|n,+\rangle\quad (C|n,+\rangle=|n,+\rangle;\, n=1,2,\cdots,2^{N/2-1})$&$+1$      &$E_{n,+}$\\ \hline
$|n,-\rangle\quad (C|n,+\rangle=|n,-\rangle;\, n=1,2,\cdots,2^{N/2-1})$&$-1$      &$E_{n,-}$\\ \hline
\end{tabular},
\end{align}
where $E_{n,+}\neq E_{n,-}$.

Since $|I\rangle$ satisfies $\Gamma_i^{(R)}|I\rangle=i\Gamma_i^{(L)}|I\rangle$, we have $H_{SYK}^{(L)}|I\rangle=H_{SYK}^{(R)}|I\rangle$, which suggests that $|I\rangle$ is expanded as
\begin{align}
|I\rangle=\sum_{n=1}^{2^{N/2}-1}\sum_{\sigma=\pm 1}a_{n,\sigma}|n,\sigma\rangle\otimes |n,\sigma\rangle.
\end{align}
Since the ground state of $H_{int}$ is non-degenerate, $a_{n,\sigma}$ can be determined uniquely by solving $\langle I|H_{int}|I\rangle=-\frac{N}{2}$.
By using the explicit expressions of $\Gamma_i^{(L)},\Gamma_i^{(R)}$ \eqref{GammaiLGammaiR} and the properties of $|n,\sigma\rangle$ under $C,\gamma_c$ we can rewrite $\langle I|H_{int}|I\rangle$ as
\begin{align}
\langle I|H_{int}|I\rangle
&=\frac{i}{2}\sum_{i=1}^N\sum_{m,n}\sum_{\sigma}a_{m,\sigma}^*a_{n,-\sigma}\langle m,\sigma|\gamma_i\gamma_c|n,-\sigma\rangle\langle m,\sigma|\gamma_i|n,-\sigma\rangle\nonumber \\
&=\frac{i}{2}\sum_{i=1}^N\sum_{m,n}\sum_{\sigma}(-\sigma)a_{m,\sigma}^*a_{n,-\sigma}\langle m,\sigma|\gamma_i|n,-\sigma\rangle\langle m,\sigma|\gamma_i|n,-\sigma\rangle\nonumber \\
&=\frac{i}{2}\sum_{i=1}^N\sum_{m,n}\sum_{\sigma}(-\sigma)a_{m,\sigma}^*a_{n,-\sigma}\langle m,\sigma|\gamma_i|n,-\sigma\rangle\langle n,-\sigma|\gamma_i|m,\sigma\rangle,
\label{tobefurthersimplified}
\end{align}
where in the first line we have used the fact that $\gamma_i$ flip the fermion number parity $\gamma_c$ and in the third line we have used the formula \eqref{Cformula}.
If we assume $a_{n,\sigma}=a_\sigma$ is independent of $n$, we can use the fact that $\{|n,\sigma\rangle\}$ is a complete orthonormal basis, i.e., $\sum_{n,\sigma'}|n,\sigma'\rangle\langle n,\sigma'|=1$, to further simplify the right-hand side of \eqref{tobefurthersimplified}:
\begin{align}
\langle I|H_{int}|I\rangle
=\frac{i}{2}\sum_{i=1}^N\sum_{m}\sum_{\sigma}(-\sigma)a_{\sigma}^*a_{-\sigma}\langle m,\sigma|\gamma_i\gamma_i|m,\sigma\rangle
=-2^{N/2-2}iN\sum_{\sigma=\pm 1}\sigma a_\sigma^*a_{-\sigma}.
\end{align}
Hence from $\langle I|H_{int}|I\rangle=-\frac{N}{2}$ we obtain the condition $2^{N/2-1}\sum_\sigma \sigma a_\sigma^*a_{-\sigma}=-i$.
Combining this with the normalization condition $\langle I|I\rangle=2^{N/2-1}\sum_\sigma |a_\sigma|^2=1$, we can determine $a_\sigma$ as $(a_+,a_-)=(2^{-N/4},-2^{-N/4}i)$ up to an overall phase.
After all, we obtain the following expression for $|I\rangle$:
\begin{align}
|I\rangle_{q=0\text{ mod }4,N=0\text{ mod 8}}=2^{-\frac{N}{4}}\sum_{n=1}^{2^{\frac{N}{2}-1}}(
|n,+\rangle
\otimes
|n,+\rangle
-i
|n,-\rangle
\otimes
|n,-\rangle
).
\end{align}

\subsubsection{$q=0\text{ mod }4$, $N=2,6\text{ mod }8$}
In this case $\gamma_c$ anti-commutes with $C$.
Hence if $|n,+\rangle$ is an eigenstate of $H_{SYK}$ with $H_{SYK}=E_n$ and $\gamma_c=+1$, $C|n,+\rangle$ is another eigenstate of $H_{SYK},\gamma_c$ with $H_{SYK}=E_n$ and $\gamma_c=-1$, i.e., the spectrum is degenerate between $\gamma_c=\pm 1$ sectors and summarized as follows
\begin{align}
\begin{tabular}{|c|c|c|}
\hline
eigenstates of $H_{SYK}$                    &$\gamma_c$&$H_{SYK}$\\ \hline
$|n,+\rangle\quad (n=1,2,\cdots,2^{N/2-1})$ &$+1$      &$E_n$\\ \hline
$C|n,+\rangle\quad (n=1,2,\cdots,2^{N/2-1})$&$-1$      &$E_n$\\ \hline
\end{tabular}.
\end{align}
Taking into account the fact that $H_{SYK}^{(L)}|I\rangle=H_{SYK}^{(R)}|I\rangle$ together with the two-site parity $\Gamma_c|I\rangle\rangle=-|I\rangle$, we pose the following ansatz:
\begin{align}
|I\rangle=\sum_{n=1}^{2^{N/2}-1}(a_{n,+}|n,+\rangle\otimes C|n,+\rangle+a_{n,-}C|n,+\rangle\otimes |n,+\rangle),
\end{align}
where we have defined $|n,-\rangle=C|n,+\rangle$.
By imposing $\langle I|H_{int}|I\rangle=-\frac{N}{2}$ and $\langle I|I\rangle=1$ we can determine $a_{n,\sigma}$ and obtain the following expressions:
\begin{align}
|I\rangle_{q=0\text{ mod }4,N=2\text{ mod 8}}=2^{-\frac{N}{4}}\sum_{n=1}^{2^{\frac{N}{2}-1}}(
|n,+\rangle
\otimes
C|n,+\rangle
-i
C|n,+\rangle
\otimes
|n,+\rangle
),\nonumber \\
|I\rangle_{q=0\text{ mod }4,N=6\text{ mod 8}}=2^{-\frac{N}{4}}\sum_{n=1}^{2^{\frac{N}{2}-1}}(
|n,+\rangle
\otimes
C|n,+\rangle
+i
C|n,+\rangle
\otimes
|n,+\rangle
).
\end{align}

\subsubsection{$q=0\text{ mod }4$, $N=4\text{ mod }8$}
In this case $\gamma_c$ commutes with $C$, hence $|n,\sigma\rangle$ and $C|n,\sigma\rangle$ have the same eigenvalues both for $H_{SYK}$ and $\gamma_c$.
In contrast to the case $N=0\text{ mod }8$, since $C^2=-1$ it is impossible to have a state $|\phi\rangle$ as $C|\phi\rangle=|\phi\rangle$.
This implies that there are two-fold degeneracy within each of $\gamma_c=\pm 1$ sector.
In summary,
\begin{align}
\begin{tabular}{|c|c|c|}
\hline
eigenstates of $H_{SYK}$                    &$\gamma_c$&$H_{SYK}$\\ \hline
$|n,+\rangle\quad (n=1,2,\cdots,2^{N/2-2})$ &$+1$      &$E_{n,+}$\\ \hline
$C|n,+\rangle\quad (n=1,2,\cdots,2^{N/2-2})$&$+1$      &$E_{n,+}$\\ \hline
$|n,-\rangle\quad (n=1,2,\cdots,2^{N/2-2})$ &$-1$      &$E_{n,-}$\\ \hline
$C|n,-\rangle\quad (n=1,2,\cdots,2^{N/2-2})$&$-1$      &$E_{n,-}$\\ \hline
\end{tabular}.
\end{align}
Taking into account $H_{SYK}^{(L)}|I\rangle=H_{SYK}^{(R)}|I\rangle$ and $\Gamma_c|I\rangle=|I\rangle$, we pose the following ansatz for $|I\rangle$:
\begin{align}
&|I\rangle\nonumber \\
&=\sum_{n=1}^{2^{N/2}-2}\sum_{\sigma=\pm}
(a_{n,\sigma}|n,\sigma\rangle\otimes |n,\sigma\rangle
+b_{n,\sigma}|n,\sigma\rangle\otimes C|n,\sigma\rangle
+c_{n,\sigma}C|n,\sigma\rangle\otimes |n,\sigma\rangle
+d_{n,\sigma}C|n,\sigma\rangle\otimes C|n,\sigma\rangle).
\end{align}
By imposing $\langle I|H_{int}|I\rangle=-\frac{N}{2}$ and $\langle I|I\rangle=1$ we can determine $a_{n,\sigma},b_{n,\sigma},c_{n,\sigma},d_{n,\sigma}$ and obtain the following expression:
\begin{align}
&|I\rangle_{q=0\text{ mod }4,N=4\text{ mod 8}}\nonumber \\
&=2^{-\frac{N}{4}}\sum_{n=1}^{2^{\frac{N}{2}-2}}\Bigl(
|n,+\rangle
\otimes
C|n,+\rangle
-C|n,+\rangle
\otimes
|n,+\rangle
-i(|n,-\rangle\otimes C|n,-\rangle
-C|n,-\rangle\otimes |n,-\rangle)
\Bigr).
\end{align}

\subsection{$q=2\text{ mod }4$}
For $q=2\text{ mod }4$, $H_{SYK}^{(L)}$ and $H_{SYK}^{(R)}$ with $J^{(L)}_{i_1\cdots i_q}=J^{(R)}_{i_1\cdots i_q}$ are written in the basis \eqref{GammaiLGammaiR} as
\begin{align}
H_{SYK}^{(L)}=H_{SYK}\otimes 1,\quad
H_{SYK}^{(R)}=-1\otimes H_{SYK},
\end{align}
with
\begin{align}
H_{SYK}=\frac{1}{2^{\frac{q}{2}}}i^{\frac{q}{2}}\sum_{i_1<\cdots <i_q}J^{(L)}_{i_1\cdots i_q}\gamma_{i_1}\cdots \gamma_{i_q}.
\end{align}
For $q=2\text{ mod }4$, $H_{SYK}$ does not commute with $C$ due to the factor $i^{\frac{q}{2}}$.
Nevertheless, since $H_{SYK}$ anti-commutes with $C$, an eigenstate of $H_{SYK}$ with eigenvalue $E_n$ transforms to another eigenstate of $H_{SYK}$ with eigenvalue $-E_n$ and $C$ is still useful to classify the eigenstates of $H_{SYK}$.

\subsubsection{$q=2\text{ mod }4$, $N=0,4\text{ mod }8$}
Since $\gamma_c$ and $C$ commutes, we obtain the following classififcation of the spectrum of single $H_{SYK}$:
\begin{align}
\begin{tabular}{|c|c|c|}
\hline
eigenstates of $H_{SYK}$                    &$\gamma_c$&$H_{SYK}$\\ \hline
$|n,+\rangle\quad (n=1,2,\cdots,2^{N/2-2})$ &$+1$      &$E_{n,+}$\\ \hline
$C|n,+\rangle\quad (n=1,2,\cdots,2^{N/2-2})$&$+1$      &$-E_{n,+}$\\ \hline
$|n,-\rangle\quad (n=1,2,\cdots,2^{N/2-2})$ &$-1$      &$E_{n,-}$\\ \hline
$C|n,-\rangle\quad (n=1,2,\cdots,2^{N/2-2})$&$-1$      &$-E_{n,-}$\\ \hline
\end{tabular}.
\label{classificationq6N04mod8}
\end{align}
There are no degeneracy for generic $J^{(L)}_{i_1\cdots i_q}$.

To write down an ansatz for $|I\rangle$ notice that $i\Gamma_i^{(L)}\Gamma_i^{(R)}|I\rangle=|I\rangle$ implies $(H_{SYK}\otimes 1)|I\rangle=-(1\otimes H_{SYK})|I\rangle$ for $q=2\text{ mod }4$.
Hence we need to pair an energy eigenstate of single site with $H_{SYK}=E$ and a different energy eigenstate with $H_{SYK}=-E$.
Taking this into account together with the classification \eqref{classificationq6N04mod8} we pose the following ansatz:
\begin{align}
|I\rangle=\sum_{n=1}^{2^{\frac{N}{2}-2}}\sum_\sigma (a_{n,\sigma}|n,+\rangle\otimes C|n,\sigma\rangle
+b_{n,\sigma}C|n,\sigma\rangle\otimes |n,\sigma\rangle).
\end{align}
Now we can determine the coefficients $a_{m,\sigma},b_{m,\sigma}$ by completely the same strategy as we have used for $q=4$, and we obtain
\begin{align}
&|I\rangle_{q=2\text{ mod }4,N=0\text{ mod 8}}\nonumber \\
&=2^{-\frac{N}{4}}\sum_{n=1}^{2^{\frac{N}{2}-2}}\Bigl(
|n,+\rangle
\otimes
C|n,+\rangle
+C|n,+\rangle
\otimes
|n,+\rangle
-i(
|n,-\rangle
\otimes
C|n,-\rangle
+C|n,-\rangle
\otimes
|n,-\rangle
)\Bigr),\nonumber \\
&|I\rangle_{q=2\text{ mod }4,N=4\text{ mod 8}}\nonumber \\
&=2^{-\frac{N}{4}}\sum_{n=1}^{2^{\frac{N}{2}-2}}\Bigl(
|n,+\rangle
\otimes
C|n,+\rangle
-C|n,+\rangle
\otimes
|n,+\rangle
-i(
|n,-\rangle
\otimes
C|n,-\rangle
-C|n,-\rangle
\otimes
|n,-\rangle
)\Bigr).
\end{align}

\subsubsection{$q=2\text{ mod }4$, $N=2,6\text{ mod }8$}
Since $\gamma_c$ and $C$ anti-commutes, we obtain the following classififcation of the spectrum of single $H_{SYK}$:
\begin{align}
\begin{tabular}{|c|c|c|}
\hline
eigenstates of $H_{SYK}$                    &$\gamma_c$&$H_{SYK}$\\ \hline
$|n,+\rangle\quad (n=1,2,\cdots,2^{N/2-1})$ &$+1$      &$E_n$\\ \hline
$C|n,+\rangle\quad (n=1,2,\cdots,2^{N/2-1})$&$-1$      &$-E_n$\\ \hline
\end{tabular}.
\end{align}
There are no degeneracy for generic $J^{(L)}_{i_1\cdots i_q}$.

By posing the following ansatz
\begin{align}
|I\rangle=\sum_{n=1}^{2^{\frac{N}{2}-1}}(a_{n,+}|n,+\rangle\otimes C|n,+\rangle
+a_{n,-}C|n,+\rangle\otimes |n,+\rangle),
\end{align}
we can obtain $|I\rangle$ as
\begin{align}
|I\rangle_{q=2\text{ mod }4,N=2\text{ mod 8}}=2^{-\frac{N}{4}}\sum_{n=1}^{2^{\frac{N}{2}-1}}(
|n,+\rangle
\otimes
C|n,+\rangle
-i
C|n,+\rangle
\otimes
|n,+\rangle
),\nonumber \\
|I\rangle_{q=2\text{ mod }4,N=6\text{ mod 8}}=2^{-\frac{N}{4}}\sum_{n=1}^{2^{\frac{N}{2}-1}}(
|n,+\rangle
\otimes
C|n,+\rangle
+i
C|n,+\rangle
\otimes
|n,+\rangle
).
\end{align}

\bibliography{JLJRindepbunken_220103Nosaka.bib}
\end{document}